\documentclass[a4paper,11pt]{article}
\pdfoutput=1 

\usepackage{jheppub}
\usepackage{color} 
\usepackage{float}
\usepackage{ulem}

\usepackage{amsmath,graphicx,amssymb,subfigure,bbm}
\usepackage[all]{xy}

\def\beq{\begin{equation}}
\def\eeq{\end{equation}}
\def\beqa{\begin{eqnarray}}
\def\eeqa{\end{eqnarray}}

\begin{document}

\title{Magnetic catalysis and the chiral condensate in holographic QCD}

\author[a]{Alfonso Ballon-Bayona,}
\author[b,c]{Jonathan P. Shock}
\author[d,e]{and Dimitrios Zoakos}
\affiliation[a]{Instituto de F\'{i}sica, Universidade
Federal do Rio de Janeiro, \\
Caixa Postal 68528, RJ 21941-972, Brazil.}       
\affiliation[b]{Department of Mathematics and Applied Mathematics,
University of Cape Town, \\
Private Bag, Rondebosch 7700, South Africa.}
\affiliation[c]{National Institute for Theoretical Physics,
Private Bag X1, Matieland, South Africa.}
\affiliation[d]{Department of Physics, National and Kapodistrian University of Athens, 15784 Athens, Greece.}
\affiliation[e]{Hellenic American University, 436 Amherst st, Nashua, NH 03063 USA}

\emailAdd{aballonb@if.ufrj.br}
\emailAdd{jonathan.shock@uct.ac.za}
\emailAdd{zoakos@gmail.com}

\abstract{We investigate the effect of a non-zero magnetic field on the chiral condensate using a holographic QCD approach. We extend the model proposed by Iatrakis, Kiritsis and Paredes in \cite{Iatrakis:2010jb} that realises chiral symmetry breaking dynamically from 5d tachyon condensation. We calculate the chiral condensate, magnetisation and susceptibilities for the confined and deconfined phases. The model leads, in the probe approximation, to magnetic catalysis of chiral symmetry breaking in both confined and deconfined phases. In the chiral limit, $m_q=0$, we find that in the deconfined phase a sufficiently strong magnetic field leads to a second order phase transition from the chirally restored phase to a chirally broken phase. The transition becomes a crossover as the quark mass increases. Due to a scaling in the temperature, the chiral transition will also be interpreted as a transition in the temperature for fixed magnetic field. We elaborate on the relationship between the chiral condensate, magnetisation and the (magnetic) free energy density. We compare our results at low and moderate temperatures with lattice QCD results.}

\maketitle

\flushbottom

\section{Introduction}

Despite our best efforts, many phenomena of strongly coupled field theories remain enigmatic. While we have understood the fundamental building blocks of QCD for some six decades and with the most complex machines in the world at our disposal, confinement, chiral symmetry breaking, phenomenology of the strongly coupled quark gluon plasma, and more remain outside the grasp of a complete mathematical description. To be able to map out the phase diagram of QCD from first principles is a  holy grail of quantum field theory. Despite the fact that we have yet to crack these issues, there are tools at our disposal which have given us key insights and allowed us to answer questions about some of these phenomena in interesting ways. Lattice QCD has lead the way for many decades, and with increasing computational tools, both algorithmic and hardware, there are surely interesting times ahead for this approach. 
Heavy Quark Effective Theory \cite{Manohar:2000dt}, chiral perturbation theory \cite{Scherer:2012xha} and the Schwinger-Dyson equations \cite{Roberts:1994dr} 
are other powerful methods which give a window into certain parameter regions of QCD.

The gauge/gravity duality, based on the AdS/CFT correspondence \cite{Maldacena:1997re}, has been the other major branch in understanding strongly
coupled quantum field theories (for a set of pedagogical introductions see \cite{Erdmenger:2007cm,CasalderreySolana:2011us,Ramallo:2013bua, Edelstein:2009iv}). While we are a long way from having a gravity 
dual of QCD it is clear that certain QCD-like phenomena do show up in simple and elegant gravity duals of 
less-realistic field theories. Meson spectra, chiral symmetry breaking, confinement/deconfinement phase transitions 
and more are all accessible in such models. 

To create the most realistic QCD gravity dual is clearly of the most important goals of the gauge/gravity duality, and so any step in this direction is worth pursuing. In the top-down approach there have been important advances, such as the breaking of supersymmetry \cite{Klebanov:2000hb,Maldacena:2000mw, Maldacena:2000yy,Witten:1998zw}, 
the addition of fundamental matter \cite{Karch:2002sh},\footnote{A pedagogical review on the addition of unquenched flavour 
in string theory is in \cite{Nunez:2010sf}, while more solutions appear in 
\cite{Ramallo:2008ew, Arean:2010hu, Jokela:2012dw, Itsios:2013uya, Filev:2014nza, Bea:2014yda}.} 
the addition of chemical potential \cite{Mateos:2007vc} and 
external magnetic field \cite{Filev:2007gb,Erdmenger:2007bn, Albash:2007bk,DHoker:2009mmn}. Those advances have provided us with models which mimic QCD-like behaviour.

 Regarding chiral symmetry breaking, in QCD we know that there are at least two effects present. In the massless case, the Lagrangian is  chirally symmetric at high energies but the vacuum breaks the chiral symmetry spontaneously at low energies. The other effect is the explicit chiral symmetry breaking due to the presence of massive quarks. 

Chiral symmetry breaking has been one of the effects that the gauge/gravity duality has been able to model for some time. In the top-down approach, this problem has been considered in the  Klebanov-Strassler \cite{Klebanov:2000hb} and Maldacena-Nunez \cite{Maldacena:2000mw, Maldacena:2000yy}, 
the dilaton-flow geometry of Constable-Myers \cite{Constable:1999ch,Babington:2003vm,Evans:2004ia} as well as the D3/D7 \cite{Karch:2002sh} and D4/D6 \cite{Kruczenski:2003uq} brane models. A model that stands out is the Sakai-Sugimoto model \cite{Sakai:2004cn, Sakai:2005yt}, that describes the breaking of a $U(N_f) \times U(N_f)$ chiral symmetry (through the addition of $N_f$ pairs of D8-$\rm \overline{D8}$-branes on the non-extremal Witten D4-brane background \cite{Witten:1998zw}). This is considered the closest holographic model for QCD  and has been studied in great detail and in many regimes over the past years. An alternative geometrical realisation of chiral symmetry breaking was introduced in \cite{Kuperstein:2008cq} by Kuperstein and Sonnenschein, through the addition of a pair of D7-$\rm \overline{D7}$-branes on the conifold geometry \cite{Klebanov:1998hh}\footnote{A generalisation of the two aforementioned models to the case of a (2+1)-dimensional gauge theory of strongly coupled fermions, was proposed in \cite{Filev:2013vka, Filev:2014bna}, via the introductin of pair of D5-$\rm {\bar D}$5-branes on the conifold.}.

In addition to the geometrical realisations described above there is an alternative holographic description of chiral symmetry breaking in terms of open string tachyon condensation developed in \cite{Casero:2007ae,Iatrakis:2010jb,Iatrakis:2010zf}. This is the approach that we will follow in this paper and will be described in detail in the next sections.

Although the top-down approach has given us a wealth of information about what sorts of backgrounds give rise to which phenomena,  it is derisable to build bottom up models which are five dimensional models that have a small set of ingredients necessary to describe nonperturbative QCD dynamics. The archetypical bottom-up constructions that incorporate chiral symmetry breaking and mesonic physics are the hard-wall model \cite{Erlich:2005qh, DaRold:2005mxj}, inspired by the Polchinski-Strassler background \cite{Polchinski:2001tt}, and the soft-wall model \cite{Karch:2006pv}\footnote{For nonlinear extensions of the soft wall model, see \cite{Gherghetta:2009ac,Chelabi:2015gpc,Ballon-Bayona:2020qpq}.}.

A more sophisticated model that captures the dynamics of QCD more accurately  is Veneziano-QCD (V-QCD)  \cite{Jarvinen:2011qe,Arean:2013tja} (see also \cite{Jarvinen:2015ofa}). It combines the model of improved holographic QCD (IHQCD) \cite{Gursoy:2007cb,Gursoy:2007er,Gursoy:2009jd,Gursoy:2010fj} for the gluon sector and a tachyonic Dirac-Born-Infeld (DBI) action proposed by Sen for the quark sector \cite{Sen:2003tm,Bigazzi:2005md}. In order for the model to match the predictions of QCD phenomenology one has to adopt a bottom-up approach. The action is generalised to one which contains several freely-defined functions. The form of those functions is chosen in a way that qualitative QCD features are reproduced and are fitted
using lattice and experimental data. In the IHQCD case this has been considered in  \cite{Gursoy:2010fj}, while in the full 
V-QCD case the detailed comparison was initiated in \cite{Arean:2013tja,Jokela:2018ers}. 

In QCD, the presence of very strong magnetic fields $e B > \Lambda^2_{QCD}$ triggers a plethora of interesting 
phenomena, among which are the Magnetic Catalysis (MC) (see e.g. \cite{Gusynin:1994re, Gusynin:1994xp}) 
and the Inverse Magnetic Catalysis (IMC)  of chiral symmetry breaking (see e.g. \cite{Bali:2011qj, Bali:2012zg, DElia:2012ems}). 
Magnetic fields with a magnitude of $e B / \Lambda^2_{QCD} \sim 5 -10$ are realised during non-central heavy
ion collisions. Even though the magnetic field strength decays rapidly after the collision, it remains very strong 
when the quark-gluon plasma (QGP) initially forms. As a consequence it affects the plasma evolution and 
the subsequent production of charged hadrons \cite{Gursoy:2014aka}. 

{\it Magnetic catalysis} is the phenomenon by which a magnetic field favours chiral symmetry breaking. This phenomenon is characterised by the enhancement of the chiral condensate and in QCD occurs at low temperatures. The physical (perturbative) mechanism behind this is that the strong magnetic field reduces the effective dynamics from (3+1) to (1+1) dimensions, since the motion of the charged particles are restricted to the lowest Landau level. As a consequence of the magnetic field,  the lowest Landau level is degenerate and that leads to an enhancement of the Dirac spectral density. This leads, via the Banks-Casher relation \cite{Banks:1979yr}, to an enhancement of the chiral condensate (magnetic catalysis).  

{\it Inverse Magnetic catalysis} is the phenomenon by which
a magnetic field disfavours chiral symmetry breaking and it is characterised by a quark condensate that decreases in the presence of a strong magnetic field. In QCD this occurs at temperatures approximately higher than $150$ MeV. IMC is a non-perturbative effect and the current understanding
is that it originates from strong coupling dynamics around the deconfinement temperature. 
A promising explanation, coming from a lattice perspective, is that IMC is due to a competition between 
{\it valence} and {\it sea} quarks in the  path integral \cite{Bruckmann:2013oba}. The 
{\it valence}  contribution is through the quark operators inside the path integral (i.e. the trace of the inverse of the  Dirac operator). The magnetic field catalyses the condensate, since it increases the spectral density of the zero energy mode of the Dirac operator.
The  {\it sea} contribution is through the quark determinant, which is responsible for the fluctuations around the 
gluon path integral.  The dependence of the determinant on $B$ and $T$ is intricate and the net result is a suppression 
of the condensate close to the deconfinement temperature. 

For recent reviews on magnetic catalysis and inverse magnetic catalysis see e.g. \cite{Andersen:2014xxa,Miransky:2015ava,Bandyopadhyay:2020zte}. Inverse magnetic catalysis appears also at finite chemical potential where there is a competition between the energy cost of producing quark antiquark pairs and the energy gain due to the chiral condensate. For a nice review of IMC at finite chemical potential see \cite{Preis:2012fh}. 

There have been several attempts to address MC and IMC in a holographic framework, including \cite{Johnson:2008vna,Filev:2009xp,Preis:2010cq,  Erdmenger:2011bw, Filev:2011mt,Ballon-Bayona:2013cta, Jokela:2013qya,Mamo:2015dea,Dudal:2015wfn,Rougemont:2015oea,Evans:2016jzo,Gursoy:2016ofp,Fang:2016cnt, Ballon-Bayona:2017dvv,Gursoy:2017wzz,Rodrigues:2017iqi,Giataganas:2017koz, Gursoy:2018ydr,Filev:2019bll,Bohra:2019ebj,He:2020fdi}. 
Here we single out the approach that was put forward in \cite{Gursoy:2016ofp}, since it allows for a consistent description of the chiral condensate, based on the VQCD approach \cite{Jarvinen:2011qe,Arean:2013tja}. They proposed a holographic model that makes manifest the competition between the valence and sea quark contributions to the chiral condensate. In that 
framework the role of the {\it valence} contribution is played by the tachyon (the bulk field dual to the quark bilinear operator), 
while the role of the {\it sea} contribution comes from the backreaction of the magnetic field on the background, the latter relevant for IMC in this scenario. Another interesting proposal was presented in \cite{Giataganas:2017koz, Gursoy:2018ydr}, suggesting 
that the cause of IMC is the anisotropy induced by the magnetic field, rather than the charge dynamics that it creates.\footnote{Anisotropic backgrounds can also be realised in traditional 
top-down holography (for a non-exhaustive list we mention the following articles 
\cite{Mateos:2011tv, Ammon:2012qs, Conde:2016hbg, Penin:2017lqt,Jokela:2019tsb}).} Lastly, when it comes to distinguishing MC from IMC, besides the chiral condensate, it was realised in \cite{Ballon-Bayona:2017dvv,Gursoy:2017wzz} that the magnetisation plays a very important role.

In this paper we will extend the model of \cite{Iatrakis:2010jb,Iatrakis:2010zf} to investigate the effect of a nonzero magnetic field on the chiral condensate. Although the model in \cite{Iatrakis:2010jb,Iatrakis:2010zf} is less realistic than constructions such as VQCD \cite{Jarvinen:2011qe,Arean:2013tja}, it has the privilege of simplicity. There are fewer parameters to fix with the lattice computations and the potentials of the tachyon action are predetermined. We will arrive at a model that includes all of the  necessary ingredients for describing the chiral condensate in the presence of a magnetic field. Besides the chiral condensate, the model allows for a consistent description of the magnetisation and provides a very interesting holographic description of magnetic catalysis. 

We present below a summary of the main results obtained in this paper and the comparison to lattice QCD.

\medskip 

{\bf Summary of the main results}

\begin{itemize}

    \item The model provides a robust description of magnetic catalysis, where the main effect of the magnetic field is to increase the chiral condensate and therefore catalyse the breaking of chiral symmetry. This effect is present in the confined and deconfined phases regardless the value of the quark  mass. 
    
    \item In the confined phase the chiral condensate $\langle \bar q q \rangle$ is a function of the quark mass $m_q$ and the magnetic field $B$, both given in units of the confinement scale $M_{KK}$.  For any fixed value of the quark mass, the chiral condensate is always a growing function of the magnetic field, signifying magnetic catalysis. In the regime of large $B$ we find that the chiral condensate grows as $B^{3/2}$.
    
     \item In the deconfined phase we find that the chiral condensate takes the form $\langle \bar q q \rangle \propto T^3 c_3 (c_1 , {\cal B})$ where $c_3$ is a (dimensionless) function of the dimensionless quark mass $c_1 \propto m_q/T$ and the dimensionless magnetic field ${\cal B} \propto B/T^2$. This allowed us to reinterpret magnetic catalysis as the enhancement of the chiral condensate at low temperatures due to the presence of a finite magnetic field.   In the chiral limit (zero quark mass) we find that a finite magnetic field leads to a second order transition from the chirally symmetric phase (high temperatures) to a chirally broken phase (low temperatures). The chiral transition becomes a crossover as the quark mass increases.  We find that the dimensionless chiral condensate $c_3$ grows as ${\cal B}^{3/2}$ in the large ${\cal B}$ regime, regardless the value of the dimensionless quark mass $c_1$. This in turn implies that, at fixed $B$, the chiral condensate reaches a constant value in the limit of zero temperature. 
     
     \item We find that the magnetisation is always an increasing function of the magnetic field in both confined and deconfined phases. We find, however, that for small magnetic fields, the transition from a confined to a deconfined phase implies a transition from a diamagnetic to a paramagnetic behaviour. In the chiral limit (zero quark mass) we find that the susceptibility is discontinuous at the critical value of the magnetic field (or critical temperature) where the second order chiral transition takes place. 
     
     \end{itemize}
     \medskip
 {\bf Highlights of the comparison to lattice QCD}   
\medskip

 In section \ref{sec:Lattice} we provide a detailed comparison of the model results against lattice QCD.  We consider subtracted quantities in order to avoid any scheme dependence. Here we briefly describe the main similarities and differences.
 
 \begin{itemize}
     
\item For the chiral condensate we compare the subtracted chiral condensate $\Delta \langle \bar q q \rangle \equiv  \langle \bar q q (B,T) \rangle - \langle \bar q q (0,T) \rangle$ against lattice QCD results.  We find a reasonable quantitative agreement in the regime of low temperatures or large magnetic fields (see Fig. \ref{Fig:subtrcondvsT}). In particular, we conclude that both the confined and deconfined phase of the IKP model provide a good description of the chiral condensate at very low temperatures, consistent with magnetic catalysis. On the other hand, we conclude that only the deconfined phase provides a description of the chiral transition in the presence of a magnetic field. 

\item In the regime of high temperatures and moderate magnetic fields, however, we find a large discrepancy between our results and lattice QCD results (see  figures \ref{Fig:subtrcondvsT} and \ref{Fig:DeltaSigmavsb}). In particular, we never reproduce the phenomenon of inverse magnetic catalysis, where the  chiral condensate becomes a decreasing function of the magnetic field. This discrepancy has to do with the absence of anisotropy effects in our model. Going beyond the probe approximation in order to incorporate  backreaction effects should  provide for a description of anisotropy and the transition from magnetic catalysis to inverse magnetic catalysis in the high temperature regime. In the regime of small magnetic fields and moderate temperatures, we also find a discrepancy between our results for the deconfined phase and the lattice QCD results. This discrepancy occurs because of the absence of a dynamical scale similar to $\Lambda_{QCD}$ in the deconfined phase.  

\item We also consider the RG invariant product of the quark mass and subtracted chiral condensate $m_q \Delta \langle \bar q q \rangle $. We find for the confined and deconfined phase that in the regime of small magnetic fields, the leading $B^2$ dependence in $m_q \Delta \langle \bar q q \rangle $ has a coefficient that varies with the quark mass in a way qualitatively similar to that found in lattice QCD (see Fig. \ref{Fig:B2Coeffvsc1}).

\item Lastly, we compare the subtracted magnetisation $\Delta \mathbb{M} = \mathbb{M}(B,T) - \mathbb{M}(B,T_0)$, with $T_0$ a fixed reference temperature, against lattice QCD results. Unfortunately, the available data from lattice QCD does not include the regime of low temperatures where our model is expected to provide a good description. As expected, we find a quantitative discrepancy between our results and lattice QCD results in the high temperature regime (see the right panel of Fig. \ref{Fig:MagvsbLattice}).
However, the lattice QCD results reveals a change in the sign of  $\Delta \mathbb{M}$, possibly related to the competition between magnetic catalysis and inverse magnetic catalysis. Our model always leads to a $\Delta \mathbb{M}<0$ and we associate this result with magnetic catalysis. 

\end{itemize}

The outline of the paper is as follows. In section \ref{section-2}, we review the holographic approach introduced in \cite{Casero:2007ae} to describe  the dynamics of chiral symmetry breaking. This includes a specific action for the tachyon field and 
the implementation of a confinement criterion. Moreover, we describe the specific gravity setup where those ideas 
were materialised \cite{Iatrakis:2010zf, Iatrakis:2010jb}. In section \ref{section-3}, we  extend the model of \cite{Iatrakis:2010zf, Iatrakis:2010jb} to describe the effects of the addition of an external magnetic field on the tachyon dynamics. We study in detail the equation of motion for the 
tachyon in the confined and the deconfined phases and the dynamical breaking of chiral symmetry. 
In section \ref{section-4}, we calculate and analyse the chiral condensate and the magnetisation.
The magnetic catalysis phenomenon is a common feature for both phases. 
In the deconfined phase at zero quark mass and for a sufficiently strong magnetic field, there is a second order phase transition from the chirally restored phase to a chirally broken phase, signifying the spontaneous breaking of chiral symmetry above a critical value. Going away from the massless limit, the chiral transition becomes a crossover. 
 We finish the paper presenting in section \ref{sec:Lattice}  a quantitative comparison of the gravity dual predictions with computations from lattice QCD at finite temperature. The main text is supplemented with two appendices. In appendix \ref{WZterm} we describe the 
Wess-Zumino (WZ) term of the tachyonic action. In appendix \ref{appendix-2} we perform the detailed IR asymptotic analysis
for the equation of motion of the tachyon. While the analysis in the deconfined case  is 
a straightforward generalisation  of \cite{Iatrakis:2010jb}, in the confined case  the IR divergence 
emerges in a systematic and non-trivial way.



\section{The setup}
\label{section-2}

In \cite{Casero:2007ae} a holographic picture was proposed which describes the dynamics of chiral symmetry breaking by open string tachyon condensation in the gravity  side. In \cite{Iatrakis:2010zf, Iatrakis:2010jb} a particular setup was developed which allows for a quantitative description of chiral symmetry breaking, and in this section we review these models in detail.

The setup proposed in \cite{Casero:2007ae} consists of a system of $N_f$ coincident D-brane anti-D-brane pairs
in a gravitational background generated by a stack of colour branes. 
This framework is an extension of the Dirac-Born-Infeld (DBI) plus Wess-Zumino (WZ) actions, which takes into account the effects of open string tachyon condensation \cite{Sen:2003tm,Garousi:2004rd}.

The tachyonic mode, $\tau$, is an open string complex scalar, which is in the spectrum of open strings stretching between the brane-antibrane pairs, and transforms in the bifundamental representation of the $U(N_f)_L \times U(N_f)_R$ flavour group. 
More specifically, $\tau$ transforms in the antifundamental of $U(N_f)_L$ and in the fundamental of $U(N_f)_R$ and vice versa for $\tau^{\dagger}$. Fixing the mass term for  $\tau$ appropriately, it naturally couples to the 4d quark bilinear operator $\bar q q$ at the boundary. Then the 4d breaking of the global chiral symmetry  is mapped to a 5d Higgs-like breaking of gauge symmetry triggered by $\tau$, as realised in \cite{Erlich:2005qh,DaRold:2005mxj}. In QCD, spontaneous breaking of chiral symmetry is associated with a nonzero vev for the quark bilinear operator. In the holographic setup of \cite{Casero:2007ae}, this is realised via a nontrivial IR behaviour for $\tau$ generated dynamically. The model also describes the explicit breaking of chiral symmetry associated with having a nonzero mass term $m_q \langle \bar q q \rangle$ in the 4d theory.  

In the case of massless QCD the global chiral symmetry  $U(N_f)_L \times U(N_f)_R$ is preserved in the UV and spontaneously broken at low energies to the diagonal subgroup $U(N_f)_V$. In the holographic setup of \cite{Casero:2007ae}, this corresponds to a vanishing tachyon at the boundary that grows as it moves away from the boundary and becomes infinite at the end of space. This process can be thought as a recombination of the brane-antibrane pair.   

In the next subsections we will describe the tachyon plus DBI and WZ actions of the above model. The physics of the DBI part yields the vacuum configuration and excitations thereon and
the WZ part is related to global anomalies and to a holographic realisation of the Coleman-Witten theorem  
\cite{Coleman:1980mx}.


\subsection{The Tachyon-DBI action} 
\label{tachyon-DBI}

The general construction consists of a system of $N_f$ overlapping pairs of  Dq-$\rm \overline{Dq}$ flavour branes in a fixed curved spacetime generated by a set of $N_c$ $Dp$ colour  branes. We will be particularly interested in the case $p=q=4$ where the colour branes generate the asymptotic $AdS_6$ cigar geometry \cite{Kuperstein:2004yf} and the flavour brane anti-branes are 5d defects associated with quark degrees of freedom \cite{Iatrakis:2010zf, Iatrakis:2010jb} .
For simplicity we focus on the Abelian case $N_f=1$, corresponding to a single pair of D4-$\rm \overline{D4}$ branes. The corresponding  DBI action can be written as (\cite{Casero:2007ae}, see also  \cite{Sen:2003tm,Garousi:2004rd})
\begin{equation}
S_{DBI} = - \int d^5 x \,  V(\tau,\tau^*) \left [ \sqrt{-E^L} + \sqrt{-E^R} \right ] = \int d^5 x \, {\cal L}_{DBI} \, , 
\label{SDBI}
\end{equation}
where
\begin{equation}
E^{L/R} \, = \, \det (E_{mn}^{L/R}) \quad \text{with} \quad
E_{mn}^{L/R} \, = \, \tilde g_{mn} + \beta \, F_{mn}^{L/R}
\end{equation}
and we have defined\footnote{The expression for the tachyon-DBI action includes also the transverse scalars that live 
on the flavour branes. As mentioned in \cite{Casero:2007ae}, these modes (that appear in a critical string theory setup) 
are ignored in a holographic QCD analysis, since they do not have an obvious QCD interpretation.}
\begin{equation}
\tilde g_{mn} \, = \, g_{mn} + h_{mn} \quad \& \quad 
F^{L/R} \, = \, dA^{L/R}  \, . 
\end{equation}
The symmetric tensor $g_{mn}$ denotes the (five-dimensional) world-volume metric whereas the antisymmetric tensors $F_{mn}^{L/R}$ denote the field strengths associated with the Abelian gauge fields $A_m^{L/R}$. The symmetric tensor $h_{mn}$ 
describes the dynamics of a complex scalar field $\tau$ (the tachyon)
\beqa
h_{mn} = \kappa \Big[ \left(D_m \tau\right)^*\left(D_n \tau\right)  + \left(D_m \tau\right)\left(D_n \tau\right)^*  \Big] \, . 
\eeqa
The covariant derivative is the one associated with a bifundamental field, i.e. $D_m \tau = \partial_m \tau + i \, a_m \tau$, where $a_m = A_m^L - A_m^R$ is the corresponding axial gauge field.  
The explicit form of $h_{mn}$ is then given by
\beqa
h_{mn} = 2 \kappa \Big[ \partial_{(m} \tau^* \partial_{n)}\tau + j_{(m} a_{n)} + a_m a_n \tau^* \tau   \Big] \,, 
\eeqa
where we have introduced the Abelian current $j_m = i  \tau \overleftrightarrow{\partial_m} \tau^*$ and we are using the symmetric tensor notation $X_{(mn)}= ( X_{mn} + X_{nm})/2$ . 
For the tachyon potential, we consider the Gaussian form 
\beqa \label{tachyon-potential}
V(\tau,\tau^*) = V_0 \exp \left [ - \frac {m_\tau^2}{2} \, \tau^* \tau \right ] \,. 
\eeqa
This form was proposed in \cite{Casero:2007ae}, inspired by the computation in flat space that was derived in boundary 
string field theory \cite{Kutasov:2000aq,Minahan:2000tf}. Remarkably, this potential leads to linear Regge 
trajectories for the mesons \cite{Casero:2007ae}, something which is otherwise hard to model \footnote{The only other approach that leads to linear linear Regge  trajectories for the mesons is the soft wall model, based on the IR constraint for the dilaton field \cite{Karch:2006pv}.  }. 

The parameters that we use in this paper are related to those defined in \cite{Iatrakis:2010jb} by
\beqa
\beta = \frac{2 \pi \alpha'}{g_V^2}\, ,  \quad
\kappa = \pi \alpha' \lambda \, , \quad
V_0 = {\cal K} \quad \& \quad m_\tau = \mu \,.   \label{dictionary}
\eeqa
The square roots in \eqref{SDBI} can be written as 
\begin{equation} \label{rootE-GenQ}
\sqrt{-E^{L (R)}} = \sqrt{- \tilde g}  \sqrt{Q^{L(R)}} \quad {\rm with} \quad
Q^{L(R)} = 1 + \frac{\beta^2}{2!} F^{mn}_{L(R)} F_{mn}^{L(R)} + \frac{\beta^4}{4!} F^{mnpq}_{L(R)} F_{mnpq}^{L(R)}
\end{equation}
and we have introduced the totally antisymmetric 4-tensor 
\beqa
F_{mnpq} = F_{mn} F_{pq} - F_{mp} F_{nq} + F_{mq} F_{np} \,. 
\eeqa
The upper indices in \eqref{rootE-GenQ} are raised using the effective metric $\tilde g_{mn}$\footnote{For more details on the derivation of the relations \eqref{rootE-GenQ} see \cite{BallonBayona:2013gx}.}.
For the analysis of the following subsection (and also in \cite{Casero:2007ae}), 
we consider a five-dimensional metric of the form
\beqa  \label{metric}
g_{mn} = {\rm diag} \left (g_{zz}(z), g_{tt}(z), g_{xx}(z), g_{xx}(z), g_{xx}(z) \right )\,.
\eeqa
This metric preserves an $SO(3)$ symmetry and the components depend solely on the radial coordinate $z$, 
as expected for a holographic QCD background. 


\subsection{Confinement criterion for dynamical chiral symmetry breaking}

In this subsection we will describe the connection between confinement and the singular behaviour of the tachyon profile 
in the IR of the geometry, which was investigated in \cite{Casero:2007ae}. Since, at leading order, it is consistent to set
the  gauge fields to zero, the equation of motion for the tachyon comes from considering solely the DBI part 
of the action. 

Following the standard procedure, we set the phase of the complex tachyon to zero and arrive at a differential equation for $\tau$ that schematically looks like

\begin{equation} \label{eom-tachyon-schematic}
\partial_z^2 \tau \, + \, \#_1 \, (\partial_z \tau)^3 \, + \, \#_2 \, \partial_z \tau \, + \, 2 \, \tau \, 
\big[\#_3 \, +( \partial_z \tau)^2 \big] \, = \, 0
\end{equation}
where $\#_1$, $\#_2$ and $\#_3$ are combinations of the metric components which can be found in \cite{Casero:2007ae}.
This is a second order non-linear differential equation and the two integration constants, via the standard AdS/CFT
dictionary, can be related to the quark bare mass and condensate. This relationship is found by studying the UV behaviour of the tachyon.

Assuming that the space is asymptotically $AdS$ and the tachyon is dual to the quark bilinear ${\bar q} q$ with conformal
dimension $\Delta=3$, we arrive at the following expression for the UV limit of the tachyon profile
\begin{equation}
\tau \, = \, c_1 z  + \ldots \, + \, c_3 z^3 +  \ldots \qquad \text{(small z)}
\end{equation}
where the source coefficient $c_1$ is proportional to the quark mass $m_q$ whereas the vev coefficient $c_3$ is related to the quark condensate $\langle {\bar q} q \rangle$.
For the IR analysis of the tachyon equation \eqref{eom-tachyon-schematic} we consider the results of 
\cite{Kinar:1998vq}, according to which a sufficient condition for a gravity background to exhibit confinement is
\begin{align}
&g_{zz}(z_{div}) \rightarrow \infty \quad {\rm with} \nonumber \\
& g_{tt}(z_{div}) \ne 0 \, , \, g_{xx}(z_{div}) \ne 0 \quad \& \quad
\partial_z g_{tt}(z_{div}) < 0 \, , \, \partial_z g_{xx}(z_{div}) < 0
\end{align}
for some value of $z=z_{div}$. Identifying the point where the divergence appears with the confinement scale, 
namely $z_{IR} = z_{div}$, and assuming that the divergence of the metric component $g_{zz}$ is a simple pole 
near $z_{IR}$, we conclude that the tachyon diverges near $z_{IR}$ as follows
\begin{equation} \label{IR-tachyon-consistency}
\tau \propto \frac{1}{\big(z_{IR}- z\big)^{\alpha}}
\quad {\rm with} \quad \alpha>0 \,.
\end{equation}
The main lesson from this analysis is that the IR consistency condition for the tachyon 
\eqref{IR-tachyon-consistency} can be used to fix $c_3$ in terms of $c_1$, which is equivalent to fixing the chiral condensate $\langle {\bar q} q \rangle$ in terms of the quark mass $m_q$. As usual, this will be implemented using a shooting technique.
  
In the seminal paper of Coleman and Witten \cite{Coleman:1980mx} it was proved that in the limit $N_c \rightarrow \infty$
and for massless quarks, the chiral symmetry of QCD is spontaneously broken from
$U(N_f)_L \times U(N_f)_R$ to $U(N_f)_V$. The main message from the analysis of the current subsection 
is that for a confining theory the tachyon has to diverge in the IR of the geometry while it goes to zero in the UV limit. Since $\tau$ transforms in the bifundamental representation of the flavour group, 
$\tau \ne 0$ means that the symmetry has been broken down to $U(N_f)_V$. Therefore, the presence of confinement implies spontaneous chiral symmetry breaking, and this is therefore a 
holographic implementation of the ideas and results of  \cite{Coleman:1980mx}.

The analysis of the WZ part of the action is related to the study of anomalies of the chiral symmetry, when there is a coupling 
between flavour currents and external sources.  A gauge transformation of the WZ part of the action produces a boundary term that is matched with the global anomaly of the dual field theory.

In \cite{Casero:2007ae}, a precise computation of the gauge variation of the 5d WZ action was performed in the case of a real tachyon $\tau = \tau^*$.  The conclusion was that the result is given by a 4d boundary term.
This boundary term precisely matches the expected anomaly for the residual $U(N_f)_V$ group after imposing the appropriate boundary conditions for $\tau$. In fact, the divergent behaviour for the tachyon (arising from the confinement criterion) in the IR, cf. \eqref{IR-tachyon-consistency}, is crucial in the match to the QCD anomaly term. The authors of \cite{Casero:2007ae} interpreted this result as a holographic realisation of the Coleman-Witten theorem.

For more details on the WZ term of the tachyon action and the currents see the discussion in appendix \ref{WZterm}.


\subsection{The Iatrakis, Kiritsis, Paredes (IKP) model}

A simple holographic model of QCD that describes chiral symmetry breaking and the associated mesonic physics was proposed in \cite{Iatrakis:2010zf,Iatrakis:2010jb}, by Iatrakis, Kiritsis and Paredes (IKP). 
It is a construction that makes explicit the ideas introduced in \cite{Casero:2007ae}, namely that chiral symmetry breaking and the physics of the flavour sector is encoded in an effective  description of a brane-antibrane system with a tachyonic field.

The quarks and  antiquarks are introduced through the brane and antibrane, 
and the physics of interest comes about by condensation of the lowest lying bifundamental scalar on the open strings connecting those branes through a tachyonic instability. 
The next important step was the choice of the holographic geometry in which these ideas can be realised. 
The background should be smooth and asymptotically AdS and consistent with confinement in the IR.  
A simple choice is the $AdS_6$ soliton geometry \cite{Kuperstein:2004yf}, which is a solution of the two derivative approximation of  subcritical string theory. While the construction in \cite{Iatrakis:2010zf,Iatrakis:2010jb} is initially top-down, in order to reproduce QCD-like features, one goes beyond the limit in which the two derivative action is a controlled low energy approximation of string theory because we are in a regime where the curvature scale is of the same order as the string length. Thus, we think of this approach as an effective, bottom-up description. 

In terms of both complexity and correctly capturing the features of QCD, the IKP model stands somewhere between the hard wall model \cite{Erlich:2005qh,DaRold:2005mxj} and the VQCD approach \cite{Jarvinen:2011qe,Arean:2013tja}. The most interesting qualitative features of this approach to QCD physics, 
as summarized in \cite{Iatrakis:2010zf,Iatrakis:2010jb}, are that

\begin{itemize}
\item Towers of excitations with $J^{PC}=1^{--}, 1^{++}, 0^{-+}, 0^{++}$ are included in the model.
\item Dynamical chiral symmetry breaking is realised through tachyon condensation
\item The excited states have Regge trajectories of the form $m^2_n \sim n$.
\item The $\rho$-meson mass increases due to the increase of the pion mass.
\end{itemize}


\section{The IKP model at finite magnetic field}
\label{section-3}

The fundamental novelty of the current work is to investigate the effects of a finite external magnetic field on the dynamics already described by the IKP model. While an apparently small addition, the extended phase space is rich. In order to do this we need to study the tachyon field along with the gauge potential which leads to the magnetic field. The ansatz for these is
\begin{equation}
\tau \, = \,  \tau^* \, =  \, \tau(z) \quad \& \quad 
A^{L/R} \, = \,   \frac{B}{2} \, \Big( - x^2 dx^1 +  x^1 dx^2 \Big)  \, . \label{MagAnsatz}
\end{equation}
Under this ansatz, the field strengths take the form $F^{L/R} =  B \, dx^1 \wedge dx^2$ \,.


\subsection{The Euler-Lagrange equation for $\tau$} 

We work with the diagonal metric $\eqref{metric}$ and one can show that the ansatz \eqref{MagAnsatz} leads to a diagonal tensor $h_{mn}$ and that the 
effective metric $\tilde g_{mn}$ takes the form
\begin{equation} \label{eff-metric}
\tilde g_{mn} = {\rm diag} \Big( g_{zz} \Theta_z , g_{tt} , g_{xx}, g_{xx}, g_{xx} \Big) \, , 
\end{equation}
where we have introduced the function
\begin{equation}
\Theta_z = 1 + 2 \kappa \, g^{zz} (\partial_z \tau)^2 \,, \label{blobs}
\end{equation}
which dresses one component of the  metric. 
Note, in particular, that the square root of the determinant of \eqref{eff-metric} can be written as
\begin{equation}
\sqrt{- \tilde g} = \sqrt{-g} \, \sqrt{ \Theta_z} \,. 
\end{equation}
It can be shown that functions $Q^{L(R)}$ of \eqref{rootE-GenQ} are given by
\begin{equation}
Q^{L(R)} =  1 + \beta^2 (g^{xx})^2 B^2 \equiv Q_0 \, . \label{Q0} 
\end{equation}
Therefore, the DBI Lagrangian given in \eqref{SDBI} and \eqref{rootE-GenQ} reduces to 
\begin{equation} \label{effDBILag}
{\cal L}_{DBI} \, = \, - \, 2 \, V(\tau) \, \sqrt{-g} \, \sqrt{Q_0 \, \Theta_z} 
\quad \text{with} \quad 
V(\tau) \,= \, V_0 \, \exp \left [ - \, \frac {m_\tau^2}{2} \, \tau^2 \right ] \, . 
\end{equation}
In appendix \ref{WZterm} we describe the Chern-Simons (Wess-Zumino) term for a general configuration of gauge fields 
and a complex tachyon. For the particular case of the ansatz in \eqref{MagAnsatz}, we obtain $j=0$, $\Omega_5^{(0)}=0$ and 
$\Omega_4^{(0)} =0$. As a consequence the WZ term in \eqref{CSterm} vanishes. 

From the Lagrangian in \eqref{effDBILag} we find 
the Euler-Lagrange equation for $\tau$ 
\begin{equation}
Q_0 \, \tau'' + \Bigg[  \frac{Q_0}{2} \left ( \frac{g_{tt}'}{g_{tt}} + \frac{g_{xx}'}{g_{xx}} \right )
+ \frac{g_{xx}'}{g_{xx}} \Bigg] \, \Theta_z \, \tau'  - \frac{Q_0}{2} \, \frac{g_{zz}'}{g_{zz}} \, \tau' 
+ \frac{Q_0}{2 \kappa} \, g_{zz}  \, m_\tau^2  \,  \Theta_z  \, \tau = 0 \, . 
\label{specTeqv2}
\end{equation}
Recalling the definition of $\Theta_z$ in \eqref{blobs}, we can split \eqref{specTeqv2} into linear and non-linear terms
\begin{align}
& Q_0 \, \tau'' + \Bigg[  \frac{Q_0}{2} \left ( \frac{g_{tt}'}{g_{tt}} + \frac{g_{xx}'}{g_{xx}} - \frac{g_{zz}'}{g_{zz}} \right )
+ \frac{g_{xx}'}{g_{xx}} \Bigg]  \, \tau'  + \frac{Q_0}{2 \kappa} \, g_{zz} \,  m_\tau^2 \,  \tau  
\nonumber \\[5pt]
& + 2 \, \kappa \,   
\Bigg[  \frac{Q_0}{2} \left ( \frac{g_{tt}'}{g_{tt}} + \frac{g_{xx}'}{g_{xx}}  \right ) + \frac{g_{xx}'}{g_{xx}} \Bigg] \, 
g^{zz} \, \tau'^3  + Q_0 \, m_\tau^2 \, \tau'^2 \, \tau \, = \, 0 \,. 
\label{specTeqv3}
\end{align}
For the case $B=0$, we have $Q_0=1$ and \eqref{specTeqv3} reduces to eq. $(3.6)$ of 
\cite{Iatrakis:2010jb}.


\medskip 

\noindent  {\bf UV asymptotic analysis}

\medskip

\noindent 
The small $z$ limit is asymptotically AdS and therefore in this region   \eqref{specTeqv3} reduces to 
\begin{equation}
\tau'' - \frac{3}{z} \, \tau' 
+   \left ( \frac{ m_\tau^2 R^2}{2 \kappa} \right)  \, \frac{1}{z^2} \, \tau \, = \, 0 \, ,
\label{TeqAdS}
\end{equation}
where the non-linear terms are sub-leading. In order that the tachyon is dual to the quark mass operator $\bar q q$ with conformal dimension $\Delta =3$ the 5d mass of the scalar field $\tau$ must be set such that one has the identification

\begin{equation}
\frac{ m_\tau^2 R^2}{2 \kappa} \, = \, 3 
\label{massdictionary}
\end{equation}
The asymptotic solution for $\tau$ takes the form
\begin{equation} \label{BC-tachyon-UV}
\tau(z) \, = \, c_1 \, z + c_3 \, z^3+ \frac{m_\tau^2}{6} \, c_1^3 \, z^3 \, \log z - \frac{\left(48 B^2 c_1-120
   c_1^2 c_3+7 c_1^5-20 c_1^5 \log z\right)}{192}  z^5+ {\cal O}(z^6) \,. 
\end{equation}
The source coefficient $c_1$ is proportional to the quark mass whereas the vev coefficient $c_3$ will be related to the chiral condensate.


\subsection{The confined phase}

Until now we have not specified the geometry, but only put constraints on what the UV and IR asymptotic limits must look like. We know that in the confined phased, there must a mass gap corresponding to some point where the geometry stops. An appropriate space-time to consider is thus the 6d cigar-geometry given by
\begin{equation} \label{cigar}
ds_6^2 = \frac{R^2}{z^2} \Bigg[ - dt^2 + d\vec{x}_3^2 + \frac{dz^2}{f_{\Lambda}(z)} + f_{\Lambda}(z) \, d \eta^2 \Bigg] 
\quad \text{where} \quad 
f_{\Lambda}(z) = 1 - \frac{z^5}{z_{\Lambda}^5} \,.
\end{equation}
The spatial coordinate $\eta$ is compact, i.e. $0 \leq \eta \leq 2 \pi R $. At the tip of the cigar $z=z_{\Lambda}$ smoothness of the geometry implies that 
\begin{equation}
2 \pi R = \frac{4 \pi}{|f'(z_{\Lambda})|} = \frac{4 \pi}{5} z_{\Lambda} = \frac{2 \pi}{M_{KK}} \, .    
\end{equation}
The mass scale $M_{KK}$ plays an important role in the description of confinement and the glueball spectrum.  The $D4-\overline{D4}$ pair of flavour branes will be located at $\eta =0$.  
It is convenient to define the dimensionless radial coordinate $u \equiv z/z_{\Lambda}$.
Moreover in order to fully eliminate the presence of $z_{\Lambda}$ and $m_\tau$ from the Lagrangian, we rescale 
the tachyon, the magnetic field, the constant $V_0$ and the field theory coordinates
in the following way
\begin{equation} \label{redefinitions}
{\cal T} \, \equiv \, m_\tau \, \tau \, , \quad 
{\cal B} \, \equiv \, \frac{\beta}{R^2} \,  B \, , \quad 
{\cal V}_0 \, \equiv \, V_0 \, R^5
\quad \& \quad  
 \Big\{t,{\vec x}\Big\} \rightarrow  z_{\Lambda} \,  \Big\{t,{\vec x}\Big\} \, . 
\end{equation}
With these rescalings, the equation of motion for the tachyon becomes
\begin{equation}
{\cal T}'' - \Bigg[ \frac{1}{2} \left (  \frac{2}{u} + \frac{5 \, u^4}{f_{\Lambda}} \right )  + \frac{2}{u \, Q_0} \Bigg] \, {\cal T}' 
+  \frac{3\,  {\cal T}}{ u^2 \, f_{\Lambda}} - \frac{2}{3} \, \Bigg[ 1 + \frac{1}{Q_0} \Bigg]  \, u \, f_{\Lambda} \left({\cal T}'\right)^3 +  
\left({\cal T}'\right)^2 {\cal T}   = 0 
\label{specTeqConfv2}
\end{equation}
where $ {\cal T}' \equiv \partial_u  {\cal T}$ and now the functions $Q_0$ and $f_{\Lambda}$ are defined as follows 
\begin{equation}
f_{\Lambda} \, = \, 1 - u^5 \quad \& \quad 
Q_0 \, = \,  1 + {\cal B}^2 u^4 \,. 
\label{fandQ0Conf}
\end{equation}
The dimensionless coordinate $u$ runs from $0$ at the AdS boundary to $1$  at the tip of the cigar. Note that the  ${\cal T}$ differential equation \eqref{specTeqConfv2} depends only on the dimensionless parameter ${\cal B}$. 


\medskip 

\noindent  {\bf IR asymptotic analysis}

\medskip 

\noindent  Near the tip of the cigar, the asymptotic behavior of the tachyon is given by a double series expansion involving a power law behaviour which has both an integer and fractional part which can be separated. This can be parameterised as
\begin{equation} \label{IRsolConf}
{\cal T}(u) \, = \, \sum_{n=0}^{\infty} \sum_{m=0}^{\infty}  g_{n,m} \, \left(1 - u\right)^{(n-1)r + m}
\quad \text{where} \quad 
r \, = \, \frac{3}{10} \, \frac{1 + {\cal B}^2}{2 + {\cal B}^2} 
\end{equation}
and $g_{n,m}$ are constant coefficients. Note from \eqref{IRsolConf} that the $n=0, m=0$ term means that ${\cal T}$ is singular at $u=1$ so long as $g_{0,0}$ does not vanish. Indeed for any non-trivial solution $g_{0,0}\ne 0$.
Plugging \eqref{IRsolConf} into \eqref{specTeqConfv2}, the latter  becomes a double series and the coefficients 
$g_{n,m}$ are obtained by solving the double series at each order. This is described in appendix \ref{AppConfIR}. 
Here we show the first coefficients
\begin{align}
g_{0,0} &= C_0 \, , \quad  
g_{0,1} = -  \frac{3}{10} \,  
\frac{6 + 5 \, {\cal B}^2 + 3 \, {\cal B}^4}{\left(2 + {\cal B}^2\right)^2} \, C_0  \, , \quad  
g_{2,0} = - \frac{13 + 8 \, {\cal B}^2}{6\, \left(1+ {\cal B}^2\right)} \, C_0^{-1}  
\nonumber \\[7pt]
g_{2,1} &  = \frac{986+507 \, {\cal B}^2+169 \, {\cal B}^4+206 \, {\cal B}^6+58 \, {\cal B}^8}{20 \left(1+{\cal B}^2\right) \,  
\left(2+{\cal B}^2\right)^2 \, \left(13+8 \, {\cal B}^2\right)} C_0^{-1} \, , \, \cdots
\end{align}
Note that for small $C_0$ one has to be careful with the radius of convergence of the series.
An important feature of the solution in \eqref{IRsolConf} is that all the coefficients depend solely on one parameter,  $C_0$.  This is a nontrivial consequence of the nonlinear terms in the differential equation 
\eqref{specTeqConfv2} arising from the particular behavior of the tachyon potential, and can be thought as an 
effective reduction of the original second order differential equation into a first order one\footnote{A similar 
mechanism occurs in bottom-up Higgs-like models for chiral symmetry breaking \cite{Gherghetta:2009ac,Chelabi:2015gpc,Ballon-Bayona:2020qpq}.}.


\medskip 

\noindent  {\bf Numerical analysis of the tachyon equation}

\medskip

\noindent In order to solve the equation of motion for the tachyon, \eqref{specTeqConfv2}, with the UV and IR behaviours given by  \eqref{BC-tachyon-UV} and 
\eqref{IRsolConf}, respectively we fix $c_1$ and use a shooting technique to numerically integrate, tuning the value of $c_3$ such that both the UV and IR asymptotics are respected.

\begin{figure}[ht] 
   \centering
   \includegraphics[width=8cm]{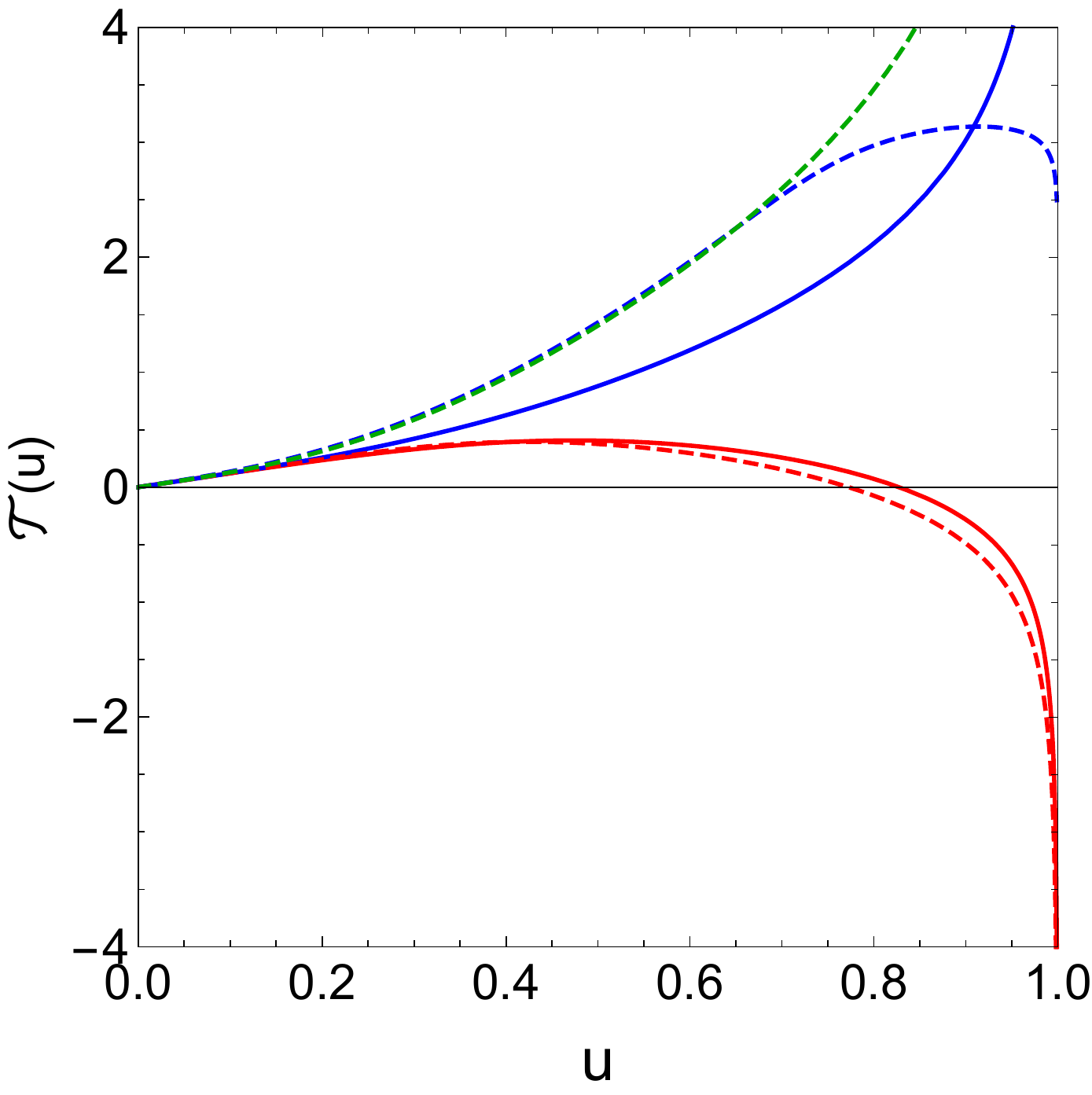}
     \caption{Plots of the tachyon profile for two different values of the magnetic field and fixed value of $c_1=1.2$. 
     When ${\cal B}=3.5$ there are two values of $c_3$ for which the tachyon diverges in the IR, namely $c_3=-0.69$ 
     (red solid line) and $c_3=2.85$ (blue solid line). When ${\cal B}=8.5$ there are three values of $c_3$ that the tachyon diverges, 
     namely $c_3=1.36$ (red dashed line), $c_3=9.23$ (blue dashed line) and $c_3=10.23$ (green dashed line). 
     Comparing the free energies of the solutions we realise that the energetically favored 
     is the one with the highest value of $c_3$.}
   \label{CP-tachyon}
\end{figure}

As happens in the ${\cal B}=0$ case of \cite{Iatrakis:2010jb}, for a fixed value of $c_1$\footnote{Since the value of 
the constant $c_1$ is related to mass of the quark, we only consider solutions with $c_1>0$.} there is more than one 
value of $c_3$ for which the tachyon diverges in the IR. For small values of ${\cal B}$ there are two values of $c_3$ while increasing the value of ${\cal B}$ above $7.5$ the behavior of the tachyon profile becomes more complex.  It is possible 
to find a single value of $c_1$ which corresponds to three (or even four) values of $c_3$. In figure \ref{CP-tachyon} we 
have plotted the different profiles for the tachyon when ${\cal B}=3.5$  and ${\cal B}=8.5$ for a fixed value of $c_1 =1.2$.

In order to find which of the two solutions is energetically favoured we must compare their free energies. This is slightly complicated by the fact that the values of $c_1$ are the same between the solutions that we are comparing, but the values of $c_3$ are different. In order to compare solution 1, described by $(c_1, c_3^{1})$ and solution 2, described by $(c_1, c_3^{2})$, we must calculate the difference
\begin{equation} \label{FE-comparison}
\quad \qquad  \frac{\Delta F_{12}}{V_4 \, {\cal V}_0}  \, = \,  {\cal W}_1 \, - \,  {\cal W}_2 \, + \, \frac{2}{3} \, c_1 \, 
\left(c_3^1 \, - \, c_3^2 \right) 
\end{equation}
where
\begin{equation}
{\cal W}_i \, = \, 2 \, \int \, du \, \frac{\exp \left[-\frac{1}{2}\, {\cal T}_i^2\right]}{u^5 \, \sqrt{1-u^5}} \, 
\sqrt{1+ {\cal B}^2 \, u^4} \, \sqrt{1+\frac{1}{3}\, u^2 \, \left(1-u^5\right) \, \dot{\cal T}_i^2} 
\end{equation}
with $i=1,2$ and where the finite term in  \eqref{FE-comparison}  is coming from the subtraction of the counterterms for each solution (the analysis of the 
holographic renormalisation is presented in section \ref{condensate}). Note that this is not the only counterterm in the free energy but since we perform the calculation in \eqref{FE-comparison} at a fixed value of $c_1$, this is the only one that survives when we look at their differences.

\begin{figure}[ht] 
   \centering
\includegraphics[width=9cm]{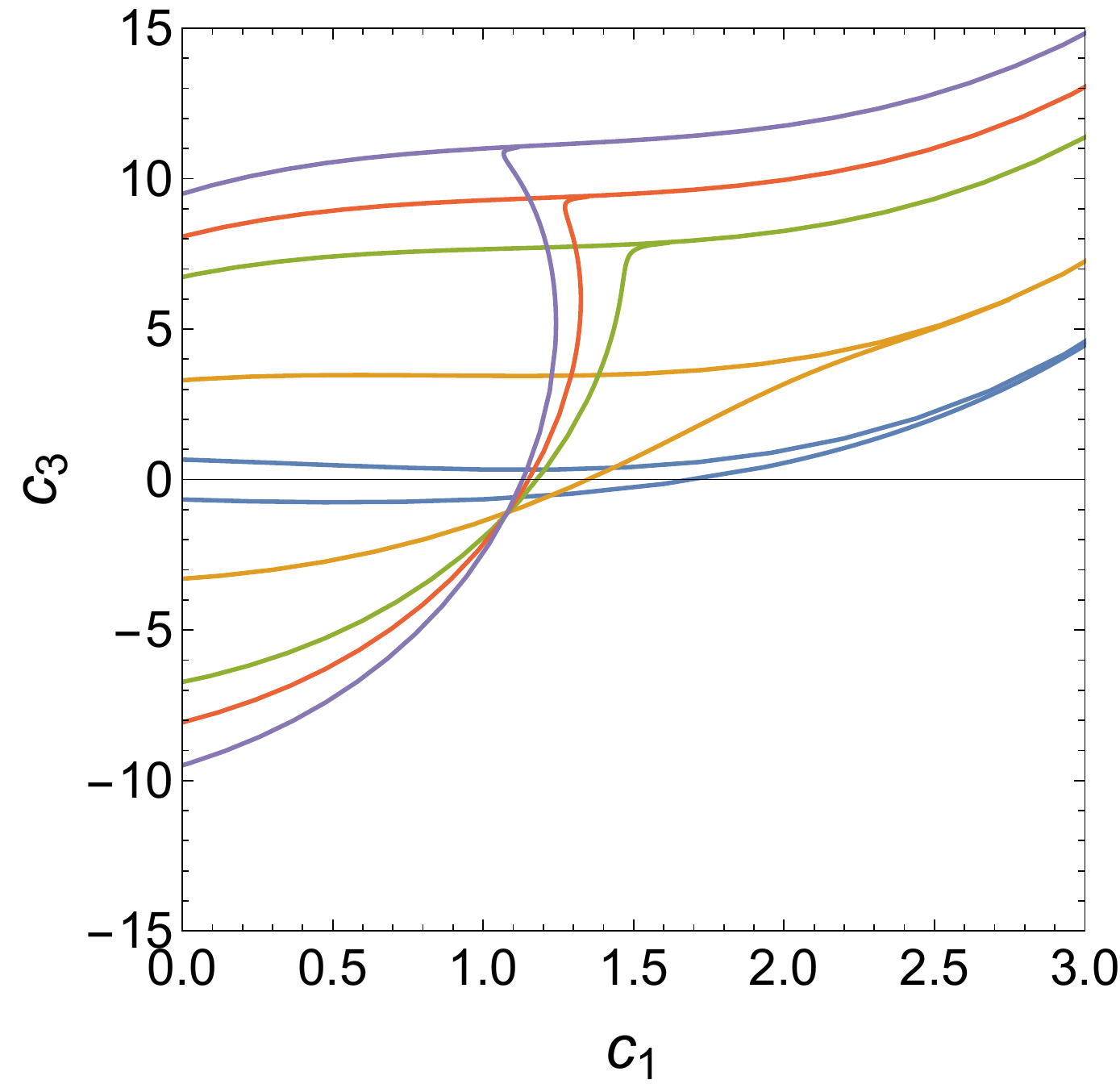}
       \caption{Plots of the values of $c_3$ determined numerically as a function of $c_1$ for ${\cal B}=0,4,7,8$ and 9 for blue, orange, green, red and purple respectively. It is clear that a single value of $c_1$ may have multiple solutions, however studying the energetics shows that the top branch is always the favoured one, giving the first indication that chiral symmetry is broken at finite magnetic field.}
   \label{CP-c3vsc1}
\end{figure}

In figure \ref{CP-c3vsc1} we present $c_3$ as a function of $c_1$ for different values of the magnetic field. In every plot 
there are two branches (for small enough $c_1$) which may become three at higher values and for large values of the magnetic field and which then become a single solution for even larger values of $c_1$.  Comparing the profiles with the use of 
\eqref{FE-comparison} it can be shown that the dominant solution is always the one with the highest value of $c_3$.

\subsection{The deconfined phase}

Having studied the confined phase, we consider the 6d black-brane for the deconfined phase. The metric in this case is given by 
\begin{equation} \label{BH}
ds_6^2 = \frac{R^2}{z^2} \Bigg[ - f_T(z) \, dt^2 + d\vec{x}_3^2 + \frac{dz^2}{f_T(z)} +  d \eta^2 \Bigg] 
\quad \text{where} \quad 
f_T(z) = 1 - \frac{z^5}{z_{T}^5} \,.
\end{equation}
The black-brane temperature is given by
\begin{equation} \label{BHTemp}
T = \frac{|f'(z_T)|}{4 \pi} = \frac{5}{4 \pi} z_T^{-1}  \,.     \end{equation}
The deconfinement transition maps to a gravitational 
 Hawking-Page transition between the cigar geometry \eqref{cigar} and the black-brane geometry \eqref{BH}\footnote{See  \cite{Mandal:2011ws} for an alternative perspective.}. This transition is first order and occurs when $z_T= z_{\Lambda}$ which corresponds to a critical temperature
\begin{equation} \label{Tc}
T_c = \frac{5}{4 \pi} z_{\Lambda}^{-1} = \frac{M_{KK}}{2 \pi}  \,.    
\end{equation}

The $D4-\overline{D4}$ pair of flavour branes is again located at $\eta =0$.  As in the confined case, it is convenient to define a new radial coordinate $v \equiv z/z_T$ and this time we rescale the quantities as 
\begin{equation} \label{redefinitionsdec}
{\cal T} \, \equiv \, m_\tau \, \tau \, , \quad 
{\cal B} \, \equiv \, \frac{\beta}{R^2} \,  B \, , \quad 
{\cal V}_0 \, \equiv \, V_0 \, R^5
\quad \& \quad  
 \Big\{t,{\vec x}\Big\} \rightarrow  z_T \,  \Big\{t,{\vec x}\Big\} \, . 
\end{equation}
With this rescaling the equation of motion for the tachyon becomes
\begin{equation}
{\cal T}'' - \Bigg[ \frac{1}{v} + \frac{5 \, v^4}{f_T}   + \frac{2}{v \, Q_0} \Bigg] \,{\cal  T}' 
+  \frac{3 \, {\cal T} }{ v^2  f_T}
-  \frac{v^2 f_T}{6} \Bigg[\frac{4}{v} + \frac{5\, v^4}{f_T}+ \frac{4}{v \, Q_0} \Bigg]  ({\cal T}')^3 +  
({\cal  T}')^2 {\cal T}   = 0 
\label{specTeqDeconfv2}
\end{equation}
where $ {\cal T}' \equiv \partial_v  {\cal T}$ and now the functions $Q_0$ and $f_T$ are defined as follows 
\begin{equation}
f_T\,  = \, 1 - v^5 
\quad \& \quad
Q_0 \, =  \, 1 + {\cal B}^2 \, v^4 \, . \label{fandQ0Deconf}
\end{equation}
The dimensionless coordinate $v$ runs from $0$ to $1$ and the
 ${\cal T}$ differential equation \eqref{specTeqDeconfv2} now depends only on the dimensionless parameter ${\cal B}$. We will show later in the paper that the dimensionless ${\cal B}$ will be proportional to the physical value of the magnetic field and inversely proportional to the square of the temperature.
 

\medskip 

\noindent {\bf IR asymptotic analysis}

\medskip 

\noindent Near the horizon ($v$ close to $1$), the tachyon field has to be regular. 
As a consequence, the asymptotic solution takes the form of an ordinary Taylor expansion
\begin{equation}
{\cal T}(v) \, = \, \sum_{n=0}^{\infty} C_n \, \left(1-v\right)^n \, . \label{IRsolDeconf}
\end{equation}
This time the differential equation \eqref{specTeqDeconfv2} becomes a simple series and the coefficients $C_n$ are 
obtained by solving the series at each order. This is described in appendix \ref{AppDeconfIR}. 
Here we show the first subleading coefficients
\begin{equation}
C_1 =  - \frac35 \, C_0
\quad \& \quad
C_2 = - \frac{3}{20}  \, C_0 \, 
\Bigg[ \frac{17}{5} - \frac{1- {\cal B}^2 }{1 + {\cal B}^2} + 
\frac{3}{10} \, C_0^2  \Bigg] \, . 
\end{equation}
We find from \eqref{IRsolDeconf} that the tachyon solution depends solely on one parameter $C_0$. 
Again, this is a consequence of the non-linearity of the differential equation \eqref{specTeqDeconfv2} that effectively 
reduces a second order differential equation to a first order one in the near horizon limit. 

The UV analysis remains the same in both the confined and deconfined cases and thus doesn't need to be treated separately here.


\medskip 

\noindent  {\bf Numerical analysis of the tachyon equation}

\medskip 

\noindent In this subsection we present the details of the numerical solution of the equation of motion for the 
tachyon in the deconfined case \eqref{specTeqDeconfv2} with UV and IR boundary conditions given in \eqref{BC-tachyon-UV} and 
\eqref{IRsolDeconf}, respectively.

\begin{figure}[ht] 
   \centering
   \includegraphics[width=7cm]{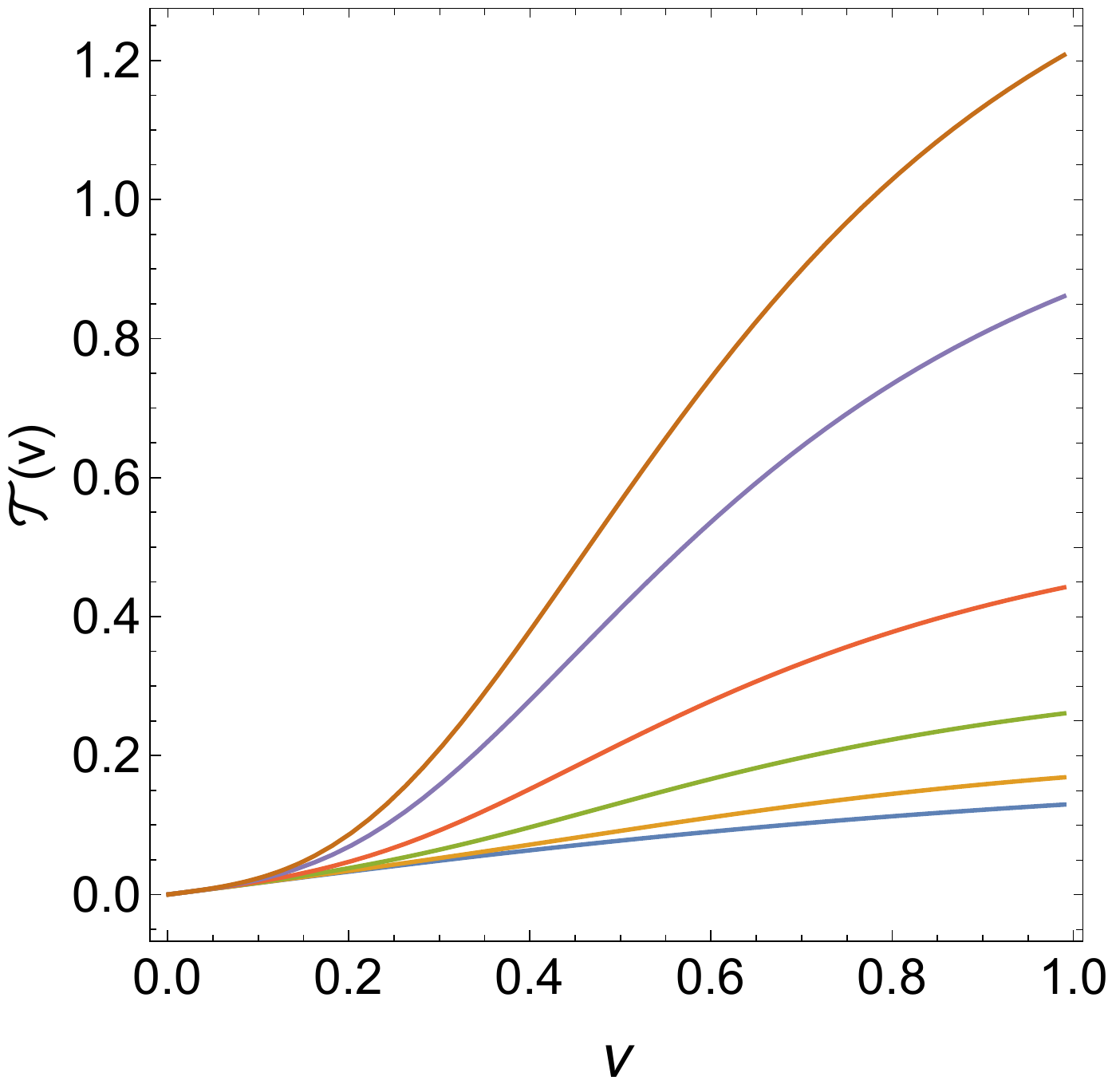}
   \hspace{0.2cm}
    \includegraphics[width=7cm]{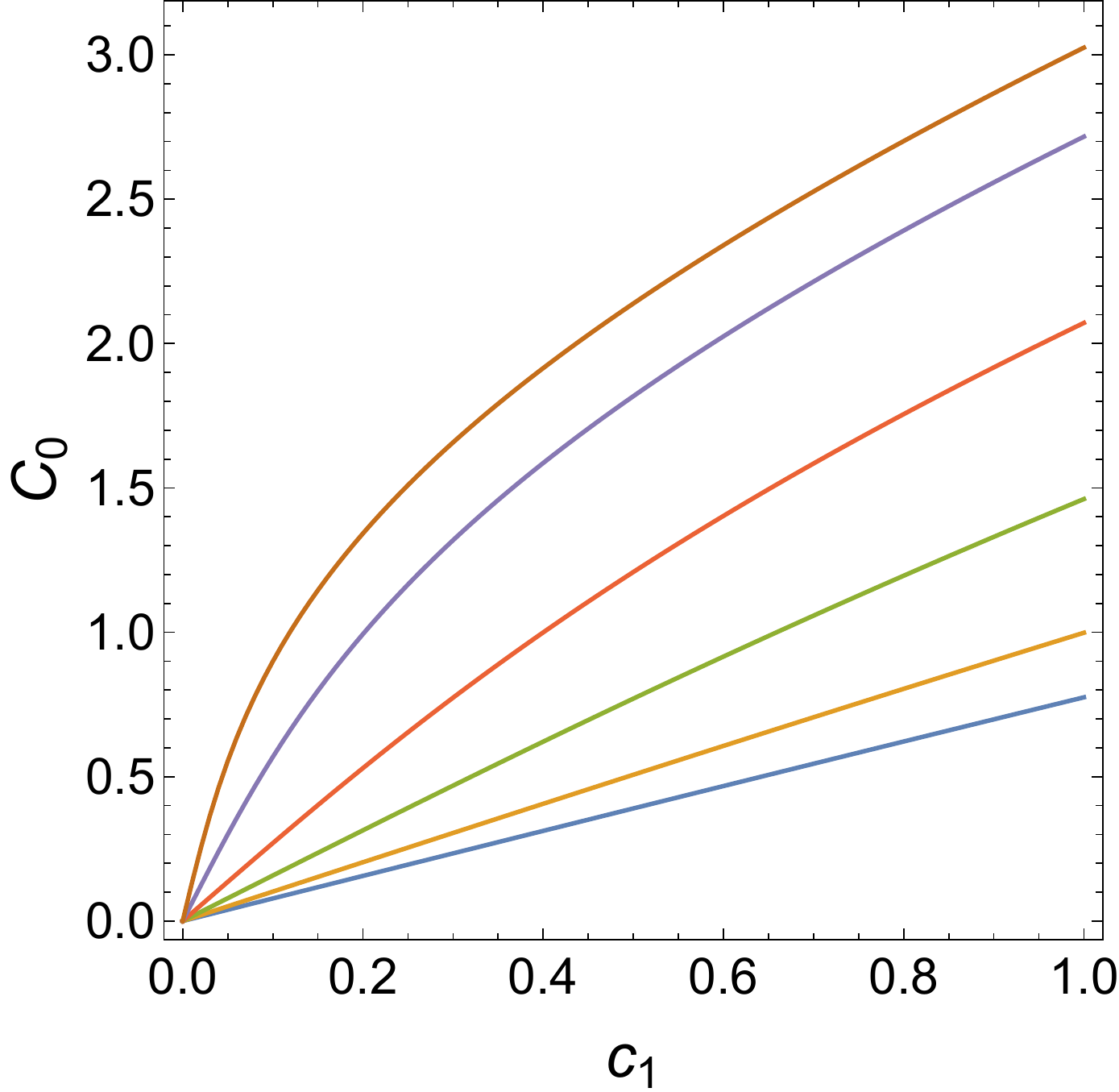}
    \hspace{0.2cm}
    \\
    \includegraphics[width=7cm]{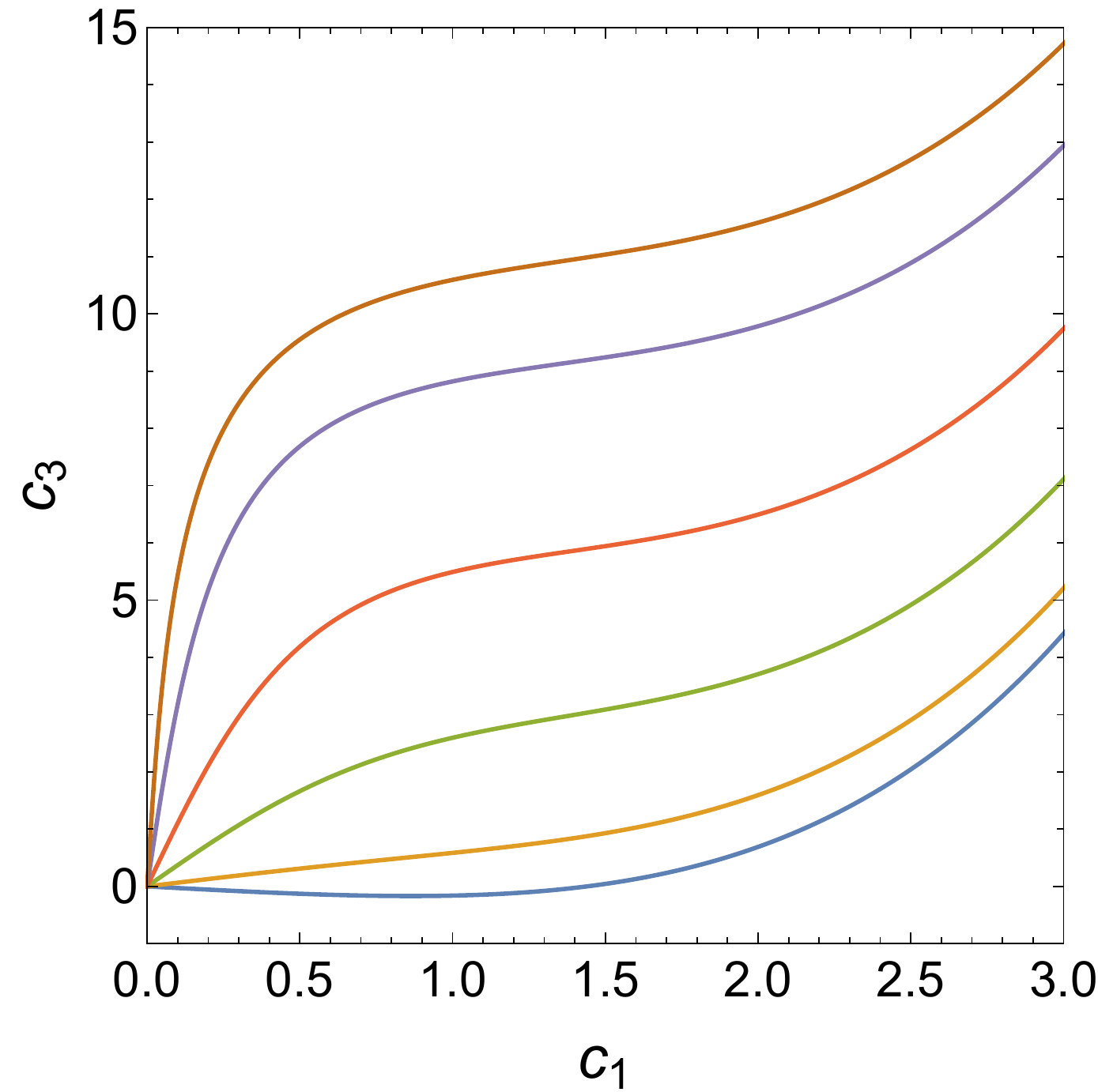}
     \caption{Plots of the tachyon profile (fixing $c_1=1/6$) as a function of $v$ (upper left panel), $C_0$
     as a function of $c_1$ (upper right panel) and $c_3$ as a function of $c_1$ (lower two panels), for 
     different values of the magnetic field $\cal B$ (lower than the critical value ${\cal B} \approx 10$). 
     The correspondence between colour and values of the magnetic field, for all the plots of this figure, is 
     Blue $=0$, Orange $=2$, Green $=4$, 
     Red$=6$, Purple $=8$ \& Brown $=9$.}
   \label{DC-below-critical-B}
\end{figure}

For small values of the magnetic field ${\cal B}$ the analysis of the tachyon equation is similar to the ${\cal B} = 0$ 
case of \cite{Iatrakis:2010jb}. In figure \ref{DC-below-critical-B} we have plotted tachyon profiles for different values of 
the magnetic field, at a fixed value of $c_1$. Increasing the value of ${\cal B}$ changes the profile in a continuous way and 
does not affect its shape. This in turn is reflected in the three other plots of figure \ref{DC-below-critical-B} that 
present $C_0$ and $c_3$ as functions of $c_1$ for different values of ${\cal B}$. 
Notice here that the non-monotonic behavior for $c_3$ that was observed in  \cite{Iatrakis:2010jb} for the ${\cal B} = 0$ 
case disappears as soon as we increase the value of ${\cal B}$.
Once we choose the value of $c_1$, the values of $C_0$ and $c_3$
are determined dynamically by the IR boundary condition, using the shooting technique to numerically solve the 
equation of motion for the tachyon. Chiral symmetry remains unbroken (spontaneously) for the range of values that we consider in 
figure \ref{DC-below-critical-B}, since for $c_1=0$ the value of $c_3$ is also zero, and as a result ${\cal T} =0$ for all $v$.
This observation was put forward also in \cite{Iatrakis:2010jb}.

\begin{figure}[ht] 
   \centering
   \includegraphics[width=9cm]{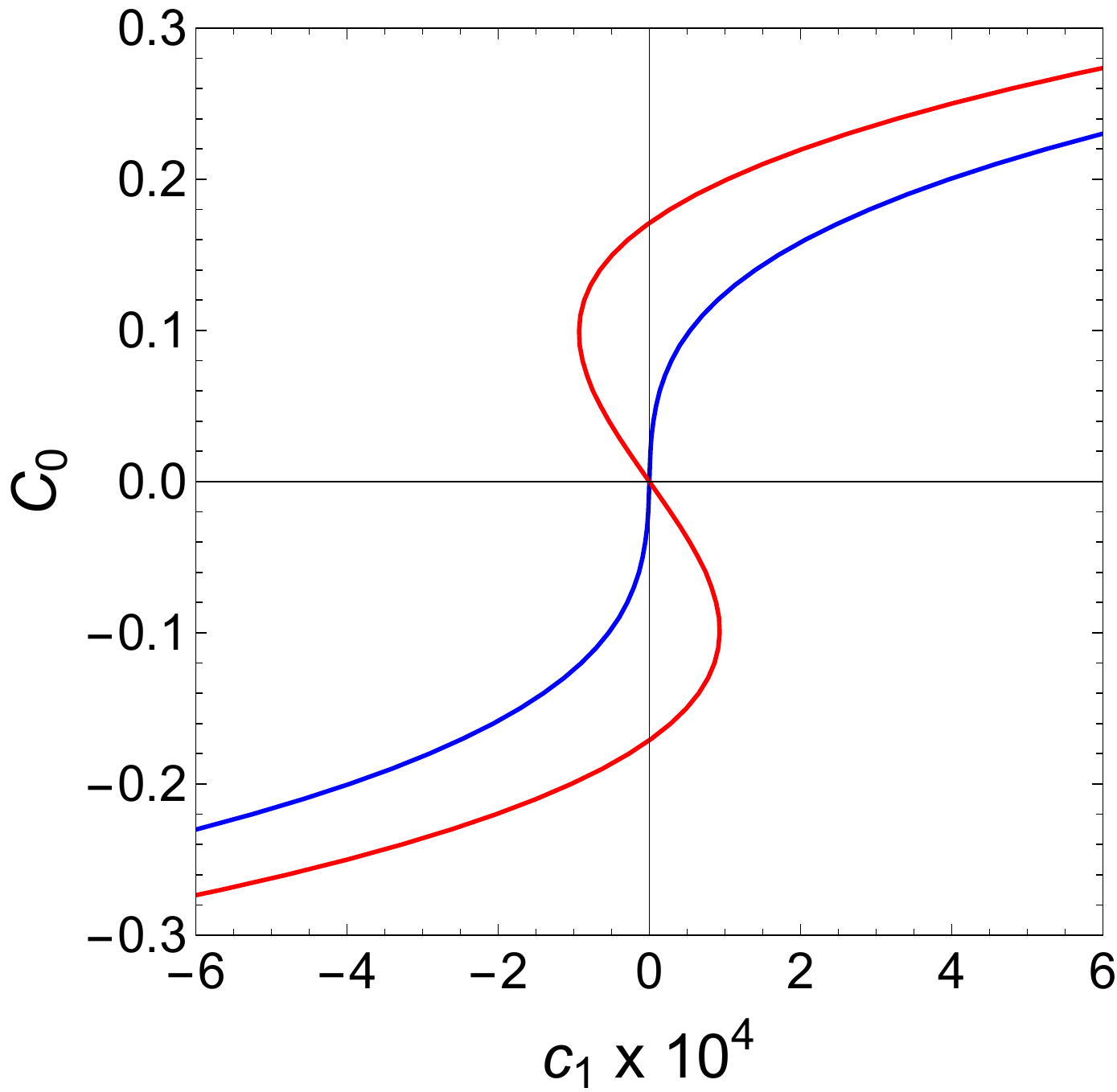}
   \caption{Plot of $C_0$ as a function of $c_1$ for two nearby values of the magnetic field, namely 
   ${\cal B} =9.99$ and  ${\cal B} =10.01$. It is clear from the behavior of $C_0$ in this plot ($c_3$ as a function 
   of $c_1$ behaves analogously) that for values of the magnetic field above the critical value ${\cal B} =10$ the function
   of $C_0$ as a function of $c_1$ becomes multivalued. There are values of $c_1$ that correspond to three 
   values of $C_0$.}
   \label{DC-critical-B}
\end{figure}

For values of ${\cal B} \ge 10 $ we see an interesting behaviour appear. In figure \ref{DC-critical-B} we plot $C_0$ as a function of $c_1$ for 
two values of the magnetic that are just below and just above the value ${\cal B} =10$, namely 
${\cal B} =9.99$ and  ${\cal B} =10.01$. From this plot one can see that when the value of the magnetic field 
exceeds the (critical) value ${\cal B} \approx 10$, $C_0$ as a function of $c_1$ becomes multivalued. 
Notice that the same behavior appears in the plot of $c_3$ as a function of $c_1$. For  ${\cal B} =10.01$ and values of $|c_1|\lesssim 10^{-4}$ there are three values of $C_0$ and consequently three different profiles. 

\begin{figure}[ht] 
   \centering
   \includegraphics[width=7.05cm]{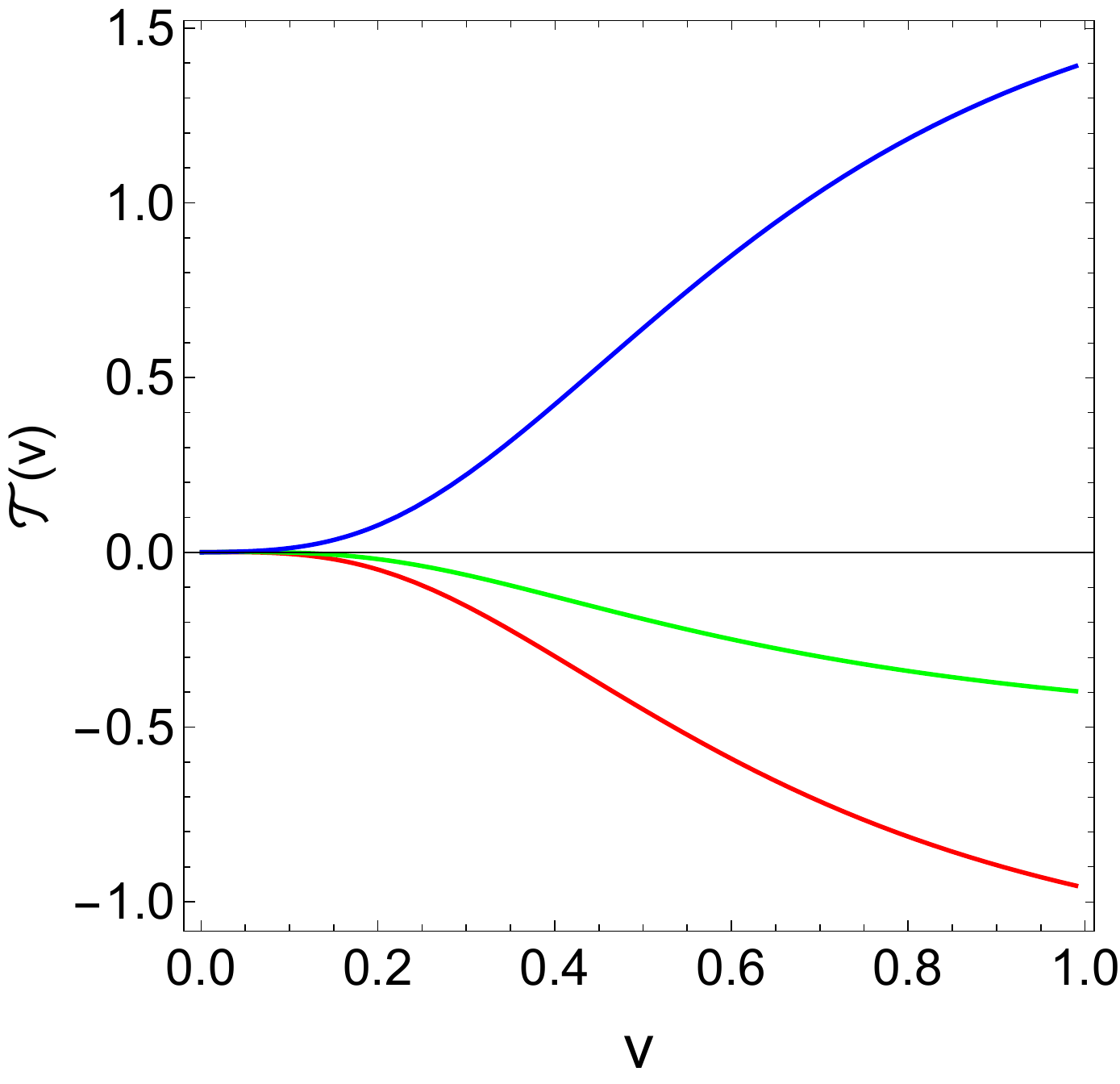}
   \hspace{0.2cm}
    \includegraphics[width=7.505cm]{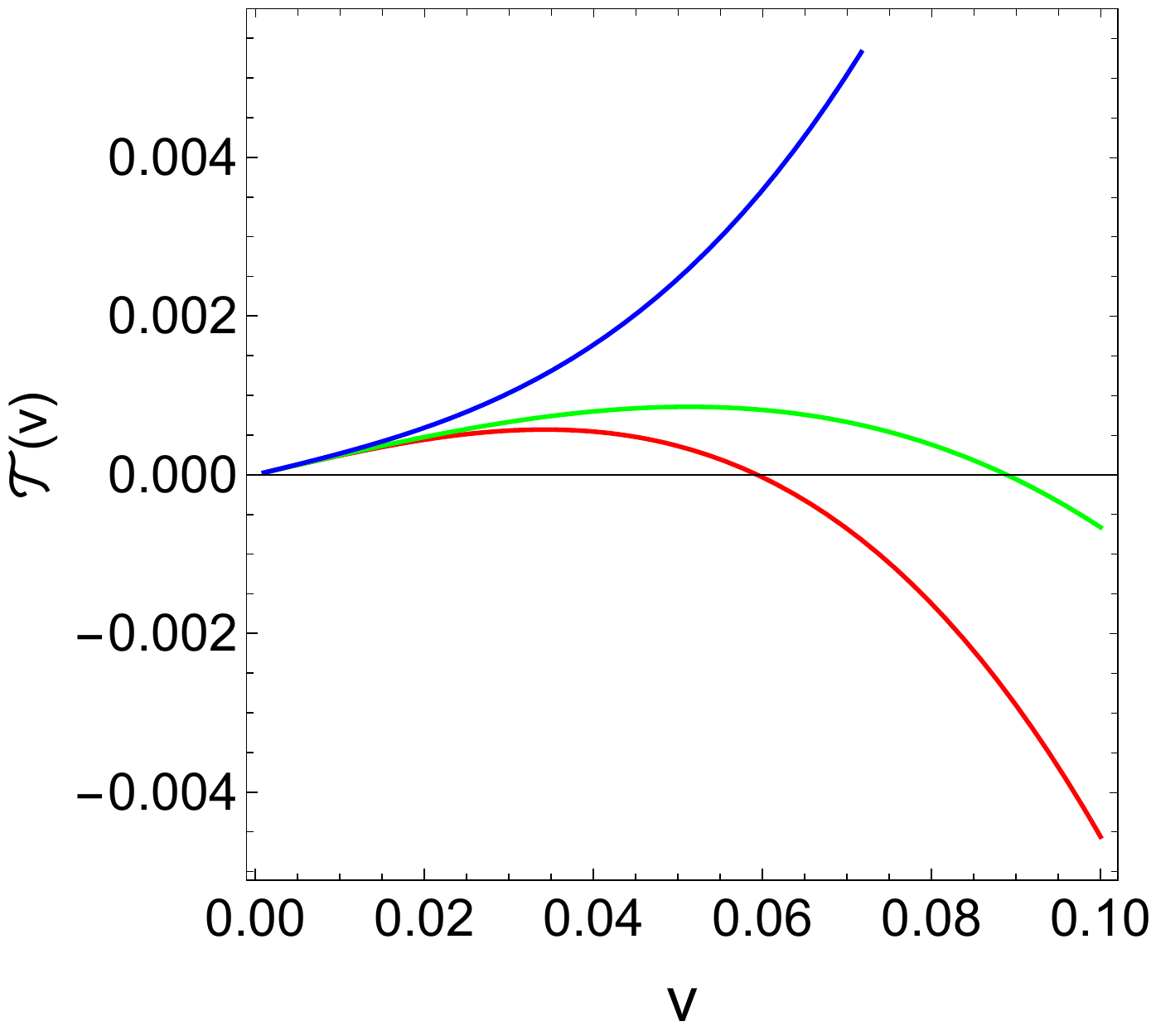}
     \caption{Plots of the tachyon profile for ${\cal B} =11$ and $c_1=1/40$. Since ${\cal B} > 10$ there is more than
     one tachyon profile. The red plot corresponds to 
     $C_0 \approx - \ 0.96$, the green plot to $C_0 \approx - \ 0.4$ and the blue plot to $C_0 \approx 1.4$. 
     On the right panel of the figure we depict a zooming of the plot close to the boundary ($v \rightarrow 0$) 
     that is on the left panel (with the same colour) to highlight
     that the tachyon profile has a monotonous behaviour only for the highest value of $C_0$ (or equivalently $c_3$).}
   \label{DC-above-critical-B-1}
\end{figure}

In the upper part of figure \ref{DC-above-critical-B-1}, and for ${\cal B} =11$, we have plotted the three tachyon profiles 
that correspond to the value $c_1=1/40$. Zooming into the first two plots (red and green) close to the boundary it can be seen that 
they do not have a monotonic behavior. It is only the last profile, which corresponds to the largest value of $C_0$, or equivalently of $c_3$ that is 
monotonic.

To distinguish between the three solutions and determine the energetically favored one, 
we have to compare the free energies of the different tachyon 
profiles for a fixed value of $c_1$. As in the confined case, we have to calculate the difference in free energies, given by \eqref{FE-comparison}. In the deconfined case 
 ${\cal W}_i$ is given by the following expression
\begin{equation}
{\cal W}_i \, = \, 2 \, \int \, dv \, \frac{\exp \left[-\frac{1}{2}\, {\cal T}_i^2\right]}{v^5} \, 
\sqrt{1+ {\cal B}^2 \, v^4} \, \sqrt{1+\frac{1}{3}\, v^2 \, \left(1-v^5\right) \, \dot{\cal T}_i^2} \, . 
\end{equation}
where $i$ indexes the different solutions.

\begin{figure}[ht] 
   \centering
   \includegraphics[width=8cm]{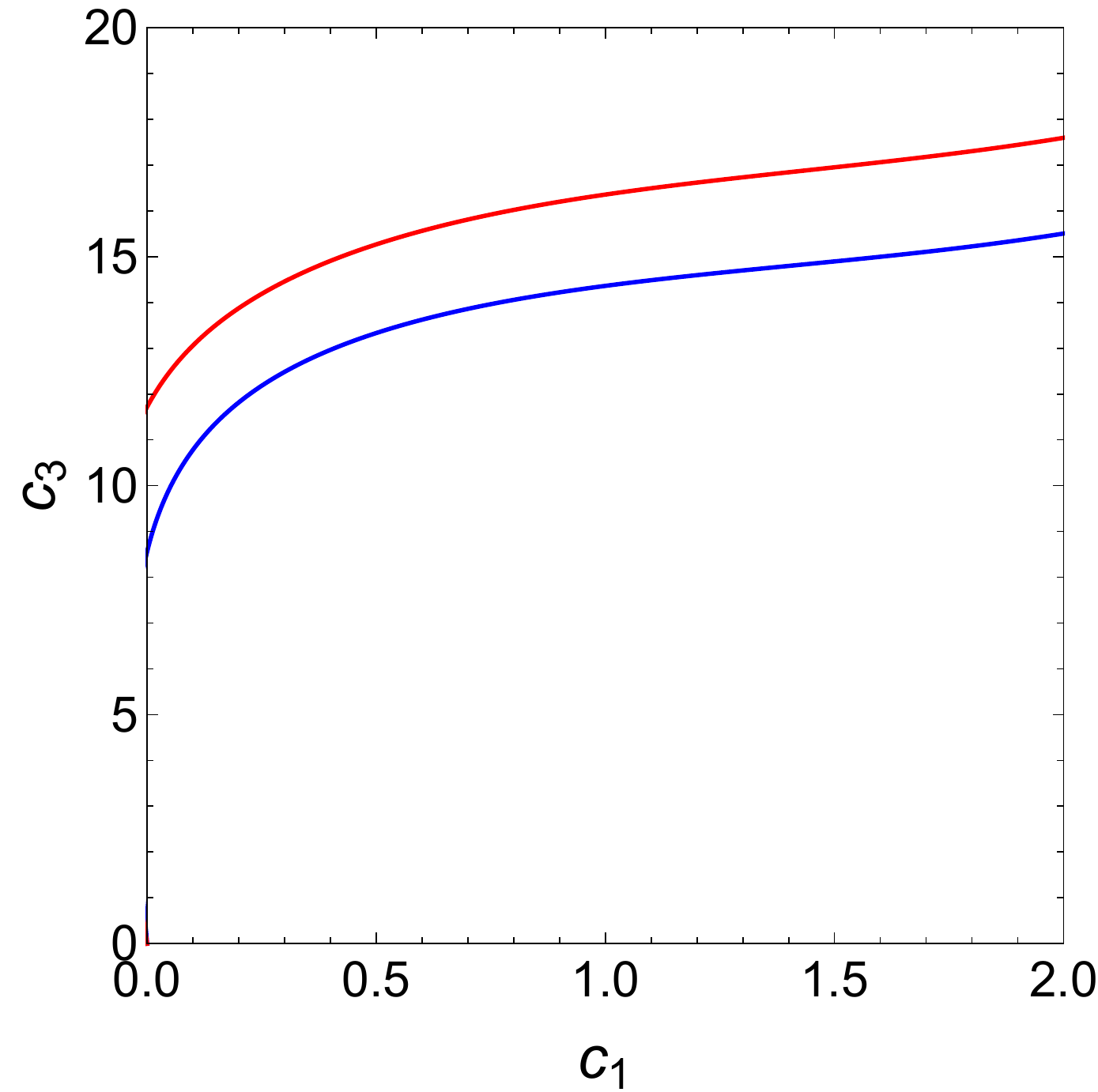}
   \caption{Plot of $c_3$ as a function of $c_1$ for two values of the magnetic field that are above the critical value
     ${\cal B} =10$. Comparing tachyon profiles with the same value of $c_1$ (for a fixed value of the magnetic field), we 
     have concluded that the profile with the highest value of $c_3$ is the energetically favored one. As a result, 
     spontaneous breaking of chiral symmetry is realised for magnetic fields above the critical value. In the 
     two cases that are presented in the plot above we have $c_3 \approx 8.6$ (with ${\cal B} =11$ - blue line) and 
     $c_3 \approx 11.8$ (with ${\cal B} =12$ - red line) for $c_1=0$.}
   \label{DC-above-critical-B-2}
\end{figure}

In figure \ref{DC-above-critical-B-2} we present $c_3$ as a function for $c_1$ for ${\cal B} =11$ and ${\cal B} =12$, 
after the comparison of the free energies of the tachyon profiles has been performed. The energetically favored 
profile is the one with the highest value of $c_3$ (or equivalently $C_0$). In this way an unexpected phenomenon arises:
For values of the magnetic field above the critical value ${\cal B} =10$ spontaneous breaking of chiral symmetry is realised.


\section{The chiral condensate and magnetisation}
\label{section-4}

Until now we have discussed the parameters which describe the UV asymptotics of the tachyon solution, $c_1$ and $c_3$ but not their corresponding field theory quantities. In this section we will connect them with the phenomenological gauge theory parameters of the quark mass and quark bilinear condensate. We then go on to study the magnetisation and the magnetic free energy density.

\subsection{The renormalised action}
\label{renormaction}

In this section we follow the analysis of \cite{Iatrakis:2010jb}, modified accordingly when there is an external magnetic field. Here we will concentrate on the confined phase, but the deconfined phase follows a very similar analysis. The DBI Lagrangian, after using the redefinitions of \eqref{redefinitions}, depends on two constants, namely ${\cal B}$ \& ${\cal V}_0$, and becomes
\begin{equation} \label{LDBIConf}
{\cal L}_{DBI} \, = \, - \, 2 \, {\cal V}_0 \, \exp \left[-\frac{1}{2}\, {\cal T}^2\right] \, \frac{1}{u^5 \, \sqrt{1-u^5}} \, 
\sqrt{1+ {\cal B}^2 \, u^4} \, \sqrt{1+\frac{1}{3}\, u^2 \, \left(1-u^5\right) \, \dot{\cal T}^2} \, . 
\end{equation}
In the deconfined phase we exchange $u$ with $v$ and the term $\sqrt{1-u^5}$ in the denominator is absent. We regularise the action by introducing a UV cut-off at $u=\epsilon$ and integrate from $\epsilon$ to the tip of the cigar at $u=1$
\begin{equation} \label{reg-action}
S_{reg} \, = \, \int_\epsilon^{1} d^4x \, du \, {\cal L}_{DBI} \, . 
\end{equation}
In the following we need the appropriate covariant counterterms that will be added to the regularised action in order to cancel the divergences at the boundary. The necessary expression is 
\begin{equation} \label{counterterms}
S_{ct} = - \frac{{\cal V}_0}{R^4}\int d^4x \sqrt{-\gamma} \, \Bigg[- \frac{1}{2} + \frac{{\cal T}^2}{3}
+\frac{{\cal T}^4}{18} \left (\log\epsilon  + \frac32 \alpha_1 \right ) 
+ \frac{\beta^2}{2} \,F^{\mu \nu} F_{\mu \nu} \, \left ( \log\epsilon + \alpha_2 \right )  \Bigg]
\end{equation}
where $\gamma$ is the induced metric at $u=\epsilon$, namely
$\sqrt{-\gamma}= {R^4 \over \epsilon^{4}}$. The finite counterterms depending on the  constants $\alpha_1$ and $\alpha_2$ capture the scheme dependence of the renormalised action. The last term in \eqref{counterterms} cancels the divergence due to the presence of the magnetic field close to the boundary and has the standard form which is known from the probe brane physics analysis \cite{Albash:2007bk,Erdmenger:2011bw}. The renormalised action is obtained from the following expression
\begin{equation} \label{ren-action}
S_{ren}=\lim_{\epsilon\to 0} \left( S_{reg} + S_{ct} \right) \, . 
\end{equation}

\subsection{The chiral condensate and magnetic catalysis}
\label{condensate}
The quark condensate is defined as usual in the following way
\begin{equation} \label{def-condensate}
\langle \bar q\,q\rangle \, = \,   \, \frac{\delta S_{ren}}{\delta m_q}
\end{equation}
To calculate the variation of the regularised action $S_{reg}$ with respect to $m_q$, we need to compute the functional derivative with respect to ${\cal T}$, since 
\begin{equation}
\frac{\delta {S_{reg}}}{\delta c_1} \, = \, \frac{\delta {\cal T}}{\delta c_1} \, 
\frac{\delta {S_{reg}}}{\delta {\cal T}} \, . 
\end{equation}
To calculate the functional derivative of $S_{reg}$ with respect to ${\cal T}$, we have to use the equation of motion for the 
tachyon and we arrive to the following expression \cite{Iatrakis:2010jb}
\begin{equation} 
\frac{\delta {S_{reg}}}{\delta {\cal T}} \, =\,  - \, \frac{\partial {\cal L}_{DBI}}
{\partial {\cal T}'}{\Big|}_{u\, = \, \epsilon} \, . 
\end{equation}
Finally, for the computation of the functional derivative of the tachyon with respect to $c_1$ we have to take into account that $c_3$ is a function of $c_1$. Putting together all the ingredients and using the UV expansion of the tachyon we arrive to following result for the  functional derivative of the renormalised action with respect to $c_1$
\begin{equation}
\frac{\delta S_{ren}}{{\delta c_1}} \, = \,  - \, \frac{{\cal  V}_0}{3} \,   \,   
\Big[- 4 \, c_3 +  c_1^3 \, \left(1+\alpha_1\right) \Big] \, . 
\end{equation}
The source coefficient $c_1$  is proportional to the quark mass $m_q$. %
\begin{equation} \label{quark-mass}
 c_1 = \zeta  m_q 
\end{equation}
where $\zeta$ is a normalisation constant, usually fixed as $\zeta = \sqrt{N_c}/2 \pi$ to satisfy large $N_c$ counting rules \cite{Cherman:2008eh}. In this way we obtain
\begin{equation} \label{quark-condensate}
 \zeta^{-1} \langle \bar q q \rangle \, = \,{\cal V}_0 \,  
\left[ \frac43 \, c_3 - \frac13 (\zeta m_q )^3 \,  \left(1+\alpha_1\right) \right] \, . 
\end{equation}
The quark mass $m_q$ in \eqref{quark-mass} and the chiral condensate in \eqref{quark-condensate} are 
dimensionless because of the redefinitions \eqref{redefinitions}. Both  $m_q$ and $\bar{q}q$ will  be redefined  in 
section \ref{sec:Lattice} in their dimensionful forms in order to compare with lattice data.
Note that the chiral condensate depends implicitly on the  magnetic field  through the vev coefficient $c_3$ and 
on the renormalisation scheme, through the parameter $\alpha_1$. 

\begin{figure}[ht]
\centering
\includegraphics[width=7cm]{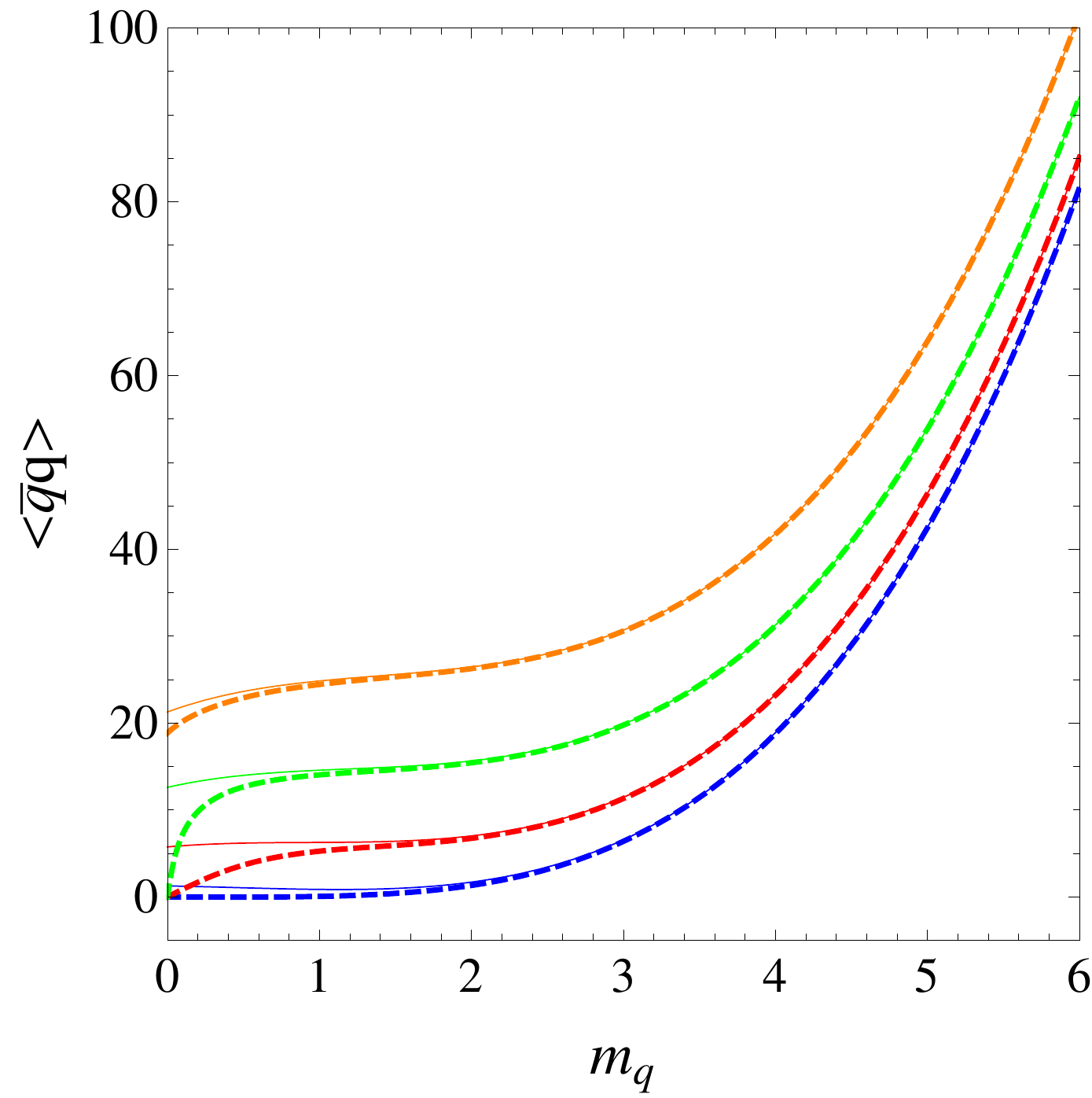}
\hfill
\includegraphics[width=7cm]{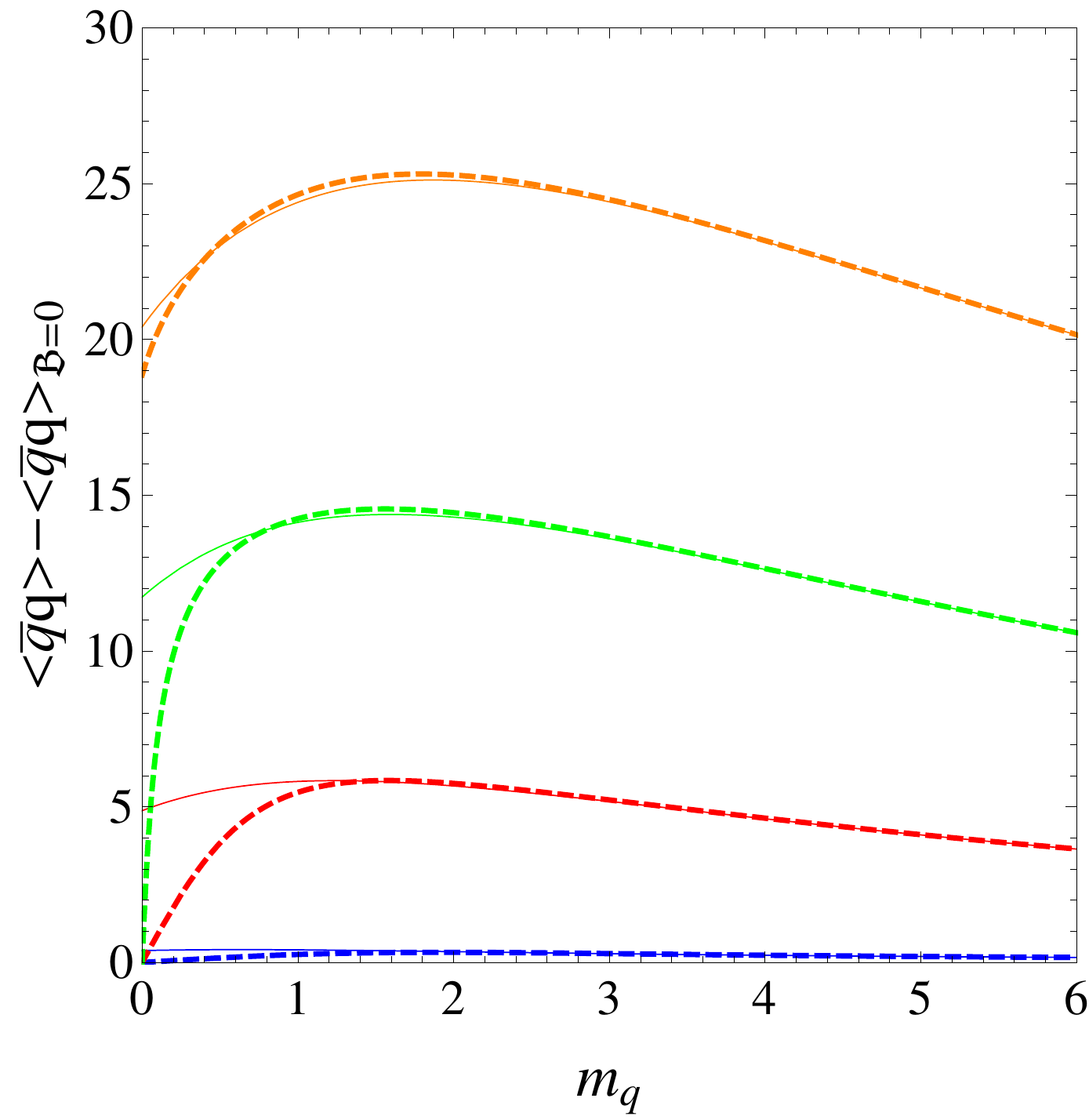}
\caption{{\bf (Right) Left panel:} (subtracted) chiral condensate as a function of the quark mass in the confined (solid) and deconfined (dashed) phases.  In both panels the magnetic field ${\cal B}$ takes the values $1$ (blue), $5$ (red), $9$ (green) and $13$ (orange).    We have used the formulae \eqref{quark-mass} and \eqref{quark-condensate} with ${\cal V}_0=1$  and $\zeta =1$. In both panels the quark mass and chiral condensate  are given in units where $z_{\Lambda}=z_T=1$. There is actually a nontrivial scaling in $M_{KK}$ and $T$ for the confined and deconfined phases, to be discussed in the next subsection. Note that the chiral condensate for the confined and deconfined phase have the same behaviour in the limit of heavy quark mass, suggesting a universal description. The chiral condensate on the left panel depends on the renormalisation scheme and  the chosen scheme was $\alpha_1=-1$. The subtracted chiral condensate on the right panel is independent of the renormalisation scheme. }
\label{Fig:Condensatevsc1}
\end{figure}

In figure \ref{Fig:Condensatevsc1} we plot the chiral condensate as a function of the quark mass 
in the confined and deconfined phases. To avoid the presence of the scheme dependent parameter we either fix 
it to $\alpha_1=-1$ (so that the chiral condensate becomes proportional to the vev coefficient $c_3$) or subtract the 
value of the chiral condensate at zero magnetic field. In this way the subtracted chiral condensate is 
independent of the renormalisation scheme.

\noindent
\noindent

\begin{figure}[ht]
\centering
\includegraphics[width=7cm]{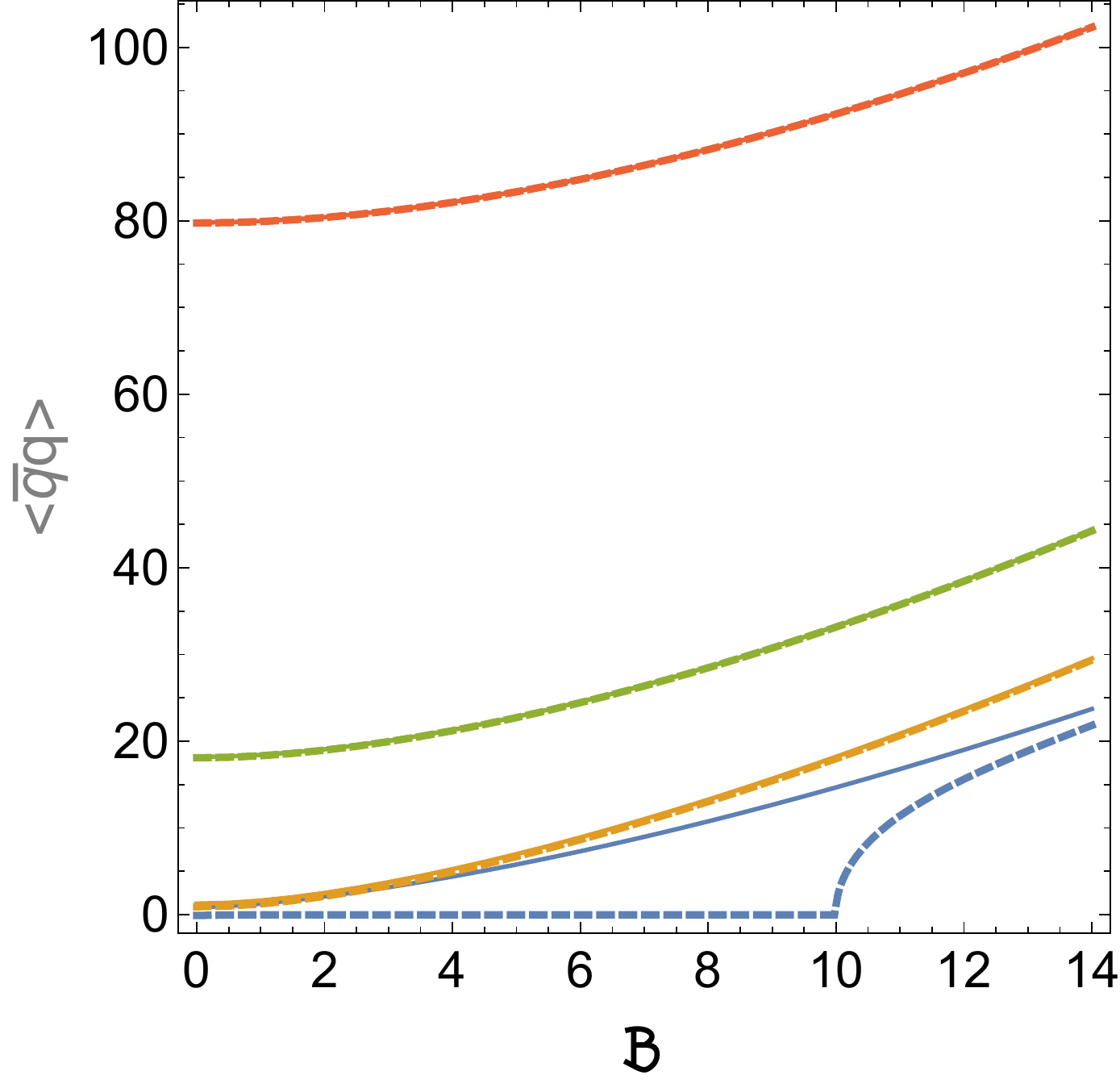}
\hfill
\includegraphics[width=7cm]{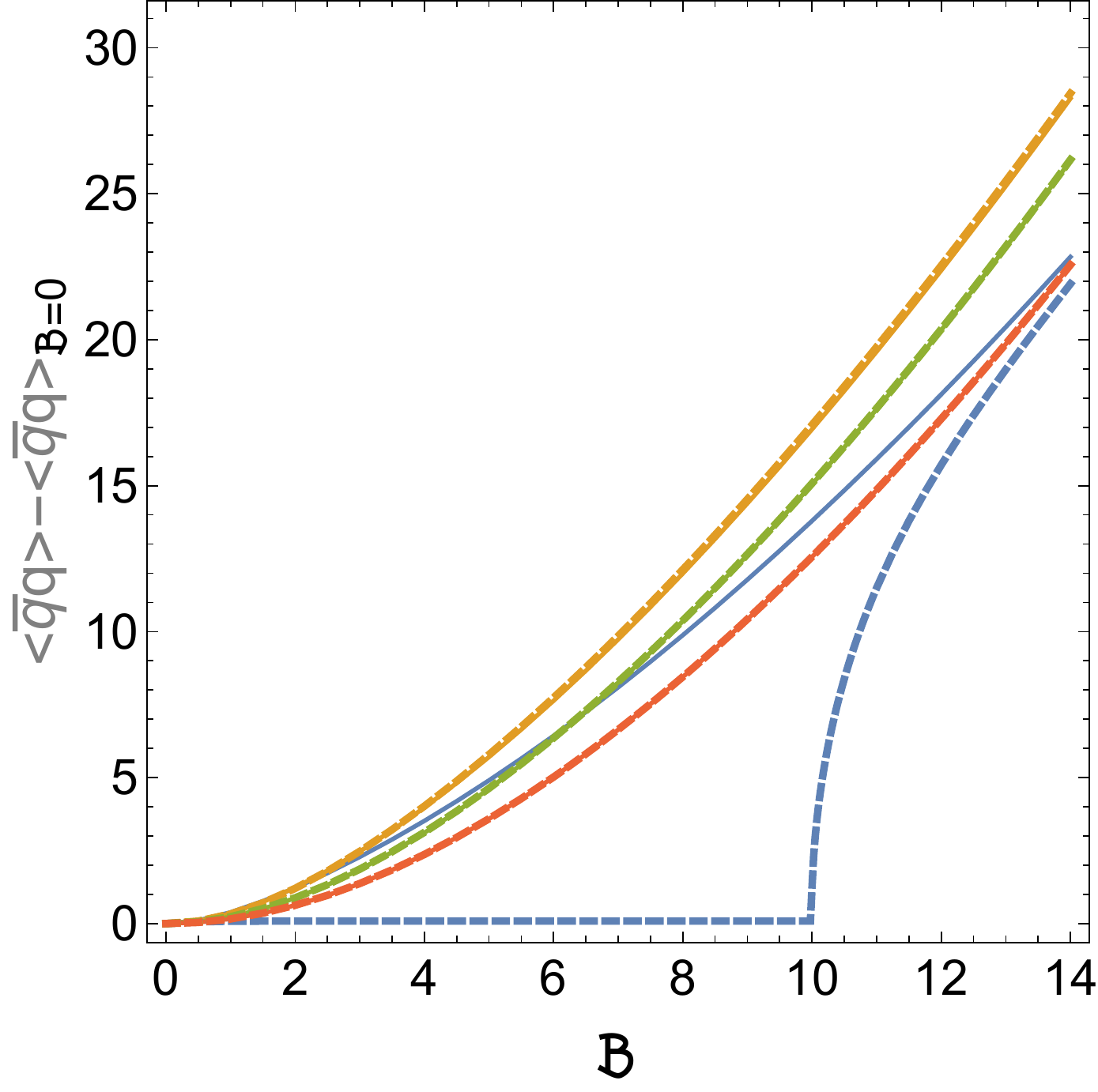}
\caption{ {\bf (Right) Left panel:} The (subtracted) chiral condensate as a function of the magnetic field in the confined (solid) and deconfined (dashed) phases.  In both panels the quark mass parameter $c_1$ takes the values $0$ (blue), $2$ (red), $4$ (green) and $6$ (orange).  We have used the formula \eqref{quark-condensate} with ${\cal V}_0=1$ and $\zeta =1$. In both panels the  chiral condensate and magnetic field are given in units where $z_{\Lambda}=z_T=1$. There is actually a nontrivial scaling in $M_{KK}$ and $T$ for the confined and deconfined phases, to be discussed in the next subsection. The chiral condensate on the left panel depends on the renormalisation scheme and  the chosen scheme was $\alpha_1=-1$. The subtracted chiral condensate on the right panel is independent of the renormalisation scheme.}
\label{Fig:SubtrCondensate}
\end{figure}

In figure \ref{Fig:SubtrCondensate} we plot the chiral condensate as a function of the magnetic field
in the confined and deconfined phases. The presence of the scheme dependent parameter is fixed/subtracted as in 
figure \ref{Fig:Condensatevsc1}. Note that for $c_1=0$ the chiral condensate changes drastically as the 
magnetic field crosses the critical value ${\cal B} =10$. This is a consequence of the spontaneous chiral symmetry 
breaking that is analysed in figures \ref{DC-critical-B} and \ref{DC-above-critical-B-2}. A common feature for both 
figures  \ref{Fig:Condensatevsc1} and \ref{Fig:SubtrCondensate} is that for large values of the mass or a very strong 
magnetic field the chiral condensate of the confined the deconfined phases almost coincide. 
This is related to the fact that in both phases we work in units where $z_{\Lambda}=z_T=1$.  We have performed a  fit in the regime of large ${\cal B}$ and found that the chiral condensate grows as $\# {\cal B}^{3/2}$ in both phases. This asymptotic behaviour will be important in the last subsection, where we compare our results against lattice QCD.
In the specific gravity model that we are working on, it seems that the IR boundary conditions do not affect significantly 
the physics in the limit of very large mass of the quarks or very strong magnetic field. Note that this phenomenon is generally obtained in holographic brane constructions where in the regime of large quark mass (or large magnetic field) the brane is always far from away from the deep IR (see e.g. \cite{Erdmenger:2007cm} ).

\noindent
\noindent

\begin{figure}[ht]
\centering
\includegraphics[width=7cm]{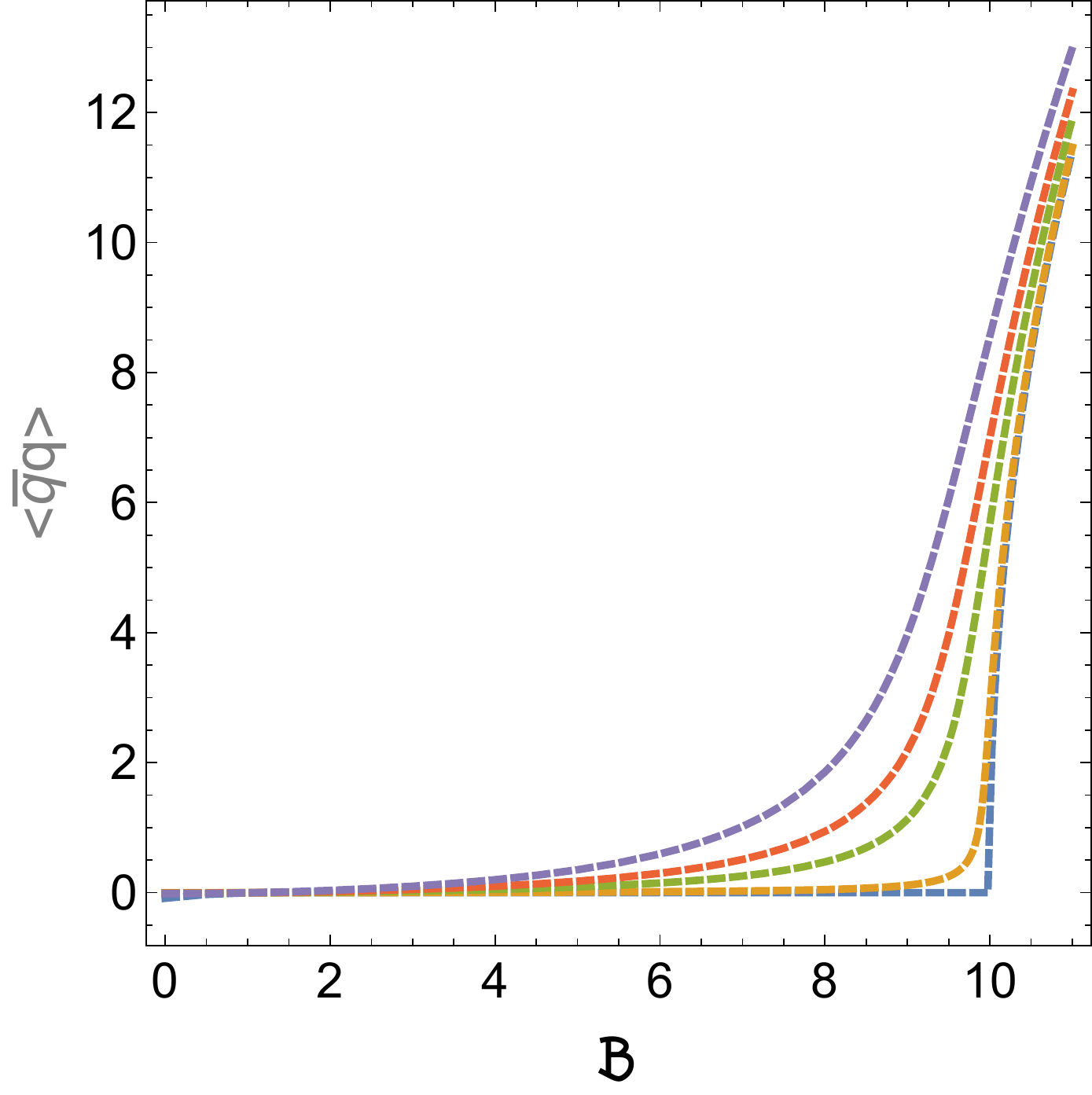}
\caption{The unsubtracted chiral condensate for $c_1=0, 0.001, 0.01, 0.02, 0.04$ in the deconfined phase. As the quark mass parameter $c_1$ increases the second order transition becomes a crossover.}
\label{Fig:CondensatevsbSmallc1}
\end{figure}

In figure \ref{Fig:CondensatevsbSmallc1} we elaborate on the spontaneous chiral symmetry 
breaking that occurs when the magnetic field exceeds the critical value ${\cal B} =10$. For that reason we plot 
the unsubtracted chiral condensate as a function of the magnetic field ${\cal B}$,  for values of the quark mass that 
are close to zero.
The analysis indicates that at zero quark mass there is a second order phase transition. 
As the quark mass increases this phase transition degenerates to a crossover. The scaling in $T$ for the deconfined phase implies that the chiral transition when varying ${\cal B}$ has actually two interpretations. Since ${\cal B}$ is proportional to $B/T^2$, either we fix $T$ and increase the dimensionful magnetic field $B$ or we fix $B$ and decrease the temperature. In the last subsection we will present a plot that describes the chiral transition when varying the temperature, at fixed values of the magnetic field.

\noindent
\noindent


\subsection{The magnetisation}
\label{magnetisation}

In this section we focus the analysis on the computation of the magnetisation. 
As we did for the condensate, we restrict the analysis to the confined phase. 
Magnetisation is defined in the usual way as 
\begin{equation} \label{def-M}
M \, =\, -  \,\left(\frac{\partial F}{\partial {\cal B}}\right)_{z_{\Lambda}} \, =  \, \left(\frac{\partial S_{ren}}{\partial {\cal B}}\right)_{z_{\Lambda}}
\end{equation}
where $F$ is the free energy
and the calculation is at fixed $z_{\Lambda}$ (the tip of the cigar). In the deconfined phase the calculation 
will be performed at fixed temperature ($z_T$ is inversely proportional to the temperature). 

Starting from \eqref{def-M} we will first compute the part of the magnetisation due to the regularised action in 
\eqref{reg-action} 
\begin{eqnarray} \label{M-I}
M_I &  = &  \left(\frac{\partial S_{reg}}{\partial {\cal B}}\right)_{z_{\Lambda}} 
= \int_\epsilon^{1} du \Bigg[\frac{\partial {\cal L}_{DBI} }{\partial {\cal T}} \, \frac{\partial {\cal T}}{\partial {\cal B}} + 
\frac{\partial {\cal L}_{DBI} }{\partial {\cal T}'} \, \frac{\partial {\cal T}'}{\partial {\cal B}} +  
\frac{\partial {\cal L}_{DBI} }{\partial {\cal B}} \Bigg]
\nonumber \\ [5pt]
&= &  \int_\epsilon^{1} du \frac{\partial {\cal L}_{DBI} }{\partial {\cal B}}  +
\Bigg[\frac{\partial {\cal L}_{DBI} }{\partial {\cal T}'} \, \frac{\partial {\cal T}}{\partial {\cal B}}\Bigg]_{\epsilon}^1
\end{eqnarray}
where in the last step we have used the equation of motion for the tachyon. That expression can be further 
simplified. The contribution from the boundary terms reads
\begin{equation} \label{M-boundary}
\frac{\partial {\cal L}_{DBI} }{\partial \tau'}\Bigg|_1= \, 0 \quad \& \quad
\frac{\partial {\cal L}_{DBI} }{\partial \tau'} \, \frac{\partial \tau}{\partial {\cal B}} \Bigg|_{\epsilon} \approx 
-{2 \over 3} \, {\cal V}_0  \, c_1 \, \frac{\partial c_3}{\partial {\cal B}} + \cdots
\end{equation}
The second ingredient of \eqref{def-M} comes from the contribution of the functional derivative of  \eqref{counterterms}
with respect to $\cal B$. As can be seen by substituting the approximate expression for the tachyon  contribution 
of that term is 
\begin{equation} \label{M-II}
M_{II}  =  - {\cal V}_0 \Big [ 2  \, {\cal B} \left ( \log \epsilon  + \alpha_2 \right ) + {2 \over 3}   \, c_1 \, \frac{\partial c_3}{\partial {\cal B}}  \Big ]\, . 
\end{equation}
Combining \eqref{M-I}, \eqref{M-boundary} and \eqref{M-II} we 
arrive to the following expression for the magnetisation
\begin{equation} \label{magnetisation}
M   = - 2 \, {\cal V}_0 \, {\cal B} \left ( \log \epsilon  + \alpha_2 \right ) + \,  \int_\epsilon^{1} du \frac{\partial {\cal L}_{DBI} }{\partial {\cal B}} \, . 
\end{equation}
It can be checked explicitly that as $u \rightarrow \epsilon$ the infinite contribution from the last integral of \eqref{magnetisation} is canceled by the logarithmic counterterm. Note that the magnetisation depends on the renormalisation scheme through the parameter $\alpha_2$.

\begin{figure}[ht]
\centering
\includegraphics[width=7cm]{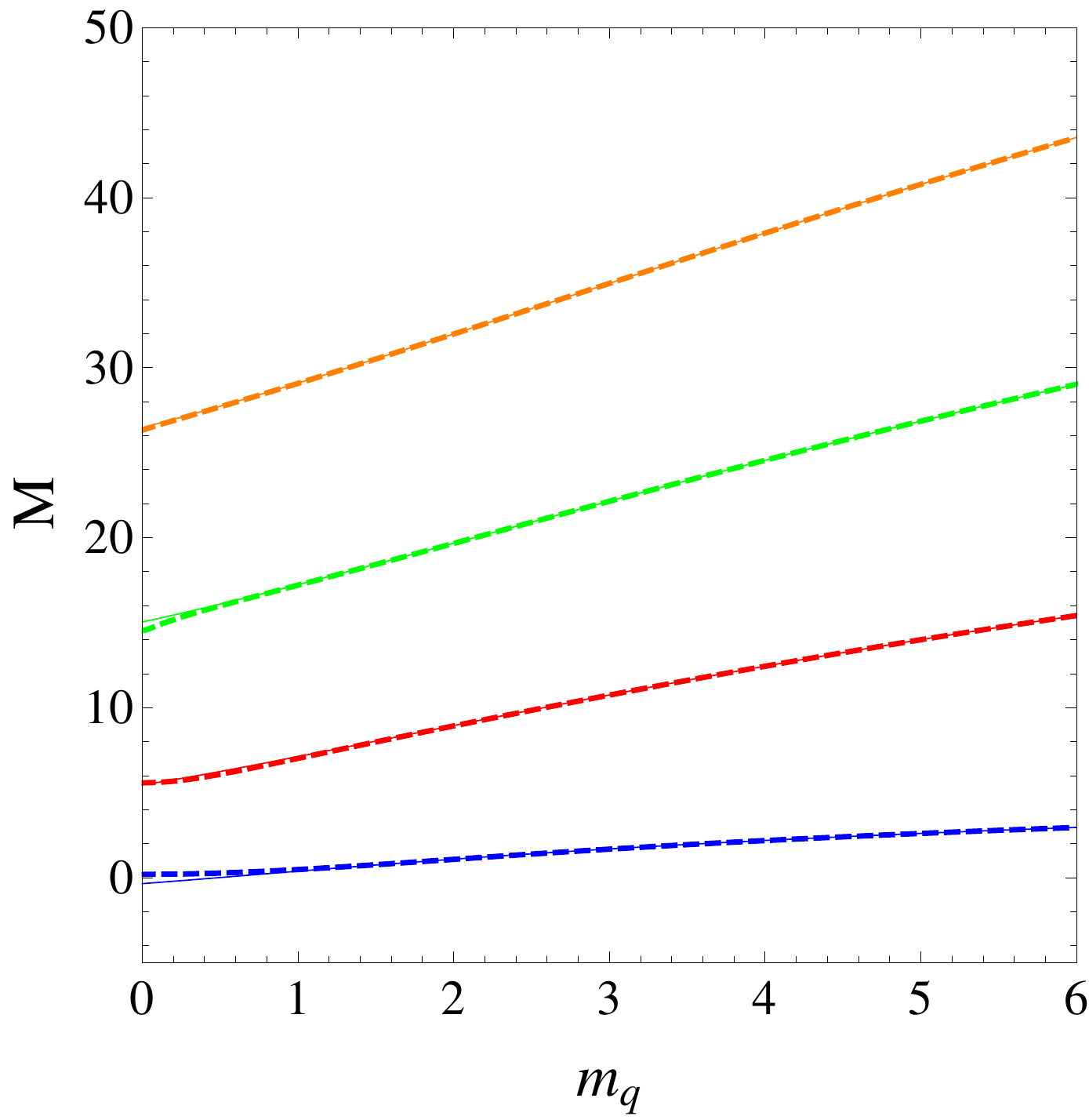}
\hfill
\includegraphics[width=7cm]{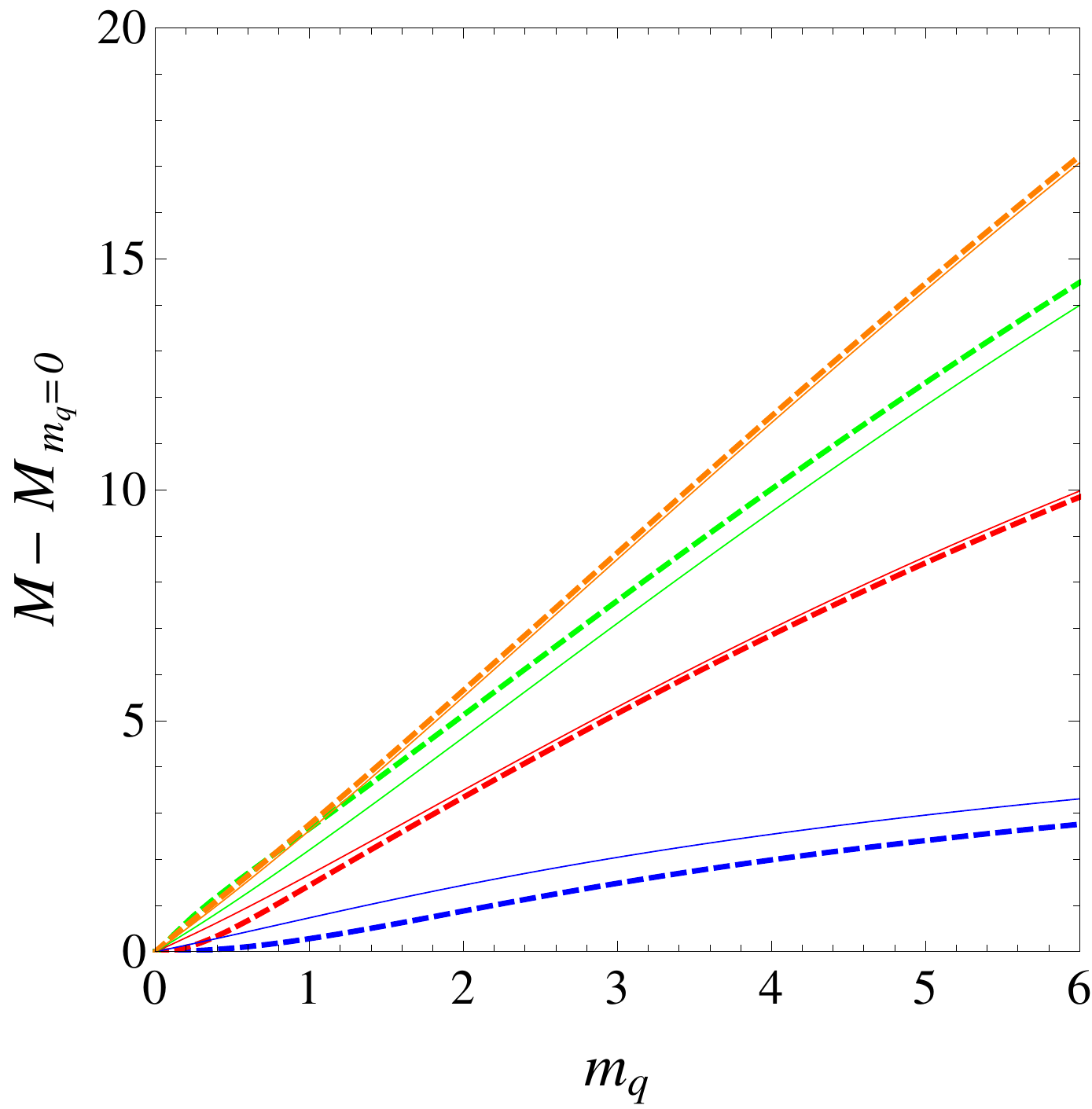}
\caption{{\bf (Right) Left panel:} The (subtracted) magnetisation as a function of the quark mass in the confined (solid) and deconfined (dashed) phases. We have used the formulas \eqref{quark-mass} and  \eqref{magnetisation} with $\zeta=1$ and ${\cal V}_0=1$. In both panels the magnetic field ${\cal B}$ takes the values $1$ (blue), $5$ (red), $9$ (green) and $13$ (orange).  In both panels the  magnetisation and quark mass are given in units where $z_{\Lambda}=z_T=1$. There is actually a nontrivial scaling in $M_{KK}$ and $T$ for the confined and deconfined phases, to be discussed in the next subsection The magnetisation on the left panel depends on the renormalisation scheme and  the chosen scheme was $\alpha_2=0$. The subtracted magnetisation on the right panel is independent of the renormalisation scheme.}
\label{Fig:Magnetisationvsc1}
\end{figure}

In figure \ref{Fig:Magnetisationvsc1} we plot the magnetisation as a function of the quark mass 
in the confined and deconfined phases. On the left panel of the figure we fix the scheme dependent parameter to
$\alpha_2=0$,\footnote{A similar renormalisation scheme was considered in a lattice QCD approach \cite{Bali:2013owa}.}  while in the right panel we plot the difference between the magnetisation from equation 
\eqref{magnetisation} and the magnetisation for zero quark mass mass. In this way the scheme dependent parameter
vanishes and the subtracted magnetisation is independent of the renormalisation scheme.

\noindent
\noindent

\begin{figure}[ht]
\centering
\includegraphics[width=7cm]{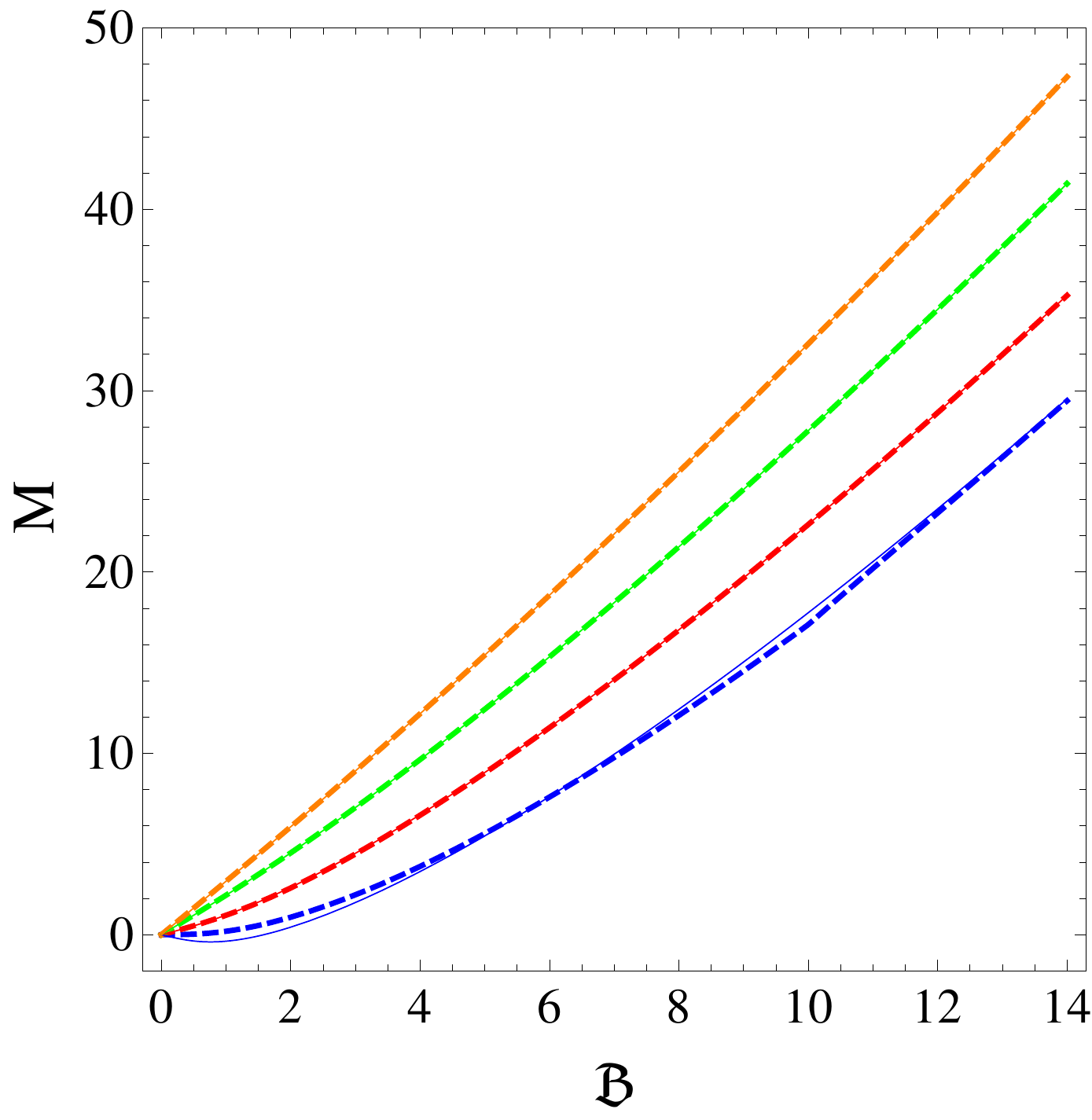}
\hfill
\includegraphics[width=7cm]{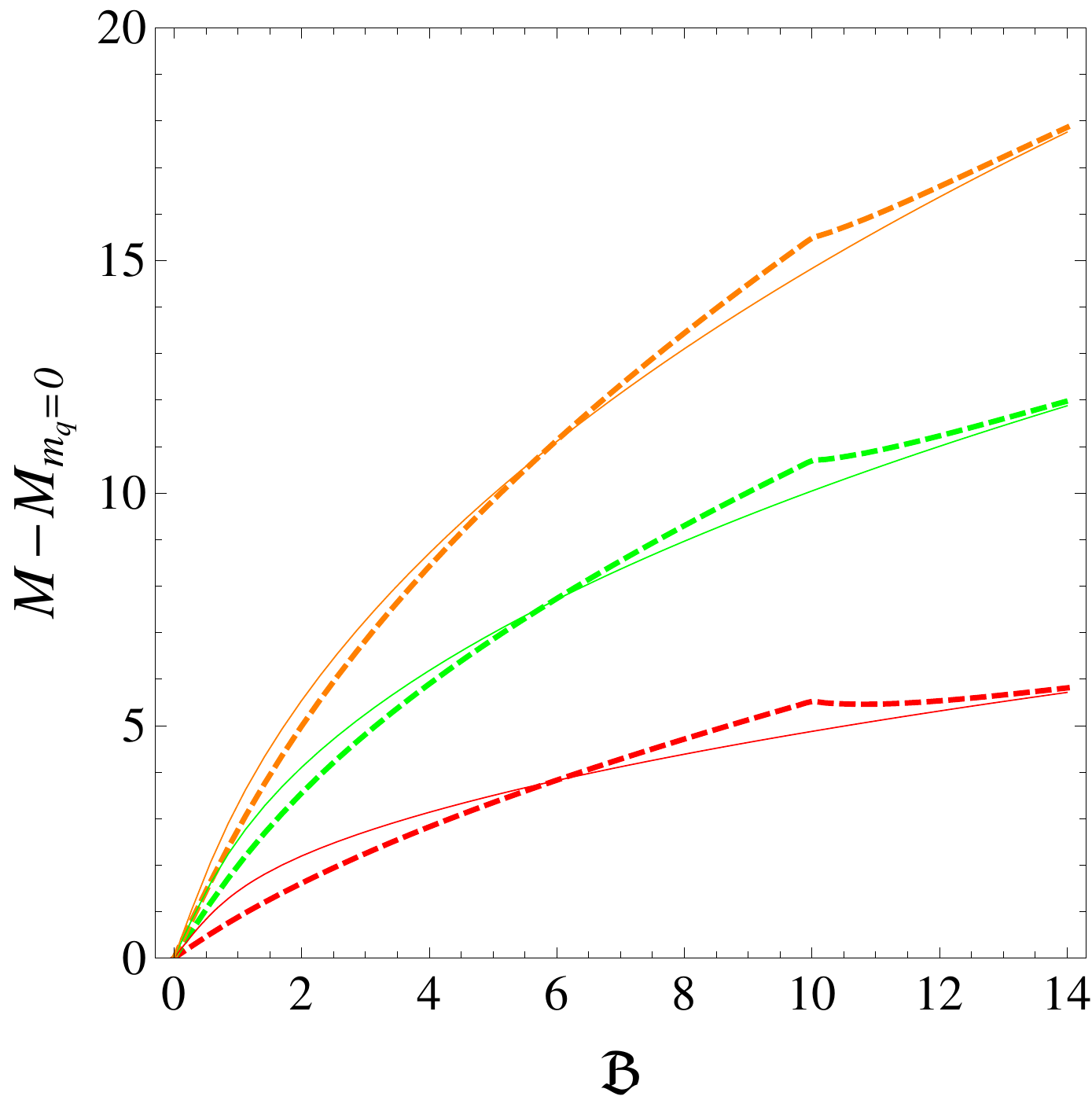}
\caption{{\bf (Right) Left panel:} Magnetisation  as a function of the magnetic field in the confined (solid) and deconfined (dashed) phases.   In both panels the quark mass parameter $c_1$ takes the values $0$ (blue), $2$ (red), $4$ (green)  and $6$ (orange).  We have used the formula \eqref{magnetisation} with ${\cal V}_0=1$.  In both panels the  magnetisation and magnetic field are given in units where $z_{\Lambda}=z_T=1$. The magnetisation on the left panel depends on the renormalisation scheme and  the chosen scheme was $\alpha_2=0$. The subtracted magnetisation on the right panel is independent of the renormalisation scheme. }
\label{Fig:Magnetisationvsb}
\end{figure}

In figure \ref{Fig:Magnetisationvsb} we plot the magnetisation as a function of the magnetic field
in the confined and deconfined phases. The presence of the scheme dependent parameter is fixed/subtracted as in 
figure \ref{Fig:Magnetisationvsc1}. In the subtracted magnetisation there is a discontinuity in the first derivative 
at ${\cal B} =10$ (it will become more evident in the plot of susceptibility), which is a consequence of the 
spontaneous chiral symmetry breaking in the deconfined case. Note that the calculation is performed in two 
(complementary) ways: First, we apply the formula \eqref{magnetisation} and in the following we calculate the 
numerical derivative of the renormalised free energy \eqref{ren-action} with respect to the magnetic field. 
The results we obtain from the two calculations are identical; this is a non-trivial confirmation of the formula we derived in \eqref{magnetisation}, especially taking into account that the renormalised free energy 
depends on the scheme dependent parameter $\alpha_1$ while the magnetisation does not.

\noindent
\noindent

Since we do not have an analytic solution for the tachyon profile for either small or large values 
of the magnetic field, we cannot approximate the magnetisation. However from the numerical analysis, 
we have verified that for large values of  $\cal B$ the magnetisation in the left panel of figure 
\ref{Fig:Magnetisationvsb} (for  $\alpha_2=0$) is approximated 
by the following expression 
\begin{equation} \label{magnetisation-largeB}
M   \approx \frac{\partial c_3}{\partial {\cal B}} \, {\cal B} + 2 \, c_1 \quad \text{for} \quad {\cal B} \gg 1 \, . 
\end{equation}

\begin{figure}[ht]
\centering
\includegraphics[width=7cm]{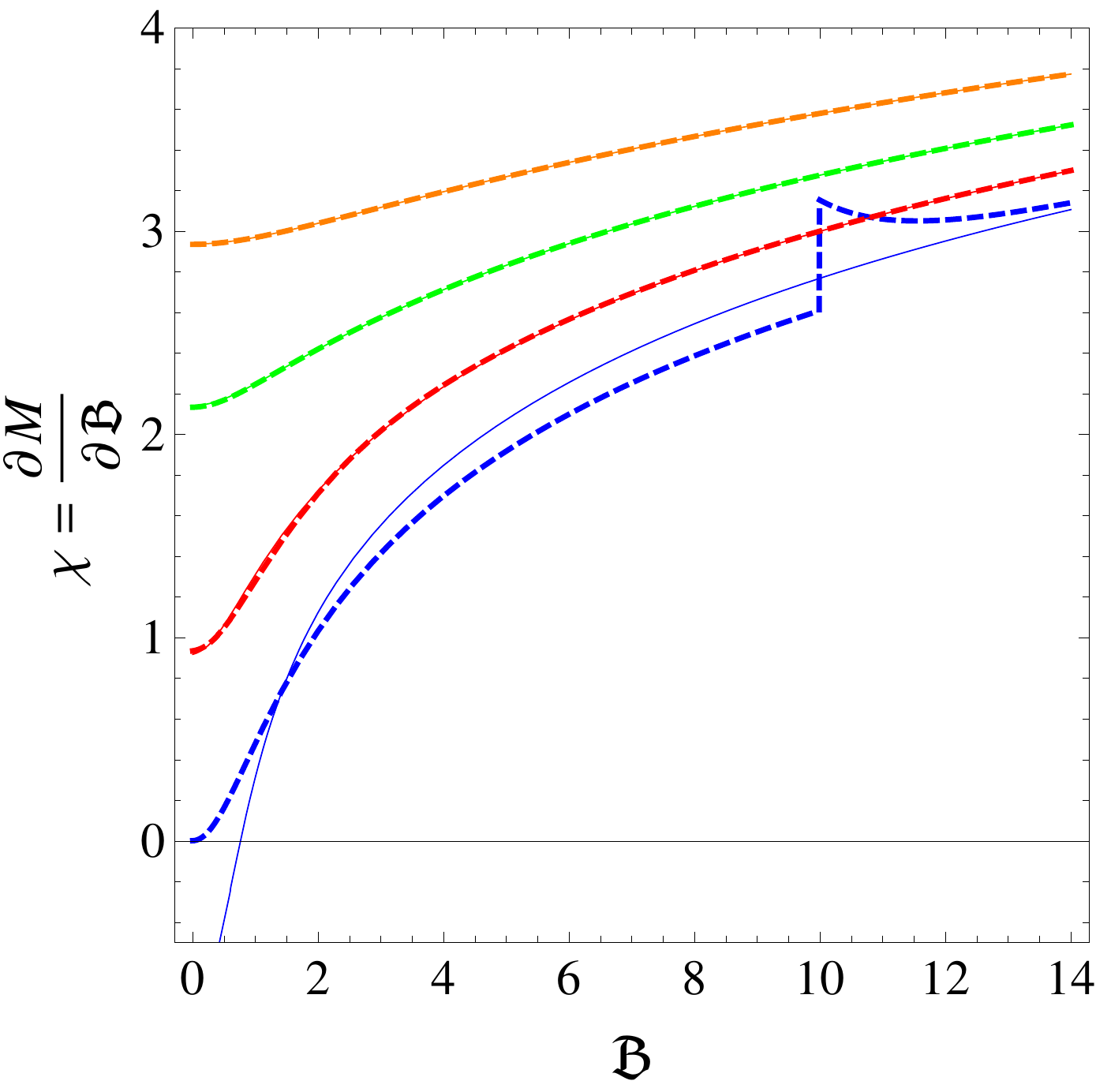}
\hfill 
\includegraphics[width=7cm]{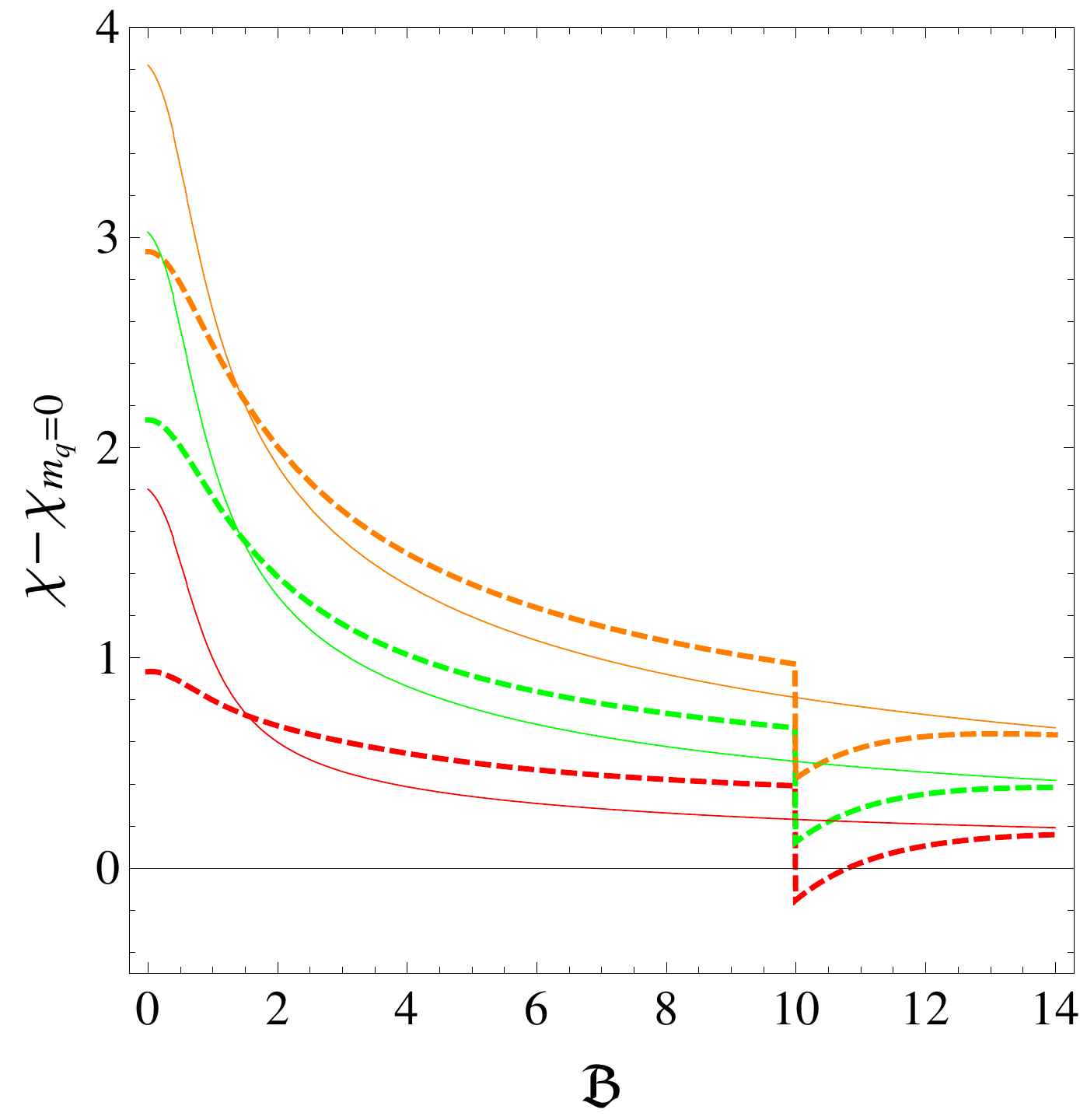}
\caption{{\bf (Right) Left panel:} The (subtracted) susceptibility  as a function of the magnetic field in the confined (solid) and deconfined (dashed) phases. The susceptibility is defined by $\chi = \partial M/ \partial {\cal B}$ where $M$ is the magnetisation. Therefore the curves in this figure correspond to the slopes of the curves in Fig. \ref{Fig:Magnetisationvsb}.  In both panels the quark mass parameter $c_1$ takes the values $0$ (blue), $2$ (red), $4$ (green)  and $6$ (orange). The quark mass and magnetic field are given in units where $z_{\Lambda}=z_T=1$. The susceptibility on the left panel depends on the renormalisation scheme which was chosen as $\alpha_2=0$. The subtracted susceptibility on the right panel is independent of the renormalisation scheme.}
\label{Fig:Susceptibilityvsb}
\end{figure}

In figure \ref{Fig:Susceptibilityvsb} we plot the susceptibility (first derivative of the magnetisation with respect 
to the magnetic field) as a function of the magnetic field in the confined and deconfined phases.
The presence of the scheme dependent parameter $\alpha_2$ is fixed/subtracted as in 
figure \ref{Fig:Magnetisationvsc1}. The jump of the susceptibility of the deconfined phase 
at the critical value of the magnetic field  ${\cal B}=10$
in the right panel of figure \ref{Fig:Susceptibilityvsb} is inherited from the right panel of figure \ref{Fig:Magnetisationvsb}.
The jump in the susceptibility that appears in the left panel of figure \ref{Fig:Susceptibilityvsb}, is due to the fact that this curve corresponds 
to zero value for the the quark mass parameter $c_1$.

\noindent
\noindent


\subsection{The condensate contribution to the magnetisation}

\begin{figure}[ht]
\centering
\includegraphics[width=7cm]{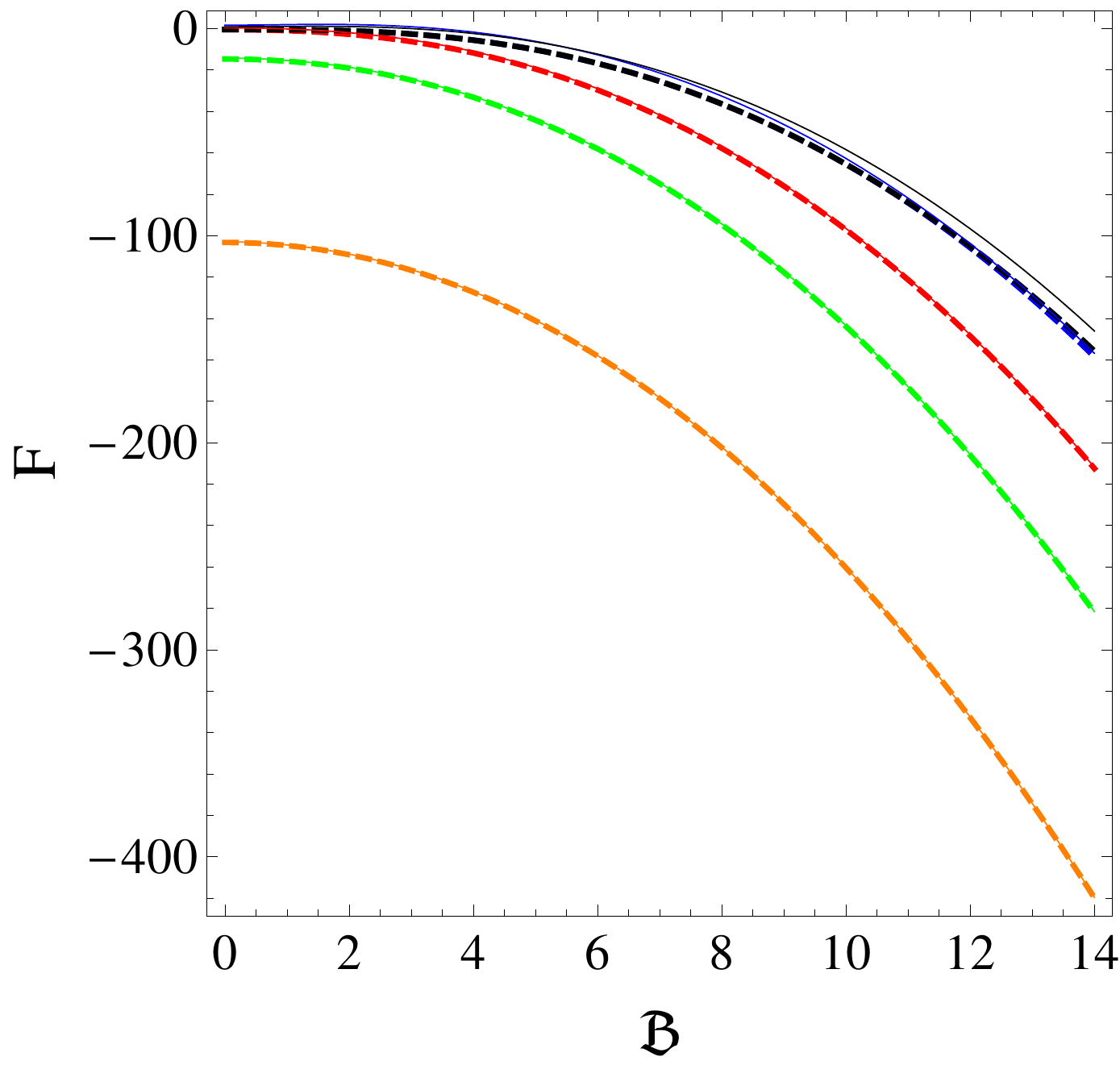}
\caption{Renormalised free energy as a function of the magnetic field in the confined (solid) and deconfined (dashed) phases.  The quark mass parameter $c_1$ takes the values $0$ (blue), $2$ (red), $4$ (green)  and $6$ (orange). The black solid (dashed) line depicts the renormalised free energy for the case ${\cal T}=0$ (pure magnetic energy) in the confined (deconfined) phase. The free energy and magnetic field are given in units where $z_{\Lambda}=z_T=1$. }
\label{Fig:FreeEn}
\end{figure}

We will extract the contribution to the free energy due to chiral symmetry breaking, which means in our framework having a nonzero tachyon. We start with the dimensionless (bare) free energy density.
In the Lorentzian signature the free energy density is identified with the Hamiltonian density, i.e. 
\begin{equation} \label{FreeEn}
F_{bare} = H_{DBI} =  - \int_{\epsilon}^1 du \, {\cal L}_{DBI} \, .
\end{equation}
In the confined phase, for instance, ${\cal L}_{DBI}$ is given in \eqref{LDBIConf}. 
We work in units where $z_{\Lambda}=z_T=1$ but there is a nontrivial scaling in $M_{KK}^4$ (vacuum energy) and $T^4$ (plasma free energy) for the confined and deconfined phases to be addressed in section \ref{sec:Lattice}. 
When the tachyon is zero the (confined) free energy reduces to 
\begin{equation} \label{FreeEnConfTauEq0}
F_{bare}^{({\cal T}=0)} =  2  \, {\cal V}_0 \int_{\epsilon}^1 du   \,  \frac{1}{u^5 \, \sqrt{1-u^5}} \, 
\sqrt{1+ {\cal B}^2 \, u^4}
\end{equation}
and the counterterms contribution becomes 
\begin{equation} \label{FreeEnctTaueq0}
 F_{ct}^{({\cal T}=0)}  =  {\cal V}_0 \Big [ - \frac12 \epsilon^{-4}  + {\cal B}^2 \ln \epsilon + \alpha_2 \, {\cal B}^2 \Big ] \,. 
\end{equation}
The renormalised free energy takes the form $F = F_{bare} + F_{ct}$. 
In figure \ref{Fig:FreeEn} we compare the full renormalised free energy $F$ against the zero tachyon free energy 
$F^{({\cal T}=0)}$. Notice the black curve (zero tachyon) and the two (solid and dashed) blue curves (i.e. $c_1=0$).
In the confined case the black and blue solid curves never coincide, since we always have spontaneous 
chiral symmetry breaking. 
However in the deconfined case and for ${\cal B}<10$ the black and 
blue dashed curves will be on top of each other, while for ${\cal B}>10$ the curves will be different. 
This last observation is hard to see in figure \ref{Fig:FreeEn}.

\noindent


Similarly, at zero tachyon, the bare magnetisation  reduces to
\begin{equation} \label{MagBareTaueq0}
M_{bare}^{({\cal T}=0)} = - 2  \, {\cal V}_0 \, {\cal B} \int_{\epsilon}^1 du   \,  \frac{1}{u \,
\sqrt{1-u^5}} \, 
\left ( 1+ {\cal B}^2 \, u^4 \right )^{-1/2}  
\end{equation}
and the counterterms contribution becomes
\begin{equation} \label{MagctTaueq0}
M_{ct}^{({\cal T}=0)} = - {\cal V}_0 \Big [ 2 {\cal B} \ln \epsilon + 2 \, \alpha_2 {\cal B} \Big ]   \, . 
\end{equation}
The renormalised magnetisation takes the form $M = M_{bare} + M_{ct}$. 
It is interesting to consider the following quantity
\begin{equation} \label{Deltamag}    
\Delta M = M - M^{({\cal T}=0)} 
\end{equation}
that contains the tachyon contribution to the renormalised magnetisation.

\begin{figure}[ht]
\centering
\includegraphics[width=7cm]{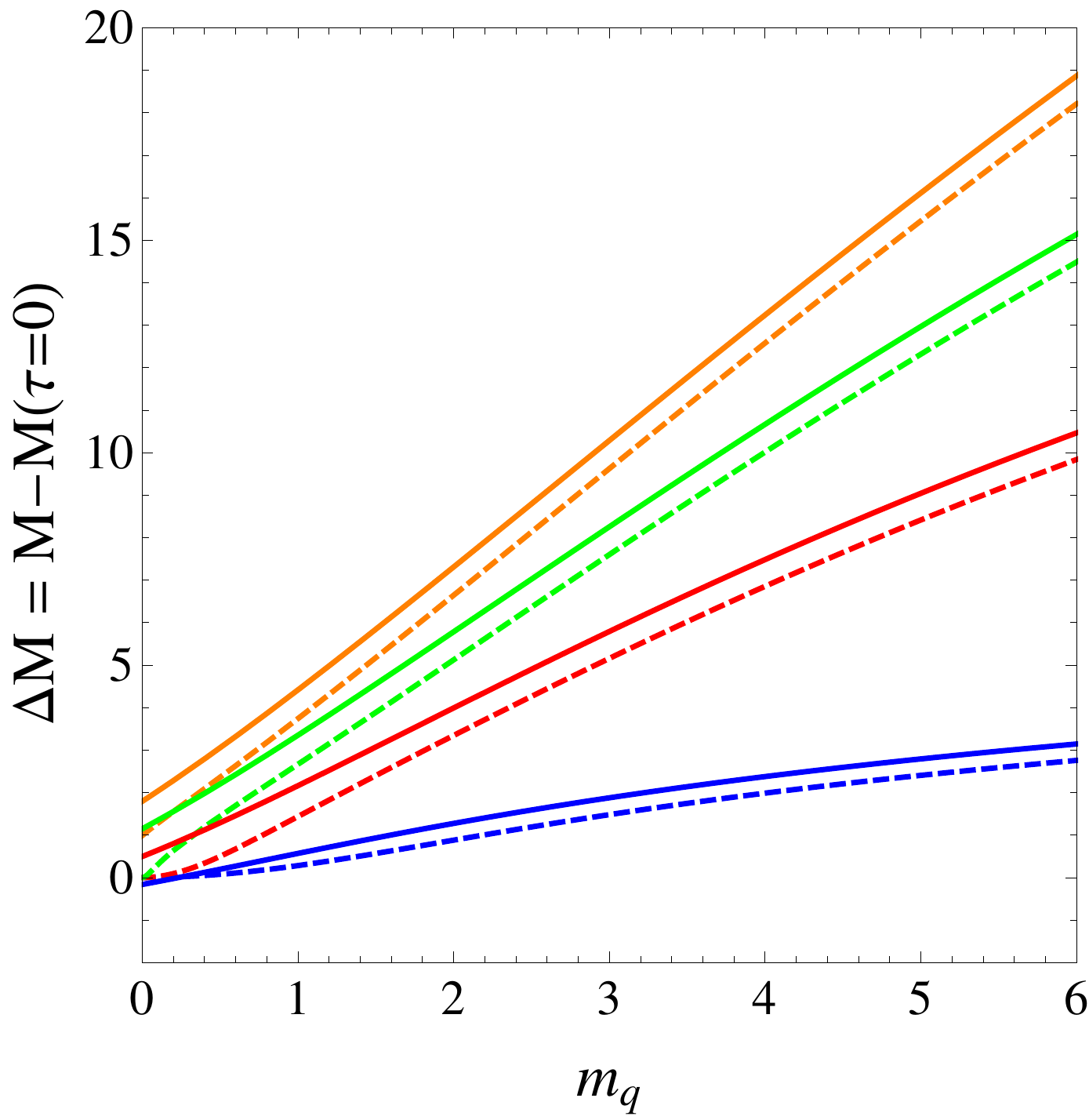}
\hfill
\includegraphics[width=7cm]{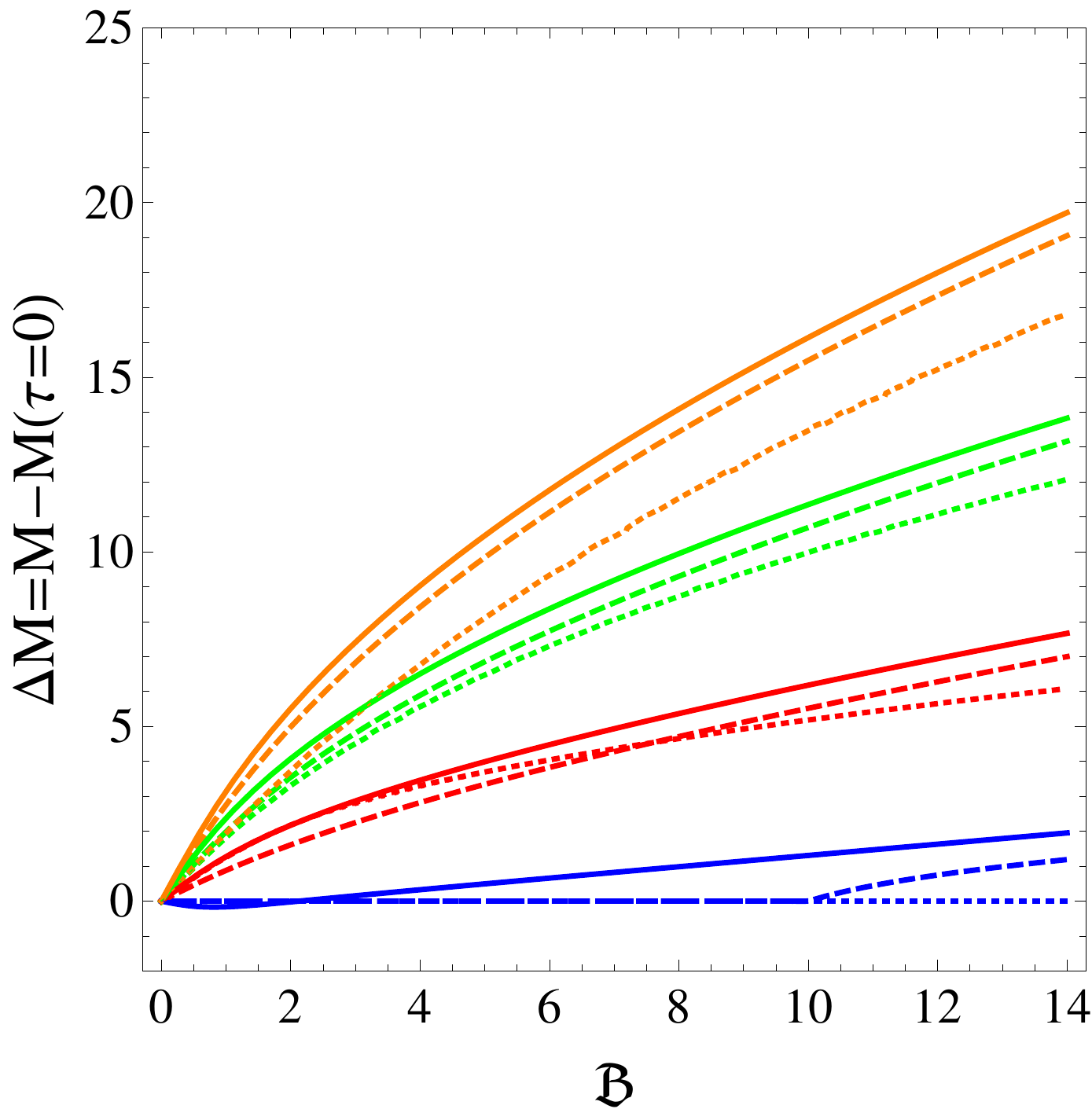}
\caption{{\bf Left panel:} Tachyon contribution to the  magnetisation as a function of the quark mass in the confined (solid) and deconfined (dashed) phases. The magnetic field ${\cal B}$ takes the values $1$ (blue), $5$ (red), $9$ (green) and $13$ (orange). {\bf Right panel:}  Tachyon contribution to the magnetisation as a function of the magnetic field in the confined (solid) and deconfined (dashed) phases.   The quark mass parameter $c_1$ takes the values $0$ (blue), $2$ (red), $4$ (green)  and $6$ (orange). The dotted  curves were obtained using the empirical formula \eqref{analyticmag}.  The magnetisation, quark mass and magnetic field are given in units where $z_{\Lambda}=z_T=1$.}
\label{Fig:DeltaMag}
\end{figure}

In figure \ref{Fig:DeltaMag} we plot the subtracted magnetisation that is defined in \eqref{Deltamag} 
as a function of the quark mass and
as a function of the magnetic field,  in the confined and deconfined phases.
For the numerical analysis we fix the renormalisation scheme to $\alpha_1 = -1$ and $\alpha_2 = 0$. 
The results can be easily extended to other renormalisation schemes.\footnote{It may be possible to redefine 
the free energy in a scheme-independent way, considering two simultaneous subtractions.} 
The main motivation behind this plot is to emphasise the contribution to the magnetisation 
coming from the chiral condensate.
  
There is  an important thermodynamic identity regarding mixed partial derivatives of the free energy 
\begin{equation}
\frac{ \partial^2 F}{ \partial {\cal B} \, \partial m_q} = \frac{ \partial^2 F}{\partial m_q \partial {\cal B}}    \, . \end{equation}
\noindent
This identity implies the following relation between the magnetisation and condensate
\begin{equation} \label{MagCond}
 \frac{ \partial \langle \bar q q \rangle}{ \partial {\cal B}}= \frac{ \partial M}{\partial m_q} = \frac{ \partial \Delta M}{\partial m_q}  \,.
\end{equation}
In the model at hand, the thermodynamic identity \eqref{MagCond} holds because the renormalised (magnetic) free energy is smooth in $m_q$ and ${\cal B}$.  We have explicitly checked this identity in the confined and deconfined phases.

The curves on the left panel of figure \ref{Fig:DeltaMag} suggest the following approximation
\begin{equation} \label{approx1}
 \frac{ \partial \langle \bar q q \rangle}{ \partial {\cal B}}= \frac{ \partial \Delta M}{\partial m_q} \approx f_1({\cal B})\end{equation}
where $f_1({\cal B})$ depends only on the magnetic field. Integrating \eqref{approx1} in $m_q$ we find that
\begin{equation} \label{ApproxDeltaM}
\Delta M \approx m_q  \frac{ \partial \langle \bar q q \rangle}{ \partial {\cal B}}  + f_0({\cal B})  \, , 
\end{equation}
where $f_0({\cal B})$ depends solely on ${\cal B}$ and can be identified with the subtracted magnetisation at 
zero quark mass, i.e. $\Delta M \vert_{m_q=0} = f_0 ({\cal B})$. Setting to zero the value of $f_0({\cal B})$
we obtain the following empirical formula 
\begin{equation} \label{analyticmag}    
\Delta M_{Emp} =  m_q \frac{\partial \langle \bar q q \rangle}{\partial {\cal B}}
\end{equation}
that provides a crude but reasonable approximation for $\Delta M$, 
as can be seen from the comparison with \eqref{ApproxDeltaM} that appears on the right panel of 
figure \ref{Fig:DeltaMag}. The formula \eqref{analyticmag} can be thought as the dominant contribution to the magnetisation arising from the chiral condensate.

\noindent
\noindent


\section{Comparison to lattice QCD and the Sakai-Sugimoto model}
\label{sec:Lattice}

The lattice QCD formalism has been a powerful method for investigating magnetic catalysis (MC) and inverse magnetic catalysis (IMC), see e.g. \cite{Bali:2012zg,Bali:2013owa,Bali:2014kia,DElia:2018xwo,Bali:2020bcn}. In this section we  compare some of our results for the condensate and magnetisation at finite temperature with lattice QCD results. In the top-down approach to holographic QCD, the Sakai-Sugimoto model \cite{Sakai:2004cn,Sakai:2005yt} stands out as the closest description of large-N QCD in the strongly coupled regime. At the end of this section we compare our results for the chiral transition at finite temperature and magnetic field against the results obtained in \cite{Johnson:2008vna} for the Sakai-Sugimoto model.

\subsection{Dimensionful and dimensionless parameters}

For all of the previous numerical analysis we have been working in dimensionless variables. However, in order to connect with lattice QCD results we have to convert to dimensionful quantities.

We remind the reader that the IR parameter of the theory in the confined background is the length scale $z_{\Lambda} = (5/2) M_{KK}^{-1}$. The free parameters of the model are $\zeta = \sqrt{N_c}/2 \pi$, ${\cal V}_0=V_0 R^5$ and $m_\tau$ which are associated with the magnitude of the tachyon potential and the tachyon mass respectively, and $\beta R^{-2}$. In \cite{Iatrakis:2010jb} it was shown that, in order to obtain the appropriate normalisation for the 2-point correlation functions, the model parameters should obey the relations\footnote{The dictionary between our notation and that in \cite{Iatrakis:2010jb} is in \eqref{dictionary}. Also our $\zeta$ corresponds to $\beta^{-1}$ in \cite{Iatrakis:2010jb}.}
\begin{equation}
{\cal V}_0 = \frac{N_c}{16 \pi^2} k \quad , \quad 
\left ( \beta R^{-2} \right )^2 = \frac{4}{3 k}  \quad , \quad 
(\zeta m_{\tau})^2 =  \frac{6}{k} \,,
\end{equation}
where $k$ is a parameter that controls the meson phenomenology. 
Without loss of generality, we can fix  $m_{\tau}$ to $1$. We also take $N_c=3$ and use the phenomenological values
\begin{equation}
k \approx\frac{18}{\pi^2} \quad , \quad z_{\Lambda}^{-1} \approx 0.55 \, {\rm GeV}
\quad , \quad c_1^* \approx 0.0094 \, ,
\end{equation}
obtained in  \cite{Iatrakis:2010jb} from a fit to the meson spectrum.  Using these phenomenological quantities we can  fix the free parameters of the model to:
\begin{equation} \label{phenom}
  {\cal V}_0 \approx 0.035  \quad , \quad
  \beta \, R^{-2} \approx 0.86 \quad , \quad 
  \zeta \approx 1.8 \, .
\end{equation}
Using these we can go between the dimensionless quantities ($c_1, c_3$ and ${\cal B}$) and the dimensionful mass, condensate and magnetic fields as 
\begin{equation}
  m_q = \frac{c_1}{z_{\Lambda}\zeta m_{\tau}} \quad , \quad 
  \langle \bar q q \rangle = \frac43 \zeta {\cal V}_0 \frac{c_3}{m_{\tau}z_{\Lambda}^3} \quad, \quad 
B = \frac{{\cal B}}{z_{\Lambda}^2\beta R^{-2}} \, .
\end{equation}
Therefore in the confined phase we find the relations
\begin{equation} \label{paramrel}
 m_q = 0.31 \, c_1 \,  ({\rm GeV}) \quad , \quad   
    \langle \bar q q \rangle = 0.014 \, c_3 \, ({\rm GeV}^3)\quad , \quad 
    B = 0.35 \, {\cal B} \, ({\rm GeV}^2) \,.
\end{equation}
The physical quark mass is given by 
\begin{equation}
 m_q^* = 0.31 \, c_1^* \approx 0.0029\,  {\rm GeV}  
 .
\end{equation}
%
In the deconfined phase we replace $z_{\Lambda}$ by $z_T= 5/(4 \pi T)$
. In this case we find the relations
\begin{equation} \label{paramrelDec}
\frac{m_q}{T} =  1.4 \, c_1 \quad , \quad 
\frac{\langle \bar q q \rangle }{T^3} = 1.3 \, c_3  \quad , \quad 
\frac{B}{T^2} = 7.4 \, {\cal B} \, .
\end{equation}
The critical temperature for the first order deconfinement transition, cf. \eqref{Tc}, becomes $T_c = 0.22 \, {\rm GeV}$.

In the confined phase the   magnetisation is given by 
\begin{equation}
\mathbb{M} =  z_{\Lambda}^{-2} \beta R^{-2} M = 0.26 M \, ( {\rm GeV}^2) \, ,
\end{equation}
where $M$ is the dimensionless  magnetisations described in the previous subsections. 
In the deconfined phase the dimensionful magnetisation takes the form 
\begin{equation}
 \mathbb{M} = z_T^{-2} \beta R^{-2} M = 5.4 \, T^2 M \, ( {\rm GeV}^2) \,.
\end{equation}

\subsection{Comparing the chiral condensate against lattice QCD}

In order to compare our results to lattice QCD, we introduce the subtracted condensate
\begin{equation} \label{subtrcondensate}
\Delta \langle \bar q q \rangle \equiv  \langle \bar q q (B,T) \rangle - \langle \bar q q (0,T) \rangle  \, .
\end{equation}
This quantity is scheme independent and therefore free of ambiguities. We  evaluate this quantity in the confined and deconfined phases. 

\subsubsection{Subtracted chiral condensate as a function of $T$}

\begin{figure}[ht]
\centering
\includegraphics[width=11cm]{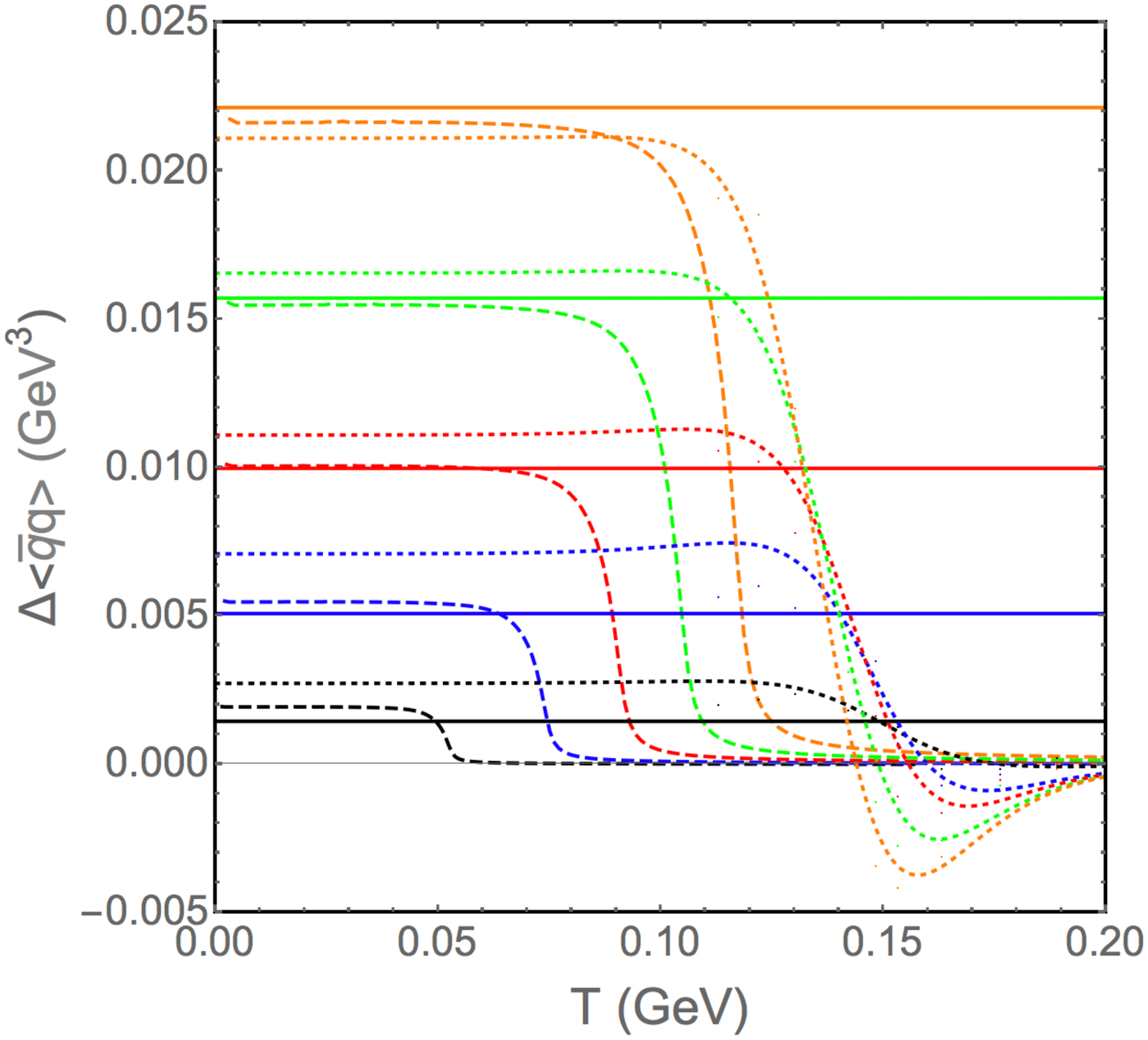}
\caption{The subtracted chiral condensate, defined in \eqref{subtrcondensate}, as a function of the temperature for fixed values of the magnetic field. The black, blue, red, green and orange curves correspond to $B=0.2, \, 0.4, \, 0.6, \,0.8$ and $1$ in ${\rm GeV}^2$ units. The solid horizontal lines represent the results for the confined phase whereas the dashed curves represent the results for the deconfined phase. The dotted curves are fits to the lattice QCD data obtained in  \cite{Bali:2013owa,Bali:2014kia}, using the empirical formula given in \cite{Miransky:2015ava}. }
\label{Fig:subtrcondvsT}
\end{figure}

We display in Fig. \ref{Fig:subtrcondvsT} our results for the subtracted chiral condensate, defined in \eqref{subtrcondensate}, as a function of the temperature for fixed values of the magnetic field. We compare our results against the lattice QCD results obtained in \cite{Bali:2013owa,Bali:2014kia}. The magnetic field varies from $B=0.2 \, {\rm GeV}$ (black lines) to $B=1 \, {\rm GeV}$ (orange lines). The  solid horizontal lines correspond to the confined phase and the dashed curved lines correspond to the deconfined phase. The dotted lines represent fits to the lattice QCD data \cite{Bali:2013owa,Bali:2014kia} using the empirical formula obtained in \cite{Miransky:2015ava}. The subtracted chiral condensate in  the confined phase is independent of the temperature because it corresponds to the  thermal extension of the QCD vacuum. The deconfined phase, on the other hand, leads to an interesting temperature dependence for the subtracted chiral condensate. The chiral transition described in subsection \ref{condensate} for a varying dimensionless magnetic field ${\cal B} \sim B/ T^2$ now is interpreted as a chiral transition for a varying temperature. As a matter of fact, the tachyon solution only depends on the dimensionless magnetic field ${\cal B} \sim B/ T^2$ and the dimensionless quark mass $c_1 \sim m_q/T$. As long as the dimensionless quark mass $c_1$ does not vary significantly one can obtain the subtracted condensate for any $B$ and $T$ from the solution found at some fixed $B$ scaling appropriately the temperature and the chiral condensate.

At very low temperatures and large magnetic fields our results for the subtracted condensate in the confined and deconfined phases agree. This can be explained by the universal scaling $c_3 = \# {\cal B}^{3/2}$, found in subsection \ref{condensate}, for the dimensionless condensate $c_3$ in the regime of large ${\cal B}$.  Regarding the comparison to lattice QCD, Fig. \ref{Fig:subtrcondvsT} shows that our results differ significantly from the lattice results in the regime of moderate and high temperatures ($T >  0.12 \, {\rm GeV}$) but there is a reasonable agreement at low temperatures ($T <  0.12 \, {\rm GeV}$). 
The agreement at low temperatures improves as the magnetic field increases. The disagreement at high temperatures is expected since our model is not able to describe anisotropy effects and therefore the transition from MC to IMC. We expect that incorporating backreaction effects in the model  would allow for such a description.  Below we provide a more detailed comparison.

\begin{itemize}
\item The confined phase in our model provides a good description of MC at low temperatures where the chiral condensate does not vary with the temperature. This is because the confined phase extends the (magnetic) vacuum to finite temperature in a trivial way so that the chiral condensate does not vary with the temperature (although it varies with the magnetic field). 
\item Note that only the deconfined phase allows for a qualitative description of a chiral condensate decreasing with the temperature, consistent with the chiral transition found in lattice QCD. Note, however, that for small magnetic fields the chiral transition in the deconfined phase of our model takes place at very low temperatures whereas in lattice QCD the chiral transition takes place at moderate temperatures (around $0.15 \, {\rm GeV}$). We suspect that this discrepancy has to do with the fact that at zero magnetic field chiral symmetry is broken in the deconfined phase of the IKP model only due to the presence of a finite quark mass (there is no dynamical scale analogous to $\Lambda_{QCD}$). Incorporating a dynamical scale in the deconfined phase would allow for a more realistic chiral transition already at small magnetic fields, absent in the present model. 
\item In both phases of our model  there is a hierarchy between the different lines in Fig. \ref{Fig:subtrcondvsT}, due to MC, consistent with lattice QCD at low temperatures. In lattice QCD, however there is an inversion of hierarchy (crossing of the dotted lines in Fig. \ref{Fig:subtrcondvsT}), corresponding to the transition from MC to IMC. As explained above, this transition is not described in our model to the absence of anisotropy effects.
\end{itemize}

\noindent
\noindent

\noindent
\begin{figure}[ht]
\centering
\includegraphics[width=7cm]{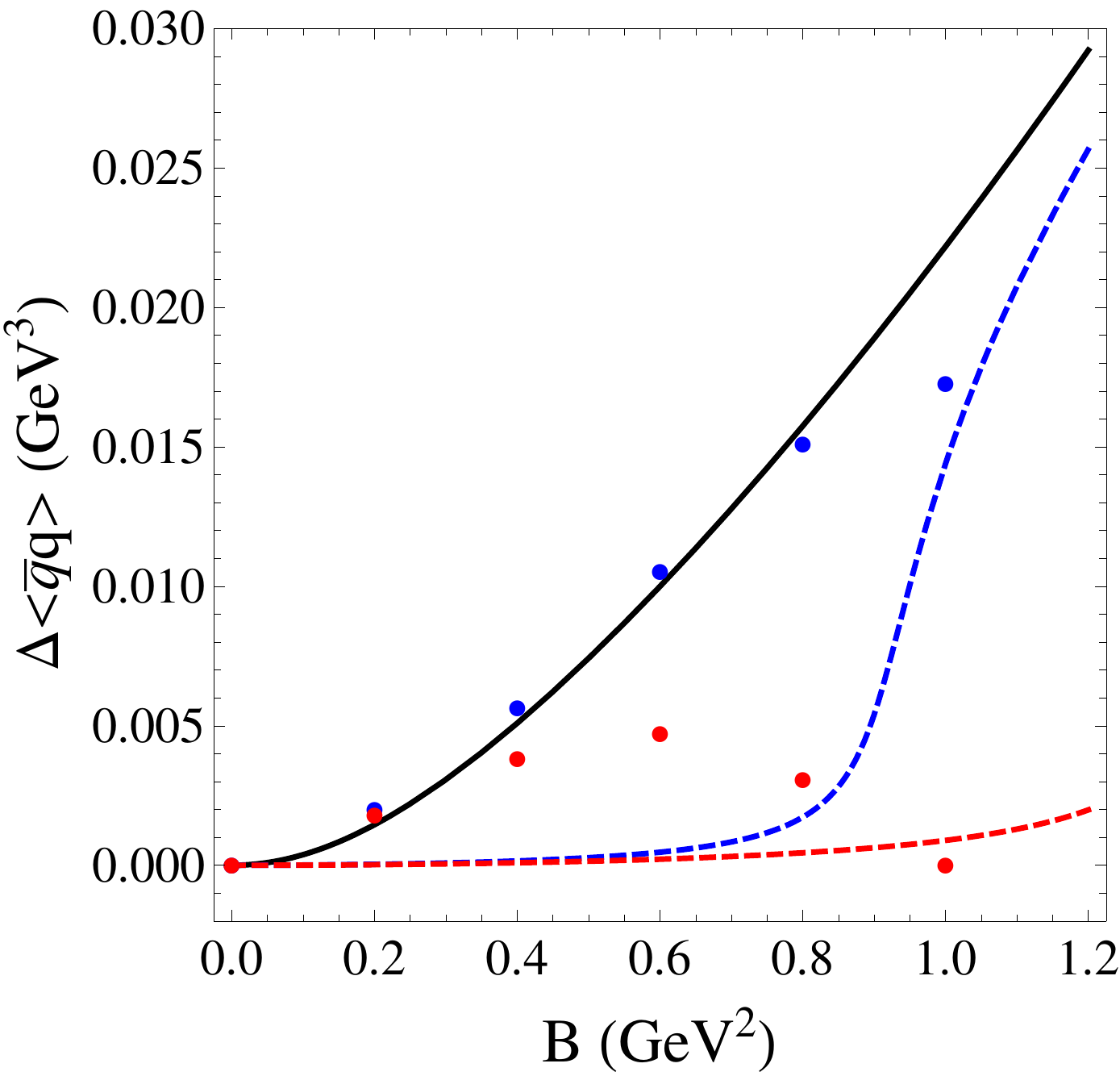}
\caption{The subtracted condensate, defined in \eqref{subtrcondensate}, as a function of the magnetic field in the confined phase and deconfined phase. The black solid line is the result for the confined phase and it is independent of the temperature. The blue and red dashed lines represent the results for the deconfined phase at $T=113 \, {\rm MeV}$  and $T=142 \, {\rm MeV}$ respectively. At those low temperatures the deconfined phase is in a metastable phase (the confined phase is thermodynamically preferred). The blue and red dots are lattice results at $T=113 \, {\rm MeV}$  and $T=142 \, {\rm MeV}$ obtained in \cite{Bali:2012zg}. The error bars in the lattice data are too small to appear in the figure. }
\label{Fig:DeltaSigmavsb}
\end{figure}
\noindent

\subsubsection{Subtracted chiral condensate as a function of $B$}

In figure \ref{Fig:DeltaSigmavsb} we compare our results for the subtracted chiral condensate $\Delta \langle \bar q q \rangle$, as a function of the magnetic field, against lattice QCD results at temperatures $T=113 \, {\rm MeV}$  and $T=142 \, {\rm MeV}$, obtained in \cite{Bali:2012zg}.  
We have set the quark mass  to the physical value  $m_q^*=2.9 \, {\rm MeV}$.

Our results always provide a subtracted condensate increasing with the magnetic field, which is interpreted as MC. The lattice results, on the other hand, show an increasing behaviour at $T=113 \, {\rm MeV}$ and a decreasing behaviour  at $T=142 \, {\rm MeV}$. This is, of course, the well known transition from MC to IMC. A clear description of these results was given in  \cite{Bruckmann:2013oba} in terms of a valence and sea contribution to the chiral condensate.  Since we work in the probe approximation, our model only describes the effect of the magnetic field on the quark mass operator and neglect magnetic effects on the gluonic vacuum (or plasma). These effects are particularly important to describe the anisotropy of the plasma, which is crucial in the description of IMC. Including backreaction it should be possible to describe these effects and therefore the transition from MC to IMC. Below we provide a more detailed comparison.

\begin{itemize}

\item The confined phase provides a good approximation for the subtracted chiral condensate  at $T=113 \, {\rm MeV}$ regardless the value of the magnetic field. As explained in the previous subsection, the confined phase lacks any temperature dependence (it is just a thermal extension of the vacuum) but provides at low temperatures ($T <  120 \, {\rm MeV}$), a good description of the magnetic dependence of the chiral condensate consistent with MC. 

\item The deconfined phase, on the other hand, provides a  poor description of lattice data in the regime of moderate temperatures and small magnetic fields. This is very clear in Fig. \ref{Fig:DeltaSigmavsb} for the temperatures $T=113 \, {\rm MeV}$ and $T=142 \, {\rm MeV}$. The reason is that in the deconfined phase at zero magnetic field, there is no dynamical scale analogous to $\Lambda_{QCD}$ and therefore chiral symmetry is weakly broken only due to a finite quark mass. The corresponding chiral condensate at zero magnetic field is extremely small. Including a scalar field, dual to the gluon condensate, would allow to dynamically generate such a scale and improve the description in the regime of small magnetic fields. 

\item We remark, however, that the deconfined phase provides a good description of the chiral condensate in the regime of low  temperatures and large magnetic fields. This feature is very clear from Fig. \ref{Fig:subtrcondvsT}, described in the previous subsection, but can not easily be seen in Fig. \ref{Fig:DeltaSigmavsb}. For $T=113 \, {\rm MeV}$ this would correspond to the last blue point, corresponding to $B=1 \, {\rm GeV}^2$, approaching the dashed blue line in Fig. \ref{Fig:DeltaSigmavsb} .
\end{itemize}

\subsubsection{The RG invariant product of the quark mass and (subtracted) condensate}

\begin{figure}[ht]
\centering
\includegraphics[width=7cm]{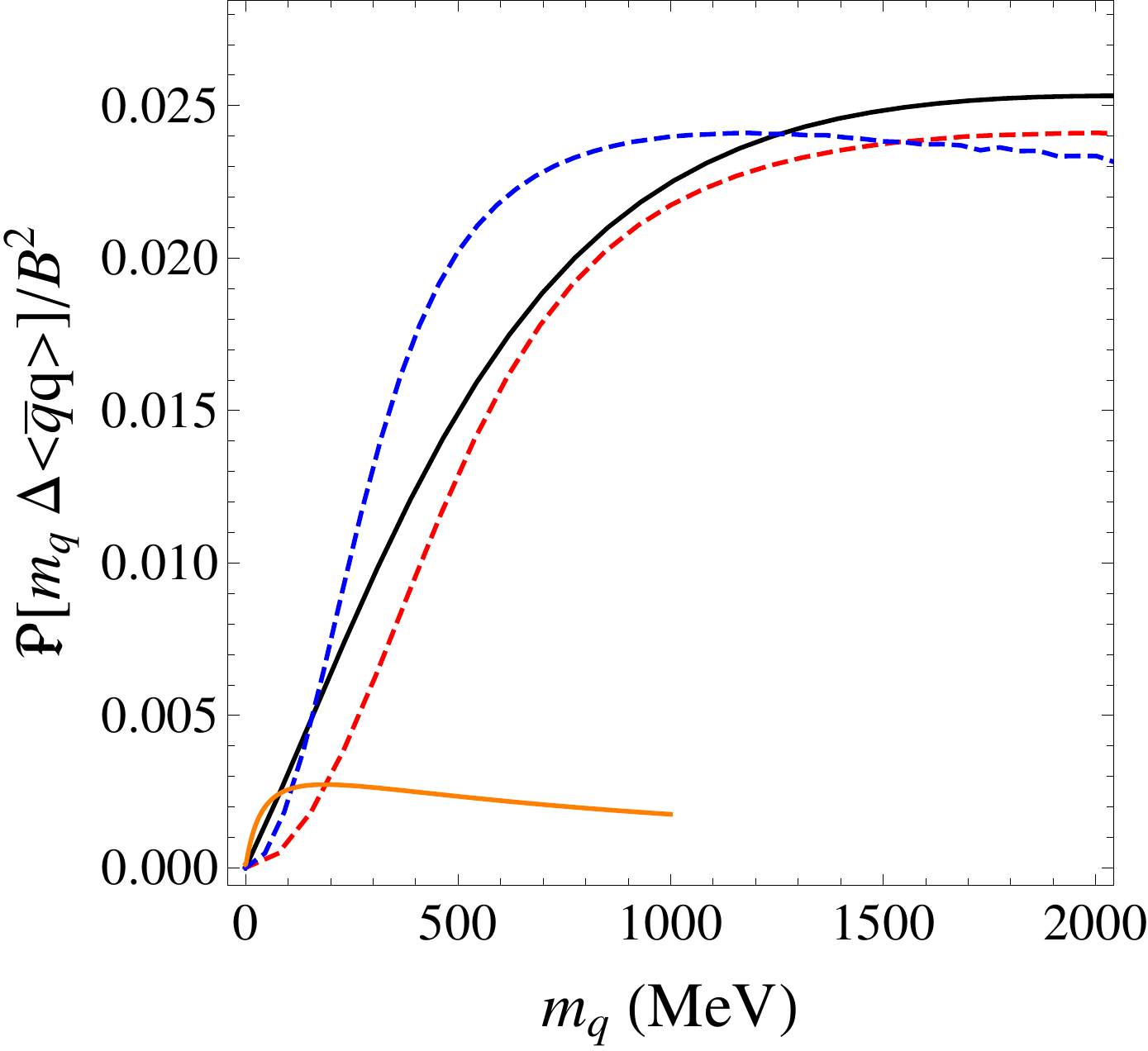}
\caption{The $B^2$ coefficient of $m_q \Delta \langle \bar q q \rangle$, with $\Delta \langle \bar q q \rangle$ defined in \eqref{subtrcondensate}, as a function of the quark mass for the confined and deconfined phase. We have used the relations  \eqref{paramrel} and \eqref{paramrelDec}  and we introduced the projector ${\cal P}[X] = B^2 \lim_{B \to 0}  B^{-2} X$. The black solid line corresponds to the confined phase and is independent of the temperature. The blue and red dashed lines correspond to the deconfined phase at $T=113 \, {\rm MeV}$ and $T=220 \, {\rm MeV}$ respectively. The orange line represents the lattice continuum extrapolation obtained in \cite{Bali:2014kia} for $T=113 \, {\rm MeV}$. See footnote 14. }
\label{Fig:B2Coeffvsc1}
\end{figure}

A nice feature of the IKP model is that we can vary the quark mass $m_q$ and go from the regime of light quarks (and mesons) to the heavy quark (heavy meson) regime. Of course, a more realistic description would imply a non-Abelian description that distinguishes the different quark flavours (up, down, strange, etc).   However, the Abelian approximation is good enough to explore the transition from light quarks to heavy quarks. In our model we find an interesting behaviour for the  RG invariant product of quark mass and (subtracted) chiral condensate, i.e.  $m_q \Delta \langle  \bar q q \rangle$, with $\Delta \langle  \bar q q \rangle $ defined in \eqref{subtrcondensate}. At small $B$ the quantity $m_q \Delta \langle  \bar q q \rangle$  can be expanded as $\# B^2 + \# B^4 + \dots$. Interestingly, the $B^2$ coefficient grows quickly with the quark mass and reaches a plateau at $m_q \sim 1 \, {\rm GeV}$, suggesting a scaling law in the heavy quark regime. This is shown in Fig.  \ref{Fig:B2Coeffvsc1} for the confined and deconfined phases. Lattice QCD results for this coefficient were obtained in  \cite{Bali:2014kia}\footnote{We have extracted this curve from \cite{Bali:2014kia} by our own fit to the lattice data points and so this is an approximation.}, represented by the orange curve in Fig. \ref{Fig:B2Coeffvsc1}.  \cite{Bali:2014kia} provided a nice weak coupling interpretation for the plateau in the heavy quark regime. In our case, we expect some scaling in the regime of large $m_q$ due to an approximate conformal symmetry for the theory at nonzero $B$ after subtracting the (conformal symmetry breaking) $B=0$ term. We suspect that the difference between the plateau we found and the plateau found in lattice QCD is associated with the fact that in the holographic QCD model at hand we are always in the strongly coupled regime whereas in real QCD there is a transition between the strongly coupled regime to the weakly coupled regime\footnote{Another important difference between our model and real QCD is that at high energies the gluon sector becomes a five dimensional theory.}.  

\noindent
\noindent

\subsection{Comparing the magnetisation against lattice QCD}

\begin{figure}[ht]
\centering
\includegraphics[width=7cm]{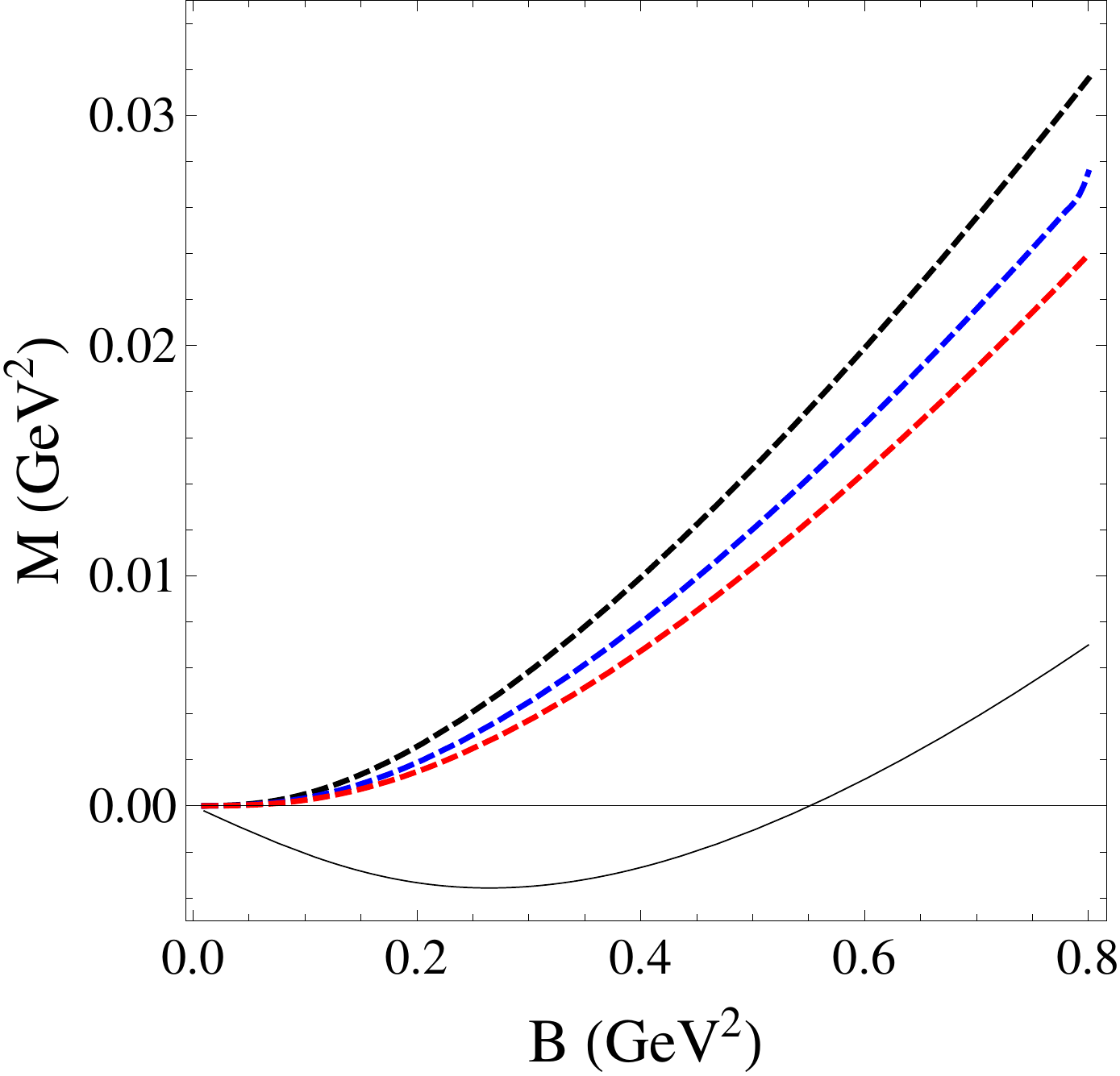}
\hfill
\includegraphics[width=7.5cm]{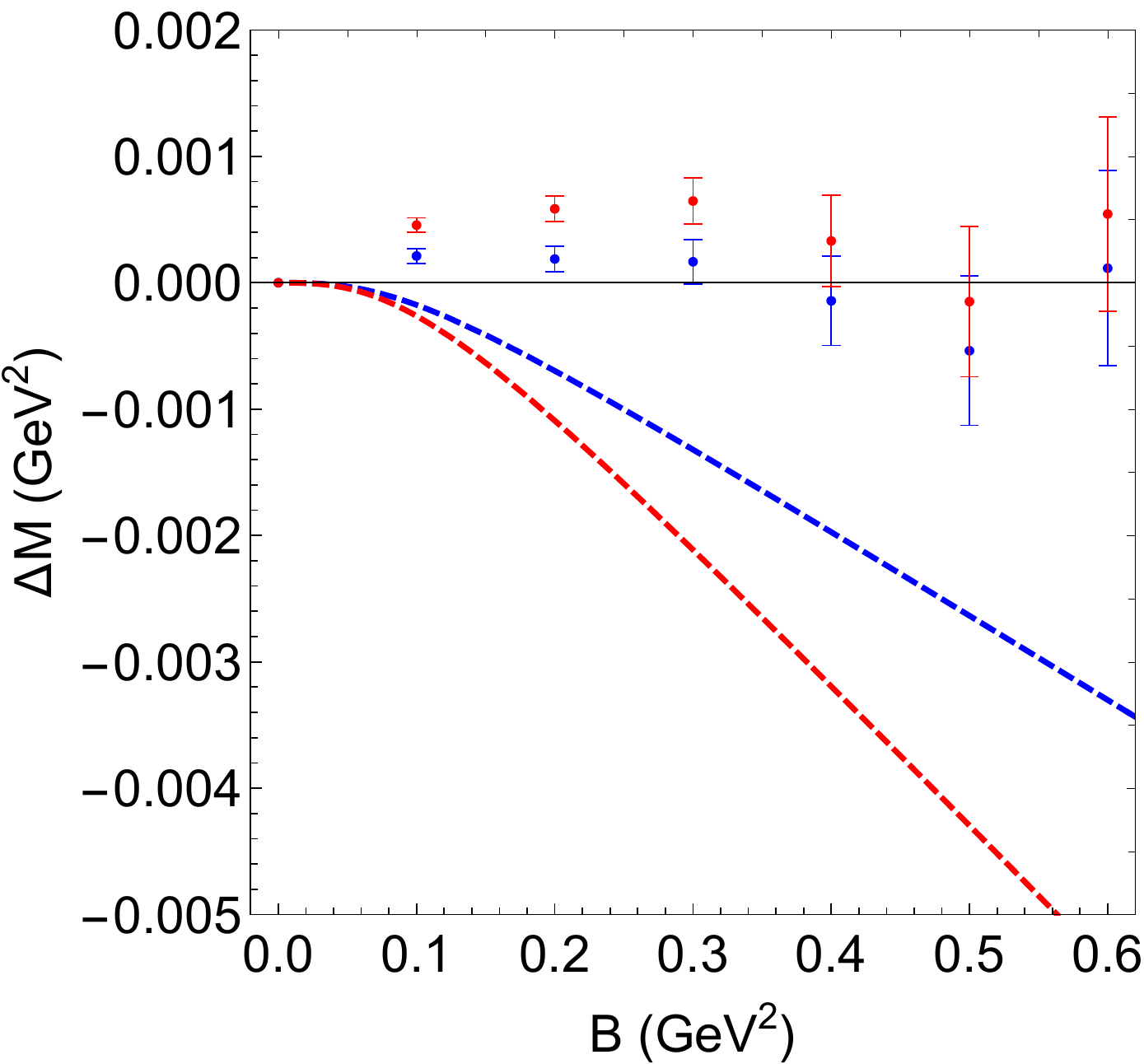}
\caption{{\bf Left Panel:} Magnetisation as a function of the magnetic field in the confined and deconfined phases. The black solid line is the result for the confined phase and it is independent of the temperature. The black, blue and red dashed lines represent the results for the deconfined phase in the IKP model at $T=114 \, {\rm MeV}$, $T=130 \, {\rm MeV}$  and $T=142 \, {\rm MeV}$ respectively.  {\bf Right Panel:} Subtracted magnetisation, defined in \eqref{subtrmagnetisation}, as a function of the magnetic field. We chose in \eqref{subtrmagnetisation} a reference temperature $T_0=114 \, {\rm MeV}$. The black solid line depicts the trivial result for the confined phase. The blue and red dashed lines represent the  results for the deconfined phase at $T=130 \, {\rm MeV}$  and $T=142 \, {\rm MeV}$ respectively.
The blue and red dots are lattice results at  $T=130 \, {\rm MeV}$  and $T=142 \, {\rm MeV}$ obtained in  \cite{Bali:2013owa,Bali:2014kia}. }
\label{Fig:MagvsbLattice}
\end{figure}

Next, we compare our results for the magnetisation against laticce QCD results. For this purpose it is convenient to work the subtracted quantity
\begin{eqnarray} \label{subtrmagnetisation}
\Delta \mathbb{M} = \mathbb{M}(B,T) - \mathbb{M}(B,T_0) \, , 
\end{eqnarray}
where $T_0$ is a reference temperature. This subtracted magnetisation is scheme independent and should therefore be free of ambiguities. 
On the left panel of Fig. \ref{Fig:MagvsbLattice}  we present our results for the (dimensionful) magnetisation $\mathbb{M}$ in the confined and deconfined phases. The black solid curve corresponds to the magnetisation in the confined phase and it is independent of the temperature.  The black, blue and red dashed lines represent the results for the deconfined phase at the temperatures $T=114 \, {\rm MeV}$, $T=130 \, {\rm MeV}$ and $T=142 \, {\rm MeV}$ respectively. At those temperatures the deconfined phase is in a metastable phase (the confined phase is thermodynamically preferred). On the right panel of Fig. \ref{Fig:MagvsbLattice} we compare our results for the subtracted magnetisation $\Delta \mathbb{M}$, defined in \eqref{subtrmagnetisation}, against the lattice QCD results obtained in  \cite{Bali:2013owa,Bali:2014kia}. Since the lattice QCD data starts at $T=114 \, {\rm MeV}$ we take that value as our reference temperature $T_0$. 
The black solid line depicts the trivial result $\Delta \mathbb{M}  =0$ for the confined phase. The blue and red dashed lines represent the results for $\Delta \mathbb{M}$ in the deconfined phase at $T=130 \, {\rm MeV}$ and $T=142 \, {\rm MeV}$. The blue and red dots (and error bars) represent the lattice QCD results obtained in \cite{Bali:2013owa,Bali:2014kia}. 
\noindent

Interestingly, the left panel of Fig. \ref{Fig:MagvsbLattice} shows that as we go from the confined phase to the deconfined phase there is a transition between a diamagnetic behaviour to a paramagnetic behaviour. This result might be particular to this model, although recent lattice QCD results indicate a similar transition \cite{Bali:2020bcn}. 
On the other hand, the right panel of Fig. \ref{Fig:MagvsbLattice}  shows significant quantitative differences between our results and the lattice QCD results. Since the lattice QCD data starts at moderate temperatures, these differences are already expected because, as explained previously, we expect backreaction effects to be important at moderate and high temperatures. 

We note, however, that the lattice QCD data reveals a variation in the sign of the subtracted magnetisation $\Delta \mathbb{M}$, possibly related to the competition between MC and IMC. In this work we have already established that our model leads to MC from the analysis of the chiral condensate and we always find $\Delta \mathbb{M}<0$. This seems to be consistent with the criterion found in \cite{Ballon-Bayona:2017dvv} for distinguishing MC from IMC. Incorporating backreaction effects in our model would allow for the description of IMC and in that scenario we expect to find a variation in the sign of $\Delta \mathbb{M}$, similar to that found in lattice QCD.

\subsection{Comparing the chiral transition against the Sakai-Sugimoto model}

One advantage of having considered the IKP model for investigating MC, compared with the Sakai-Sugimoto model, is that the IKP model allows for a full description of the chiral condensate. In the IKP model, chiral symmetry breaking occurs due to tachyon condensation and brane-antibrane recombination whereas in the Sakai-Sugimoto model it is due to a geometrical merging of the brane-antibrane pairs. The geometric realisation of chiral symmetry breaking in the Sakai-Sugimoto model requires embedding the branes and antibranes in different locations of an extra spatial dimension. This makes the study of the chiral condensate very subtle, see e.g \cite{Bergman:2007pm,Aharony:2008an}. 

\begin{figure}[ht]
\centering
\includegraphics[width=8.5cm]{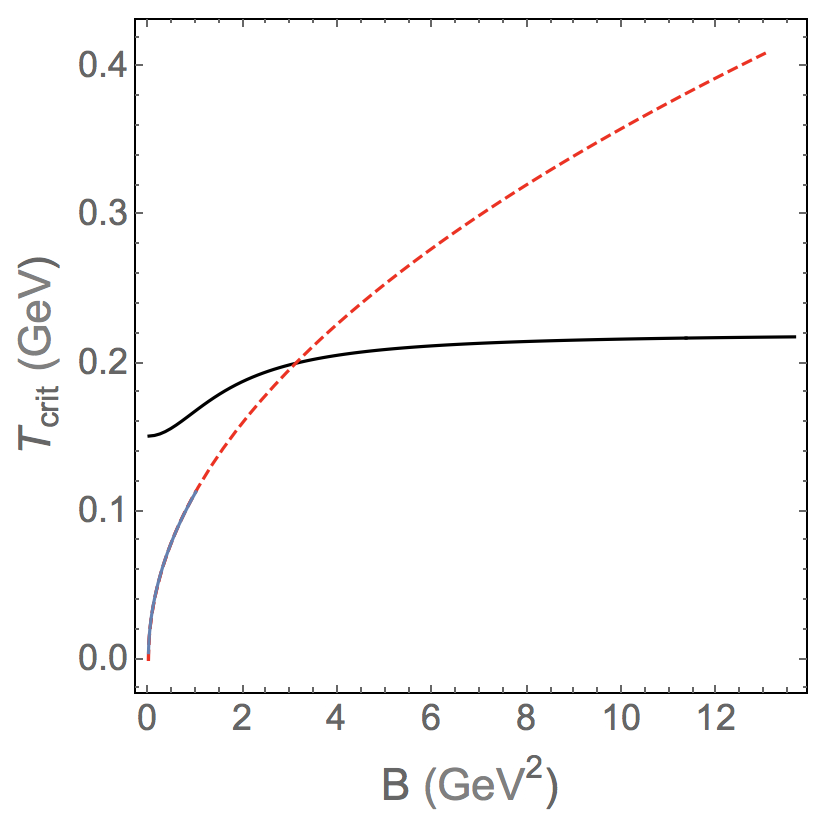}
\caption{A comparison between the Sakai-Sugimoto (black) line of chiral symmetry breaking first order transitions, found in \cite{Johnson:2008vna}, and the cross-over transitions between chirally restored and chirally broken phases in the IKP model (blue line) found in this work at $m_q^*$. The critical and pseudo-critical temperatures are displayed as a function of the magnetic field. The red dashed line is the square root extrapolation using the scaling behaviour at $T>>m_q^*$. The Sakai-Sugimoto model has a free parameter, $L$, which is fixed to set the critical temperature at zero magnetic field to the lattice QCD result, $0.15 \, {\rm GeV}$. There is no plateau in the IKP model which is seen in the Sakai-Sugimoto model. The IKP results are calculated from figure \ref{Fig:subtrcondvsT} by taking the inflection point as the ciritical temperature for each magnetic field.}
\label{Fig:sscomparison}
\end{figure}

In the Sakai-Sugimoto model the transition between the two phases is calculated by looking at the differences in free energy, which leads to a first order phase transition. In the IKP model the signature of chiral symmetry breaking is in the turning on of a condensate, which, at finite quark masses occurs gradually and thus a cross-over is apparent rather than a strict phase-transition.  We compare in Fig. \ref{Fig:sscomparison} our results for the pseudo-critical temperature for the chiral transition in the IKP model against the critical temperature for the chiral transition in the Sakai-Sugimoto model found in \cite{Johnson:2008vna}. In both models the temperature where the chiral transition takes place increases with the magnetic field, a scenario consistent with magnetic catalysis. We remark, however, that recent lattice QCD results indicate that the pseudo-critical temperature for chiral transition actually decreases with the magnetic field as a consequence of inverse magnetic catalysis. This can be seen from Fig. \ref{Fig:subtrcondvsT} where the dotted lines display the lattice QCD results for the chiral condensate. Fig. \ref{Fig:sscomparison} shows that in the Sakai-Sugimoto model the transition as a function of magnetic field saturates at some critical temperature. This is not the case in the IKP model which, due to the explicit scaling apparent in the equations of motion, has a square root behaviour for temperatures and magnetic fields much larger than the quark mass (which in the case of the physical quark mass is very low).

\noindent
\noindent


\section{Conclusions}

In this paper we have studied the effects of a non-zero magnetic field on the chiral condensate of a QCD-like theory using a  holographic QCD model. As emphasised, this model, with chiral symmetry breaking using a tachyon, has been studied in detail before in the absence of a magnetic field, but here we have shown a variety of behaviours in both the confined as well as deconfined phases with a magnetic field present. As expected in the quenched approximation, the addition of the magnetic field has given us catalysis of chiral symmetry breaking, whereby the value of the condensate goes up with increasing magnetic field. There is one caveat to this that in the case of zero quark mass and in the deconfined case, there is a critical value of the magnetic field below which there is no chiral symmetry breaking, and above which it is induced, signifying spontaneous chiral symmetry breaking. This second order phase transition only exists in the chiral limit, and at any non-zero quark mass it becomes a cross-over phase transition. 

For large quark masses we have seen that the behaviour of the confined and deconfined phases converge, as expected when the mass scale of the constituents is greater than the dynamical and thermal mass scales of the theory. The universal asymptotic behaviour in the regime of large quark mass suggest some approximate conformal symmetry in the dual field theory, after the subtraction of the (conformal symmetry breaking) mass term and it seems to be related to the AdS asymptotics of the gravity dual. It should be noted however that at large energies the gluon dynamics of these theories are not only conformal but also 4+1 dimensional. As noted in subsection 4.2, this phenomenon is generally obtained in holographic brane constructions.

In addition to the spontaneous symmetry breaking we have been able to study the magnetisation of this theory, where in the deconfined phase the second order phase transition is again apparent in both the magnetisation as well as the susceptibility. Due to the explicit nature of the tachyon in the DBI action we have been able to extract the condensate contribution to the magnetisation. We arrived at a simple empirical formula relating the magnetisation and chiral condensate. Since both quantities are important order parameters for MC and IMC, our formula could be useful for unveiling the physical mechanisms behind those phenomena.

As noted, we are here working in the quenched approximation where, in this model, we wouldn't expect to see anything but the magnetic field catalysing chiral symmetry breaking. A clear extension to this work would be to go beyond the probe approximation and allow for back-reaction on the geometry by the tachyon field. This would allow us to also investigate IMC, but this calculation will be an order of magnitude more complicated, particularly as the equations of motion would involve a divergent tachyon backreacting on the geometry when confinement is present.

With or without backreaction, several other phenomena could still be investigated in this model. As noted earlier, this model is particularly interesting as it gives rise to realistic Regge trajectories for the mesons, and so the effects of the magnetic field on these trajectories would be extremely interesting to investigate. Given this one could also study the Gellman-Oakes-Renner \cite{Gellman:1968} relation between the quark mass and condensate and the pion mass in the presence of a magnetic field. Investigating the fluctuations on top of this background would also allow for an explicit construction of the chiral effective theory from the 5d flavour action, whereby the Gasser-Leutwyler coefficients \cite{Gasser:1984} could be compared with lattice data. Such calculations have been performed before \cite{Evans:2004ia}, but this model would likely give results closer to those of QCD.

Extending the model in \cite{Iatrakis:2010zf,Iatrakis:2010jb}
to the non-Abelian case would also be a natural next step. This would allow for a more realistic description of chiral and flavour symmetry breaking as well as the meson phenomenology. Although the original proposal in \cite{Casero:2007ae} describes the non-Abelian tachyonic DBI and WZ terms, there are some subtleties when describing  spontaneous chiral symmetry  breaking and the QCD anomalies in the non-Abelian case. 

Another interesting future direction could be the addition of baryons and the further study of the holographic model. 
First in the probe approximation and then taking into account the backreaction of the baryon in the geometry. 
That would give access to the low temperature and high density region of the phase diagram. 
In a top-down framework 
the baryon vertex corresponds to a D-brane wrapping an internal sphere and connecting to the boundary with 
$N_c$ fundamental strings \cite{Witten:1998xy} (see also \cite{Seo:2009kg,Evans:2012cx}). In the Sakai-Sugimoto model, which is the closest holographic model for QCD, baryons appear as 5d instantons in the non-Abelian flavour sector \cite{Hata:2007mb,Bolognesi:2013nja}. Interestingly, these 5d instantons are the holographic dual of 4d skyrmions dressed by vector mesons.  The interplay between baryon density and magnetic field has also been investigated in  \cite{Preis:2011sp} for the Sakai-Sugimoto model, in which inverse magnetic catalysis was also observed.
In a bottom-up scenario, a baryon solution exists both in AdS/QCD \cite{Pomarol:2008aa} and  V-QCD \cite{Ishii:2019gta}.


\section*{Acknowledgments}
The authors are grateful to Matthias Ihl for his valuable work during the early stages of this project. The authors would also like to acknowledge Gunnar Bali, Gergely Endr\"odi, Luis Mamani and Carlisson Miller for useful conversations.  The work of A.B-B is partially funded by Conselho Nacional de Desenvolvimento Cientifico e Tecnologico (CNPq), grants No. 306528/2018-5 and No. 434523/2018-6. The work of D.Z has received funding from the Hellenic Foundation for Research and Innovation (HFRI) and the General Secretariat for Research and Technology (GSRT), under grant agreement No 15425.


\appendix


\section{The Tachyon-WZ term}
\label{WZterm}

It was shown in \cite{Casero:2007ae} that the Wess-Zumino (WZ) action can be written as a 5d Chern-Simons action. For the Abelian case it takes the form 
\begin{equation}
S_{CS} = i \gamma \int \Omega_5 \, , \label{CSterm}    
\end{equation}
where $\gamma$ is a constant proportional to $N_c$ and $\Omega_5$ is a $5$-form satisfying  the equation
\beqa
d \Omega_5   &=& - \frac{1}{6} V(\tau,\tau^*) \Big \{   i \, d \Omega_5^{(0)} +  \frac{m_\tau^2}{2} D\tau \wedge (D\tau)^* \wedge d \Omega_3^{(0)} \Big \}   \nonumber \\[5pt] 
&=& \frac{i}{6} V(\tau,\tau^*) \Big \{ - d \Omega_5^{(0)} +  \frac{m_\tau^2}{2}   d (\tau^* \tau) \wedge \left ( \Omega_5^{(0)} - d \Omega_4^{(0)} \right ) + \frac{m_\tau^2}{4}   d j \wedge d \Omega_3^{(0)}  \Big \} \, .\label{dOmega5}
\eeqa
We have introduced a set of forms 
\begin{align}
\Omega_5^{(0)} &\equiv  A_L \wedge F_L^2 - A_R \wedge F_R^2 \quad , \quad 
\Omega_4^{(0)} \equiv   A_L \wedge A_R \wedge (F_L + F_R)  \, ,
\nonumber \\[7pt]
\Omega_3^{(0)} & \equiv A_L \wedge F_L + \frac12 \left ( A_L \wedge F_R + A_R \wedge F_L \right ) + A_R \wedge F_R  \, , 
\end{align}
such that
\begin{align}
d \Omega_4^{(0)} &= \left ( A_R \wedge F_L - A_L \wedge F_R \right ) \wedge \left ( F_L + F_R \right ) \, , 
\nonumber \\[5pt] 
d \Omega_5^{(0)} &= F_L^3 - F_R^3 \quad , \quad d \Omega_3^{(0)} = F_L^2 + F_L \wedge F_R + F_R^2 \, .  
\end{align}
We have  used the definition of the 1-form covariant derivative $D\tau = d \tau + i (A^L - A^R) \tau$ and the 1-form current $j = i \left(\tau d \tau^* - \tau^* d\tau\right)$. 

There are two possible solutions for $\Omega_5$ related to each other by a total derivative. The simpler solution is 
\beqa
\Omega_5^{I} = \frac{i}{6} V(\tau,\tau^*) \Big \{ -\Omega_5^{(0)} + d \Omega_4^{(0)}  + \frac{m_\tau^2}{4}  \, dj \wedge \Omega_3^{(0)}  \Big \} \label{SimpleOmega5}
\eeqa
An alternative solution for $\Omega_5$ was given in \cite{Casero:2007ae} for the case of a real tachyon. For a complex tachyon it takes the form 
\begin{equation}
\Omega_5^{II} \equiv \Omega_5 + \Delta \Omega_5 \, ,     
\end{equation}
where $\Delta \Omega_5$ is a total derivative given by
\beqa
\Delta \Omega_5 &=& - \frac{i}{6} d \left [ V(\tau,\tau^*) \left ( 1 + \frac{m_\tau^2}{2} \tau^* \tau \right ) \Omega_4^{(0)} \right ] \cr 
&=& - \frac{i}{6} V(\tau,\tau^*) \Big \{  \left ( 1 + \frac{m_\tau^2}{2} \tau^* \tau \right ) d\Omega_4^{(0)}  - \frac{m_\tau^4}{4} \tau^* \tau d (\tau^* \tau) \wedge \Omega_4^{(0)} \Big \} \,. \label{DeltaOmega5}
\eeqa
From  \eqref{SimpleOmega5} and \eqref{DeltaOmega5} we obtain the explicit form
\begin{align} \label{CKPOmega5}
\Omega_5^{II} &= \frac{i}{6} V(\tau,\tau^*) \Big \{ -\Omega_5^{(0)} -  \frac{m_\tau^2}{2} \tau^* \tau \, d\Omega_4^{(0)}  \nonumber \\
&+  \frac{m_\tau^4}{4} \tau^* \tau d (\tau^* \tau) \wedge \Omega_4^{(0)} + \frac{m_\tau^2}{4}  \, dj \wedge \Omega_3^{(0)} \Big \} \,.  
\end{align}
This is the form that appears in \cite{Casero:2007ae} for the case of a real tachyon ($\tau=\tau^*$ and $j=0$). In that case we can easily find that under the residual gauge transformation $U(1)_V$, i.e. $A_{L/R} \to A_{L/R} + d \alpha$, the Chern-Simons form in \eqref{CKPOmega5} transforms as 
\begin{align}
\delta \Omega_5^{II}    &=
 \frac{i}{6} d \Big \{ d \Big [ V( \tau) \left ( 1 + \frac{m_\tau^2}{2} \tau^2 \right )   \Big ] \wedge \omega_3^{(0)} \nonumber \\
&+ V(\tau) \left ( 1 + \frac{m_\tau^2}{2} \tau^2 \right )
 \alpha \left (  F_R^2 -  F_L^2 \right ) \Big \} \, ,
\label{Omega5CKPvar}
\end{align}
where we have introduced the 3-form
\begin{equation}
\omega_3^{(0)} = \alpha ( A_R - A_L) 
\wedge (F_L + F_R)  \, . 
\end{equation}
The variation in \eqref{Omega5CKPvar} is a boundary term. Imposing the boundary conditions for $\tau$; namely a vanishing tachyon near the boundary and a divergent tachyon at the end of space we find that the first term in \eqref{Omega5CKPvar} vanishes whilst the second term reduces to a 4d anomaly term for the residual $U(1)_V$ symmetry, as expected in QCD. The description of the full $U(1)_L \times U(1)_R$ anomaly term is more subtle because it requires a very careful analysis of the variation of \eqref{CKPOmega5}. A first look at the problem suggests that a correction to \eqref{CKPOmega5} is required in order to describe the full QCD anomaly term. 


\section{IR asymptotic analysis}
\label{appendix-2}

\subsection{IR asymptotic analysis in the confined phase}
\label{AppConfIR}

To find the asymptotic solution near the tip of the brane $u=1$, it is convenient to work with the variable $x \equiv 1-u$. 
The ansatz for $T$ will be a series expansion in powers of $x$. First we write \eqref{specTeqConfv2} as
\begin{equation}
\Big\{ (x \, \partial_x)^2 
+ \left[ {\cal P}(x) - 1 \right]  \, x \, \partial_x 
+ {\cal Q}(x) \, x \, \Big \} \, {\cal T} 
+ (x \, \partial_x {\cal T})^2 \, \Big [ {\cal R}(x) \, (x \, \partial_x {\cal T}) +   {\cal T} \Big] \, = \, 0 \, ,
\label{specTeqConfv3}
\end{equation}
where we have defined the following quantities
\begin{eqnarray}
&& {\cal P}(x) = \frac12 \left [ \frac{2 \, x}{1-x} + 5 \, \frac{(1-x)^4}{g(x)} \right ] + \frac{2  \, x}{(1-x) \, Q_0(1-x) }  
= \sum_{n=0}^{\infty} {\cal P}_n \, x^n \, ,
\nonumber \\[5pt]
&& {\cal Q}(x) =  \frac{3}{(1-x)^2 \, g(x)} = \sum_{n=0}^{\infty} {\cal Q}_n \, x^n \, , \quad 
Q_0(1-x) = 1 + {\cal B}^2 \, (1 - x)^4\, , 
\\[5pt]
&& {\cal R}(x) = \frac{2\, (1-x)}{3}\, g(x) \Bigg[ 1 + \frac{1}{Q_0 \, (1-x)} \Bigg]  = \sum_{n=0}^{\infty} {\cal R}_n \, x^n 
 \quad \& \quad g(x) = \frac{1}{x} \, f_{\Lambda}(1-x) \, . 
 \nonumber 
\end{eqnarray}
The advantage of writing the ${\cal T}$ differential equation in terms of the operators $x \partial_x$ is that these operators can act on powers without changing the exponents. 
The functions ${\cal P}(x)$, ${\cal Q}(x)$ and ${\cal R}(x)$  have  nonzero values at $x=0$ and can be (Taylor) expanded in powers of $x$. We consider the series ansatz 
\begin{equation}
{\cal T}(x) = \sum_{n=0}^{\infty} x^{\alpha_n} \,  g_{n}(x)   \, ,
\label{IRTachyonConf}
\end{equation}
where $\alpha_n$ are real numbers satisfying the inequality $\alpha_0 < \alpha_1 < \dots < \alpha_n < \alpha_{n+1} < \dots$ and $g_n(x)$ are analytic functions of $x$.  Plugging the ansatz \eqref{IRTachyonConf} into \eqref{specTeqConfv3}, we obtain
\begin{align}
&\sum_{n=0}^{\infty}  x^{\alpha_n} \Big \{  {\cal O}_n^2 + [{\cal P}(x) - 1 ] {\cal O}_n + {\cal Q}(x) x \Big \} g_n 
\nonumber \\[5pt]
&+ \sum_{n=0}^{\infty} \sum_{m=0}^{\infty} \sum_{\ell=0}^{\infty}  x^{\alpha_n + \alpha_m + \alpha_{\ell}} ( {\cal O}_n g_n ) ( {\cal O}_m g_m ) \Big[ R(x) ({\cal O}_{\ell} g_{\ell}) + g_{\ell} \Big] = 0 \,, 
\label{specTeqConfv4}
\end{align}
where we have introduced the operator ${\cal O}_n \equiv x \partial_x + \alpha_n$. The quantity $({\cal O}_n g_n)$ is the function obtained when the operator ${\cal O}_n$ has already acted on the function $g_n$. 
Since the functions $g_n$ admit a Taylor expansion around $x=0$ we find that $({\cal O}_n g_n)$ becomes $\alpha_n$ when $x \to 0$. Since the exponents $\alpha_n$ are non-integer, at each order in the series we obtain differential equations for the coefficients $g_n(x)$.  

Let us focus on the first exponent $\alpha_0$. There are three cases: $\alpha_0 = 0$, $\alpha_0 > 0$ and $\alpha_0 <0$. 

\medskip 

\noindent \underline{Case I: $\alpha_0=0$}

\medskip 

\noindent

\noindent  The first term in the series \eqref{specTeqConfv4} is of order $x^0$ and we find
\begin{equation}
\Big \{  {\cal O}_0^2    + [ {\cal P}(x) - 1 ] {\cal O}_0 + {\cal Q}(x) x \Big \} g_0
+ ( {\cal O}_0 g_0 )^2  \Big[ R(x) ({\cal O}_0 g_0) + g_0 \Big] = 0 \, .
\end{equation}
This is a non-linear equation for $g_0$ and in the limit $x \to 0$ it holds automatically, since
${\cal O}_0 g_0 =\alpha_0=0$ in that limit. 

\medskip 

\noindent \underline{Case II: $\alpha_0>0$}

\medskip 

\noindent

\noindent The first term in the series \eqref{specTeqConfv4} is $x^{\alpha_0}$ and we obtain
\begin{equation}
\Big \{  {\cal O}_0^2 + [ {\cal P}(x) - 1 ] {\cal O}_0 + {\cal Q}(x) x \Big \} g_0 = 0 \,.
\end{equation}
This is a linear equation for $g_0$ and in the limit $x \to 0$ we get the relation
 $\alpha_0 - 1 + {\cal P}(0) = 0$. Since ${\cal P}(0)=1/2$ we find $\alpha_0 = 1/2$.
 
 \medskip 

\noindent \underline{Case III: $\alpha_0<0$}

\medskip 

\noindent

\noindent The physically interesting case is when the solution is singular at $x=0$ (which is a good property according 
to the anomaly story in the IKP framework). Now the first term in the series \eqref{specTeqConfv4} is of 
order $x^{3 \alpha_0}$ and we obtain the equation 
\begin{equation}
( {\cal O}_0 g_0)^2 \Big[ R(x) ({\cal O}_0 g_0) + g_0 \Big] = 0 \,. 
\label{g0eq}
\end{equation}
We have two situations:  $({\cal O}_0 g_0) = 0$ and $({\cal O}_0 g_0) = - \, g_0/R(x)$. 
Since $({\cal O}_0 g_0) = x \partial_x g_0 + \alpha_0 g_0$ the equations are first order and can be solved. In the first case we find $g_0 \sim x^{- \alpha_0}$ which contradicts the assumption that $g_0$ is regular. 
In the second case the equation can be written as 
\begin{equation}
d g_0 + g_0 \, (a_0 dx) = 0 \quad \text{with} \quad  
a_0(x) = \frac{1}{x} \, \Big [ \alpha_0 + \frac{1}{R(x)} \Big ] \, . \label{g0eqv2}
\end{equation}
Multiplying \eqref{g0eqv2} by an integrating factor $u_0(x)$ such that $du_0 = u_0 (a_0 dx)$, the l.h.s. of \eqref{g0eqv2} becomes an exact differential $d(g_0 u_0)$ and we find
\begin{equation}
g_0(x) = g_0(0) \exp \Big \{ -  \int_0^x d x' a_0(x') \Big \} \,. \label{g0sol}
\end{equation}
In the limit $x \to 0$ we have that $x \partial_x g_0 \to 0$ and $g_0 \to 1$ so we find that
\begin{equation}
\alpha_0 = - \frac{1}{R(0)} =  - \frac{3}{10} \, \frac{1 + {\cal B}^2}{2 + {\cal B}^2} \equiv - r \,, \label{alpha0} 
\end{equation}
where we have also introduced the positive real number $r = - \alpha_0$. Note that $R(0) = 1/r$. 
The function $g_0(x)$ can be Taylor expanded as
\begin{equation}
g_0(x)=\sum_{n=0}^{\infty} g_{0, \bar n} \, x^{\bar n} 
\quad \text{where} \quad g_{0,0}=g_0(0) \equiv C_0
\end{equation}
and the following coefficient in the series expansion is
\begin{equation}
g_{0,1} \, = \, g_0'(0) =  \, - \, a_0(0) \, g_0(0) \, 
= \, - \,  \frac{3}{10} \, \frac{6 + 5 \, {\cal B}^2 + 3 \, {\cal B}^4}{(2 +  {\cal B}^2)^2} \, C_0  \,.
\end{equation}

Now we make the following assumption:  the exponents $\alpha_n$ are integer powers of $r$, namely $\alpha_n = (n-1)\, r$, 
which is compatible with the case $n=0$. The strategy now is to find equations for $g_n$ at each different order in the series \eqref{specTeqConfv4}. We have already found the first equation \eqref{g0eq} at order $x^{-3r}$. At the next order $x^{-2r}$ we find the equation
\begin{equation}
 ( {\cal O}_0 g_0 )^2 \Big [ R(x) ({\cal O}_1 g_1) + g_1 \Big] = 0 \, , \label{g3eq}
\end{equation}
where we choose the trivial solution $g_1(x) = 0$. 

At the next order $x^{-r}$, we find
\begin{equation}
 \Big \{  {\cal O}_0^2  + [ {\cal P}(x) - 1 ] {\cal O}_0 + {\cal Q}(x) x \Big \} \, g_0 
+   ( {\cal O}_0 g_0)^2 \Big [ R(x) ( {\cal O}_2 g_2 ) + g_2 \Big ]  = 0 \,. \label{g2eq}
\end{equation}
Taking the limit $x \to 0$ in \eqref{g2eq}, we find that 
\begin{equation}
g_2(0) = - \frac12 \left ( 1 + \frac{1}{2r} \right) C_0^{-1} = - \frac{13 + 8 \, {\cal B}^2}{6\, (1+ {\cal B}^2)} \, C_0^{-1}  \,.
\end{equation}
The differential equation \eqref{g2eq} allows us to find $g_2(x)$ given $g_0(x)$, the latter found in \eqref{g0sol}.  
It can be put in the canonical form
\begin{equation}
dg_2 + \Big [ g_2 a_2 - b_2 \Big ] dx = 0  \, , \label{g2eqv2}
\end{equation}
where
\begin{equation}
a_2(x) = \frac{1}{x} \, \Big [ \alpha_2 + \frac{1}{R(x)} \Big ] 
\quad \& \quad
b_2(x) = - \frac{ {\cal O}_0^2  + [ {\cal P}(x) - 1 ] \, {\cal O}_0 + {\cal Q}(x) \, x}{({\cal O}_0 g_0)^2 \, R(x) \, x} \, g_0 \, ,
\end{equation}
and we remind the reader that $\alpha_2 = r$. Following a procedure similar to the one used for $g_0(x)$ we find
\begin{equation}
g_2(x) = g_2(0) \, e^{- \int_0^x dx'a_2(x') } \Bigg[ 1 + \int_0^x dx' b_2(x') e^{\int_0^{x'} d x'' a_2(x'') } \Bigg] \, . 
\label{g2sol}
\end{equation}
The solution can be Taylor expanded as
\begin{equation}
g_2(x) \, = \, \sum_{n=0}^{\infty} g_{2, \bar n} \, x^{\bar n} 
\quad \text{where} \quad 
g_{2,0}=g_2(0)
\end{equation}
and the following coefficient in the series expansion is
\begin{equation}
g_{2,1} = g_2'(0) =  
\frac{986+507 \, {\cal B}^2+169 \, {\cal B}^4+206 \, {\cal B}^6+58 \, {\cal B}^8}{20 \left(1+ {\cal B}^2\right) \, 
\left(2+ {\cal B}^2\right)^2 \, \left(13+8 \, {\cal B}^2\right)} \,  C_0^{-1} \,,
\end{equation}
from considering the equation $g_2' = - a_2 g_2 + b_2$. In the limit $B \to 0$ we obtain 
$g_{2,1}/g_{2,0} = -\frac{1479}{3380}$ which agrees with the result in \cite{Iatrakis:2010jb}.

The method described above extends in a straightforward manner and we can extract higher order terms in the expansion.


\subsection{IR asymptotic analysis in the deconfined phase}
\label{AppDeconfIR}

To find the asymptotic solution near the horizon $v=1$, we redefine the radial coordinate as $y \equiv 1-v$. and 
write \eqref{specTeqDeconfv2} as
\begin{equation}
\Big  \{ (y \partial_y)^2 + [ {\cal P}(y) - 1 ]  y \partial_y 
+ {\cal Q}(y) y \Big \} {\cal T} + (y \partial_y {\cal T})^2 \Big [ y^{-1} {\cal R}(y) (y \partial_y {\cal T}) +   {\cal T} \Big ]= 0 \, ,
\label{specTeqDeconfv3}
\end{equation}
where we have defined the following quantities
\begin{eqnarray}
&& {\cal P}(y) =   \frac{y}{1-y} + 5 \frac{(1-y)^4}{g(y)}  + \frac{2  \, y}{(1-y) \, Q_0(1-y) }  
= \sum_{n=0}^{\infty} {\cal P}_n \, y^n \, , 
\nonumber \\[5pt]
&& {\cal Q}(y) =  \frac{3}{(1-y)^2 \, g(y)} 
= \sum_{n=0}^{\infty} {\cal Q}_n \, y^n \, ,  
\nonumber \\[5pt]
&& {\cal R}(y) = \frac23 \, y\, (1-y) \, g(y) + \frac56 \, (1-y)^6 
+ \frac13 \, \frac{y\, (1-y)\,  g(y)}{Q_0(y)} 
= \sum_{n=0}^{\infty} {\cal R}_n \, y^n  , 
\nonumber \\[5pt]
&& Q_0(1-y) = 1 + {\cal B}^2 (1 - y)^4 \quad \& \quad 
g(y) = \frac{1}{y} \, f_T(1-y) \, .
\end{eqnarray}
The functions ${\cal P}(y)$, ${\cal Q}(y)$ and ${\cal R}(y)$  have  nonzero values at $y=0$ and can be (Taylor) expanded in powers of $y$. 
We consider again the general ansatz
\beqa
{\cal T}(y) = \sum_{n=0}^{\infty}  y^{\alpha_n} \, C_n  \,g_{n}(y)   \, ,
\label{IRTachyonDeconf}
\eeqa
where $\alpha_n$ are real numbers satisfying the inequality 
$\alpha_0 < \alpha_1 < \dots < \alpha_n < \alpha_{n+1} < \dots$ and $g_n(y)$ are analytic functions of $y$.

Plugging the ansatz \eqref{IRTachyonDeconf} into \eqref{specTeqDeconfv3}, we obtain
\begin{align}
&\sum_{n=0}^{\infty} y^{\alpha_n}  \Big \{  {\cal O}_n^2   + [{\cal P}(y) - 1 ] {\cal O}_n + {\cal Q}(y) y \Big \} g_n 
\nonumber \\[5pt]
&+ \sum_{n=0}^{\infty} \sum_{m=0}^{\infty} \sum_{\ell=0}^{\infty} 
y^{\alpha_n + \alpha_m + \alpha_{\ell}}   ( {\cal O}_n g_n ) ( {\cal O}_m g_m ) 
\Big[ y^{-1} R(y) ({\cal O}_{\ell} g_{\ell}) + g_{\ell} \Big] = 0 \,, 
\label{specTeqDeconfv4}
\end{align}
where ${\cal O}_n \equiv y \partial_y + \alpha_n$. 

Let us find the first exponent of the series, namely $\alpha_0$. There are 4 possible cases: 

\medskip 

\noindent \underline{Case I: $\alpha_0 <0$ \& $0 < \alpha_0 < 1/2$}

\medskip 

\noindent The first term in the series \eqref{specTeqDeconfv4} is of order $y^{3 \alpha_0 - 1}$ and we obtain the equation ${\cal O}_0 g_0 = 0$ with solution $g_0 \sim x^{- \alpha_0}$ which contradicts the assumption $g_0(0)=1$.

\medskip 

\noindent \underline{Case II: $\alpha_0=1/2$}

\medskip 

\noindent The first term in \eqref{specTeqDeconfv4} is of order $y^{1/2}$ and we obtain 
\beqa
\Big \{ {\cal O}_0^2 + [ P(y) - 1 ] {\cal O}_0 + Q(y) y \Big \} g_0 
+ R(y) ({\cal O}_0 g_0)^3 = 0 \, . 
\eeqa
Taking the limit $y \to 0$ and using the results $P(0)=1$ and $R(0)=5/6$ we find the condition $g_0(0)^2 = - 12/5$ which is not a valid (real) solution. 

\medskip 

\noindent \underline{Case III: $\alpha_0 > 1/2$}

\medskip 

\noindent The first term in \eqref{specTeqDeconfv4} is of order $y^{\alpha_0}$ and we obtain
\beqa
\Big \{ {\cal O}_0^2 + [ P(y) - 1 ] {\cal O}_0 + Q(y) y \Big \} g_0  = 0 \,.
\eeqa
Taking the limit $y \to 0$ and using $P(0)=1$ and $Q(0)=3/5$ we find
$g_0(0)=0$ which is not a valid solution. 

\medskip 

\noindent \underline{Case IV: $\alpha_0 =0$}

\medskip 

\noindent The first term in \eqref{specTeqDeconfv4} is of order $y^{-1}$ and we obtain the equation ${\cal O}_0 g_0 = y \partial_y g_0 = 0$ with solution $g_0(y)= {\rm const}$.  

We conclude from this analysis that $\alpha_0=0$. Let us now find the subleading exponent $\alpha_1$.  We will prove that $\alpha_1$ is an integer and the series \eqref{IRTachyonDeconf} actually reduces to an ordinary Taylor expansion. The proof is by contradiction. Assuming that $\alpha_1$ is not an integer we have 3 possibilities: $0 < \alpha_1 < 1/2$, $\alpha_1 = 1/2$ 
and $\alpha_1 > 1/2$. 

\medskip 

\noindent \underline{Case I: $0 < \alpha_1 < 1/2$}

\medskip 

\noindent  The second term in the series \eqref{specTeqDeconfv4} is of order $y^{2 \alpha_1 -1}$ and we obtain the equation ${\cal O}_1 g_1 =0$ with solution $g_1 \sim x^{- \alpha_1}$ that contradicts $g_1(0)=1$. 

\medskip 

\noindent \underline{Case II: $\alpha_1 = 1/2$}

\medskip 

\noindent The second term in the series  \eqref{specTeqDeconfv4} is of order $y^{1/2}$ and we obtain 
\beqa
\Big \{ {\cal O}_1^2 + [ P(y) - 1 ] {\cal O}_1 + Q(y) y \Big \} g_1 
+  R(y) ({\cal O}_1 g_1)^3 = 0 \, . 
\eeqa
Taking the limit $y \to 0$ and using $P(0)=1$ and $R(0)=5/6$ we obtain the condition $g_1(0)^2 = -12/5$ which is not a valid (real) solution.

\medskip 

\noindent \underline{Case III: $\alpha_1 > 1/2$}

\medskip 

\noindent The second term in \eqref{specTeqDeconfv4} is of order $y^{\alpha_1}$ and we obtain 
\beqa
\Big \{ {\cal O}_1^2 + [ P(y) - 1 ] {\cal O}_1 + Q(y) y \Big \} g_1  = 0 \,.
\eeqa
Taking the limit $y \to 0$ and using $P(0)=1$ and $Q(0)=3/5$ we find
$g_1(0)=0$ which is not a valid solution. 

From the analysis above we conclude that $\alpha_0=0$ and $\alpha_1$ is a (positive) integer. Therefore, it is reasonable to assume that the series \eqref{IRTachyonDeconf} reduces to an ordinary Taylor expansion
\begin{equation}
{\cal T}(y) = \sum_{n=0}^{\infty}  y^n \, C_n  \,.
\label{IRTachyonDeconfv2}
\end{equation}
Had we found a noninteger solution for $\alpha_1$ such that $0 < \alpha_1 <1$ that would have corresponded to the spurious case where ${\cal T}$ is not singular but has a singular derivative. 

Plugging the ansatz \eqref{IRTachyonDeconfv2} into eq. \eqref{specTeqDeconfv3} and using the Taylor expansions for $P(y)$, $Q(y)$ and $R(y)$ we obtain
\begin{align}
&\sum_{n=0}^{\infty}  y^n \, C_n \Bigg \{  n^2   + \Big [ \sum_{i=0}^{\infty}  {\cal P}_i \, y^i - 1 \Big ] n + \sum_{i=0}^{\infty} {\cal Q}_i \, y^{i+1} \Bigg \} \nonumber \\[5pt]
&+ \sum_{n=0}^{\infty} \sum_{m=0}^{\infty} \sum_{\ell=0}^{\infty} 
y^{n + m + \ell} \, C_n C_m C_{\ell} \,  n m \Bigg [  \ell  \sum_{i=0}^{\infty}  {\cal R}_i \, y^{i-1}  + 1 \Bigg ] = 0 \, . 
\label{specTeqDeconfv5}
\end{align}
The terms of order $y^{-1}$ and $y^0$ in \eqref{specTeqDeconfv5} vanish automatically whereas the the term of order $y$ leads to the condition
\begin{equation}
C_1 \, = \, - \,  Q_0 \, C_0 \, = \, - \,  \frac35 \, C_0 \,. 
\end{equation}
The next term in \eqref{specTeqDeconfv5} is of order $y^2$ and vanishes if
\begin{eqnarray}
C_2 &=& \frac14 \, Q_0 \, C_0 \, \Bigg [ P_1 + Q_0 - \frac{Q_1}{Q_0}
+ \Big ( R_0 - \frac{1}{Q_0} \Big ) \, Q_0^2 \, C_0^2  \Bigg] 
\nonumber \\[5pt]
&=& - \, \frac{3}{20} \, C_0 \Bigg [ \frac{17}{5} - \frac{1-{\cal B}^2 }{1+ {\cal B}^2 } + \frac{3}{10} \, C_0^2  \Bigg] \,,
\end{eqnarray}
where we have used the results $Q_0 = 3/5$, $Q_1/Q_0 = 4$, $R_0 = 5/6$ and $P_1 = (1 - {\cal B}^2)/(1 + {\cal B}^2)$.



\begin{thebibliography}{}

\bibitem{Iatrakis:2010jb} 
  I.~Iatrakis, E.~Kiritsis and A.~Paredes,
  ``An AdS/QCD model from tachyon condensation: II,''
  JHEP {\bf 1011}, 123 (2010)
  [arXiv:1010.1364 [hep-ph]].
  
  
 \bibitem{Manohar:2000dt}
A.~V.~Manohar and M.~B.~Wise,
``Heavy quark physics,''
Camb. Monogr. Part. Phys. Nucl. Phys. Cosmol. \textbf{10}

\bibitem{Scherer:2012xha}
S.~Scherer and M.~R.~Schindler,
Lect. Notes Phys. \textbf{830}, pp.1-338 (2012)

\bibitem{Roberts:1994dr}
C.~D.~Roberts and A.~G.~Williams,
``Dyson-Schwinger equations and their application to hadronic physics,''
Prog. Part. Nucl. Phys. \textbf{33}, 477-575 (1994)
[arXiv:hep-ph/9403224 [hep-ph]].

\bibitem{Maldacena:1997re}
J.~M.~Maldacena,
``The Large N limit of superconformal field theories and supergravity,''
Int. J. Theor. Phys. \textbf{38} (1999), 1113-1133
[arXiv:hep-th/9711200 [hep-th]].


\bibitem{Erdmenger:2007cm}
J.~Erdmenger, N.~Evans, I.~Kirsch and Threlfall,
Eur. Phys. J. A \textbf{35}, 81-133 (2008)
[arXiv:0711.4467 [hep-th]].

\bibitem{CasalderreySolana:2011us}
J.~Casalderrey-Solana, H.~Liu, D.~Mateos, K.~Rajagopal and U.~A.~Wiedemann,
``Gauge/String Duality, Hot QCD and Heavy Ion Collisions,''
[arXiv:1101.0618 [hep-th]].

\bibitem{Ramallo:2013bua}
A.~V.~Ramallo,
``Introduction to the AdS/CFT correspondence,''
Springer Proc.\ Phys.\  \textbf{161} (2015), 411-474
[arXiv:1310.4319 [hep-th]].


\bibitem{Edelstein:2009iv}
J.~D.~Edelstein, J.~P.~Shock and D.~Zoakos,
``The AdS/CFT Correspondence and Non-perturbative QCD,''
AIP Conf.\ Proc.\  \textbf{1116} (2009) no.1, 265-284
[arXiv:0901.2534 [hep-ph]].



\bibitem{Klebanov:2000hb}
I.~R.~Klebanov and M.~J.~Strassler,
``Supergravity and a confining gauge theory: Duality cascades and chi SB resolution of naked singularities,''
JHEP \textbf{08} (2000), 052
[arXiv:hep-th/0007191 [hep-th]].




\bibitem{Maldacena:2000mw}
J.~M.~Maldacena and C.~Nunez,
``Supergravity description of field theories on curved manifolds and a no go theorem,''
Int. J. Mod. Phys. A \textbf{16} (2001), 822-855
[arXiv:hep-th/0007018 [hep-th]].


\bibitem{Maldacena:2000yy}
J.~M.~Maldacena and C.~Nunez,
``Towards the large N limit of pure N=1 superYang-Mills,''
Phys. Rev. Lett. \textbf{86} (2001), 588-591
[arXiv:hep-th/0008001 [hep-th]].

\bibitem{Witten:1998zw}
E.~Witten,
``Anti-de Sitter space, thermal phase transition, and confinement in gauge theories,''
Adv. Theor. Math. Phys. \textbf{2} (1998), 505-532
[arXiv:hep-th/9803131 [hep-th]].

\bibitem{Karch:2002sh}
A.~Karch and E.~Katz,
``Adding flavor to AdS / CFT,''
JHEP \textbf{06} (2002), 043
[arXiv:hep-th/0205236 [hep-th]].


\bibitem{Nunez:2010sf}
C.~Nunez, A.~Paredes and A.~V.~Ramallo,
``Unquenched Flavor in the Gauge/Gravity Correspondence,''
Adv. High Energy Phys. \textbf{2010} (2010), 196714
[arXiv:1002.1088 [hep-th]].


\bibitem{Ramallo:2008ew}
A.~V.~Ramallo, J.~P.~Shock and D.~Zoakos,
``Holographic flavor in N=4 gauge theories in 3d from wrapped branes,''
JHEP \textbf{02} (2009), 001
[arXiv:0812.1975 [hep-th]].

\bibitem{Arean:2010hu}
D.~Arean, E.~Conde, A.~V.~Ramallo and D.~Zoakos,
``Holographic duals of SQCD models in low dimensions,''
JHEP \textbf{06} (2010), 095
[arXiv:1004.4212 [hep-th]].


\bibitem{Jokela:2012dw}
N.~Jokela, J.~Mas, A.~V.~Ramallo and D.~Zoakos,
``Thermodynamics of the brane in Chern-Simons matter theories with flavor,''
JHEP \textbf{02} (2013), 144
[arXiv:1211.0630 [hep-th]].

\bibitem{Itsios:2013uya}
G.~Itsios, V.~G.~Filev and D.~Zoakos,
``Backreacted flavor in non-commutative gauge theories,''
JHEP \textbf{06} (2013), 092
[arXiv:1304.5211 [hep-th]].

\bibitem{Filev:2014nza}
V.~G.~Filev and D.~Zoakos,
``Multiple backreacted flavour branes,''
JHEP \textbf{12} (2014), 186
[arXiv:1410.2879 [hep-th]].

\bibitem{Bea:2014yda}
Y.~Bea, N.~Jokela, M.~Lippert, A.~V.~Ramallo and D.~Zoakos,
``Flux and Hall states in ABJM with dynamical flavors,''
JHEP \textbf{03} (2015), 009
[arXiv:1411.3335 [hep-th]].



\bibitem{Mateos:2007vc}
D.~Mateos, S.~Matsuura, R.~C.~Myers and R.~M.~Thomson,
``Holographic phase transitions at finite chemical potential,''
JHEP \textbf{11} (2007), 085
[arXiv:0709.1225 [hep-th]].


\bibitem{Filev:2007gb}
V.~G.~Filev, C.~V.~Johnson, R.~Rashkov and K.~Viswanathan,
``Flavoured large N gauge theory in an external magnetic field,''
JHEP \textbf{10} (2007), 019
[arXiv:hep-th/0701001 [hep-th]].


\bibitem{Erdmenger:2007bn}
J.~Erdmenger, R.~Meyer and J.~P.~Shock,
``AdS/CFT with flavour in electric and magnetic Kalb-Ramond fields,''
JHEP \textbf{12} (2007), 091
[arXiv:0709.1551 [hep-th]].



\bibitem{Albash:2007bk} 
  T.~Albash, V.~G.~Filev, C.~V.~Johnson and A.~Kundu,
  ``Finite temperature large N gauge theory with quarks in an external magnetic field,''
  JHEP {\bf 0807}, 080 (2008)
  [arXiv:0709.1547 [hep-th]].
  

\bibitem{DHoker:2009mmn}
E.~D'Hoker and P.~Kraus,
``Magnetic Brane Solutions in AdS,''
JHEP \textbf{10}, 088 (2009)
[arXiv:0908.3875 [hep-th]].

\bibitem{Constable:1999ch}
N.~R.~Constable and R.~C.~Myers,
``Exotic scalar states in the AdS / CFT correspondence,''
JHEP \textbf{11} (1999), 020
[arXiv:hep-th/9905081 [hep-th]].

\bibitem{Evans:2004ia}
N.~J.~Evans and J.~P.~Shock,
``Chiral dynamics from AdS space,''
Phys. Rev. D \textbf{70}, 046002 (2004)
[arXiv:hep-th/0403279 [hep-th]].

\bibitem{Kruczenski:2003uq}
M.~Kruczenski, D.~Mateos, R.~C.~Myers and D.~J.~Winters,
``Towards a holographic dual of large N(c) QCD,''
JHEP \textbf{05}, 041 (2004)
[arXiv:hep-th/0311270 [hep-th]]

\bibitem{Sakai:2004cn}
T.~Sakai and S.~Sugimoto,
``Low energy hadron physics in holographic QCD,''
Prog. Theor. Phys. \textbf{113} (2005), 843-882
[arXiv:hep-th/0412141 [hep-th]].


\bibitem{Sakai:2005yt}
T.~Sakai and S.~Sugimoto,
``More on a holographic dual of QCD,''
Prog. Theor. Phys. \textbf{114} (2005), 1083-1118
[arXiv:hep-th/0507073 [hep-th]].

\bibitem{Babington:2003vm}
J.~Babington, J.~Erdmenger, N.~J.~Evans, Z.~Guralnik and I.~Kirsch,
``Chiral symmetry breaking and pions in nonsupersymmetric gauge / gravity duals,''
Phys. Rev. D \textbf{69} (2004), 066007
[arXiv:hep-th/0306018 [hep-th]].

\bibitem{Kuperstein:2008cq}
S.~Kuperstein and J.~Sonnenschein,
``A New Holographic Model of Chiral Symmetry Breaking,''
JHEP \textbf{09} (2008), 012
[arXiv:0807.2897 [hep-th]].

\bibitem{Klebanov:1998hh}
I.~R.~Klebanov and E.~Witten,
``Superconformal field theory on three-branes at a Calabi-Yau singularity,''
Nucl. Phys. B \textbf{536} (1998), 199-218
[arXiv:hep-th/9807080 [hep-th]].

\bibitem{Filev:2013vka}
V.~G.~Filev, M.~Ihl and D.~Zoakos,
``A Novel (2+1)-Dimensional Model of Chiral Symmetry Breaking,''
JHEP \textbf{12} (2013), 072
[arXiv:1310.1222 [hep-th]].

\bibitem{Filev:2014bna}
V.~G.~Filev, M.~Ihl and D.~Zoakos,
``Holographic Bilayer/Monolayer Phase Transitions,''
JHEP \textbf{07} (2014), 043
[arXiv:1404.3159 [hep-th]].

\bibitem{Iatrakis:2010zf} 
  I.~Iatrakis, E.~Kiritsis and A.~Paredes,
  ``An AdS/QCD model from Sen's tachyon action,''
  Phys.\ Rev.\ D {\bf 81}, 115004 (2010)
  [arXiv:1003.2377 [hep-ph]].
  


\bibitem{Casero:2007ae} 
  R.~Casero, E.~Kiritsis and A.~Paredes,
  ``Chiral symmetry breaking as open string tachyon condensation,''
  Nucl.\ Phys.\ B {\bf 787}, 98 (2007)
  [hep-th/0702155].
 


\bibitem{Erlich:2005qh}
J.~Erlich, E.~Katz, D.~T.~Son and M.~A.~Stephanov,
``QCD and a holographic model of hadrons,''
Phys. Rev. Lett. \textbf{95} (2005), 261602
[arXiv:hep-ph/0501128 [hep-ph]].

  
  
\bibitem{DaRold:2005mxj}
L.~Da Rold and A.~Pomarol,
``Chiral symmetry breaking from five dimensional spaces,''
Nucl. Phys. B \textbf{721} (2005), 79-97
[arXiv:hep-ph/0501218 [hep-ph]].



\bibitem{Polchinski:2001tt}
J.~Polchinski and M.~J.~Strassler,
``Hard scattering and gauge / string duality,''
Phys. Rev. Lett. \textbf{88} (2002), 031601
[arXiv:hep-th/0109174 [hep-th]].

\bibitem{Karch:2006pv}
A.~Karch, E.~Katz, D.~T.~Son and M.~A.~Stephanov,
``Linear confinement and AdS/QCD,''
Phys. Rev. D \textbf{74} (2006), 015005
[arXiv:hep-ph/0602229 [hep-ph]].

\bibitem{Gherghetta:2009ac} 
  T.~Gherghetta, J.~I.~Kapusta and T.~M.~Kelley,
  ``Chiral symmetry breaking in the soft-wall AdS/QCD model,''
  Phys.\ Rev.\ D {\bf 79}, 076003 (2009)
  [arXiv:0902.1998 [hep-ph]].

\bibitem{Chelabi:2015gpc}
K.~Chelabi, Z.~Fang, M.~Huang, D.~Li and Y.~Wu,
``Chiral Phase Transition in the Soft-Wall Model of AdS/QCD,''
JHEP \textbf{04}, 036 (2016)
[arXiv:1512.06493 [hep-ph]].

\bibitem{Ballon-Bayona:2020qpq}
A.~Ballon-Bayona and L.~A.~Mamani,
``Nonlinear realisation of chiral symmetry breaking in holographic soft wall models,''
[arXiv:2002.00075 [hep-ph]].

  
  
\bibitem{Jarvinen:2011qe}
M.~Jarvinen and E.~Kiritsis,
``Holographic Models for QCD in the Veneziano Limit,''
JHEP \textbf{03} (2012), 002
[arXiv:1112.1261 [hep-ph]].

\bibitem{Arean:2013tja}
D.~Arean, I.~Iatrakis, M.~Jarvinen and E.~Kiritsis,
``The discontinuities of conformal transitions and mass spectra of V-QCD,''
JHEP \textbf{11}, 068 (2013)
[arXiv:1309.2286 [hep-ph]].  

\bibitem{Jarvinen:2015ofa}
M.~Jarvinen,
``Massive holographic QCD in the Veneziano limit,''
JHEP \textbf{07} (2015), 033
[arXiv:1501.07272 [hep-ph]].

\bibitem{Gursoy:2007cb}
U.~Gursoy and E.~Kiritsis,
``Exploring improved holographic theories for QCD: Part I,''
JHEP \textbf{02} (2008), 032
[arXiv:0707.1324 [hep-th]].


\bibitem{Gursoy:2007er}
U.~Gursoy, E.~Kiritsis and F.~Nitti,
``Exploring improved holographic theories for QCD: Part II,''
JHEP \textbf{02} (2008), 019
[arXiv:0707.1349 [hep-th]].

\bibitem{Gursoy:2009jd}
U.~Gursoy, E.~Kiritsis, L.~Mazzanti and F.~Nitti,
``Improved Holographic Yang-Mills at Finite Temperature: Comparison with Data,''
Nucl.\ Phys.\ B \textbf{820} (2009), 148-177
[arXiv:0903.2859 [hep-th]].

\bibitem{Gursoy:2010fj} 
  U.~Gursoy, E.~Kiritsis, L.~Mazzanti, G.~Michalogiorgakis and F.~Nitti,
  ``Improved Holographic QCD,''
  Lect.\ Notes Phys.\  {\bf 828}, 79 (2011)
 [arXiv:1006.5461 [hep-th]].
 
\bibitem{Bigazzi:2005md}
F.~Bigazzi, R.~Casero, A.~Cotrone, E.~Kiritsis and A.~Paredes,
``Non-critical holography and four-dimensional CFT's with fundamentals,''
JHEP \textbf{10} (2005), 012
[arXiv:hep-th/0505140 [hep-th]].

\bibitem{Sen:2003tm} 
  A.~Sen,
  ``Dirac-Born-Infeld action on the tachyon kink and vortex,''
  Phys.\ Rev.\ D {\bf 68}, 066008 (2003)
  [hep-th/0303057].
  
\bibitem{Jokela:2018ers}
N.~Jokela, M.~Jarvinen and J.~Remes,
``Holographic QCD in the Veneziano limit and neutron stars,''
JHEP \textbf{03} (2019), 041
[arXiv:1809.07770 [hep-ph]].
 
  
  
  
\bibitem{Gusynin:1994re}
V.~Gusynin, V.~Miransky and I.~Shovkovy,
``Catalysis of dynamical flavor symmetry breaking by a magnetic field in (2+1)-dimensions,''
Phys.\ Rev.\ Lett.\  \textbf{73} (1994), 3499-3502
[arXiv:hep-ph/9405262 [hep-ph]].

\bibitem{Gusynin:1994xp}
V.~Gusynin, V.~Miransky and I.~Shovkovy,
``Dimensional reduction and dynamical chiral symmetry breaking by a magnetic field in (3+1)-dimensions,''
Phys.\ Lett.\ B \textbf{349} (1995), 477-483
[arXiv:hep-ph/9412257 [hep-ph]].

\bibitem{Bali:2011qj} 
  G.~S.~Bali, F.~Bruckmann, G.~Endrodi, Z.~Fodor, S.~D.~Katz, S.~Krieg, A.~Schafer and K.~K.~Szabo,
  ``The QCD phase diagram for external magnetic fields,''
  JHEP {\bf 1202}, 044 (2012)
  [arXiv:1111.4956 [hep-lat]].
  
  
\bibitem{Bali:2012zg}
G.~Bali, F.~Bruckmann, G.~Endrodi, Z.~Fodor, S.~Katz and A.~Schafer,
``QCD quark condensate in external magnetic fields,''
Phys.\ Rev.\ D \textbf{86} (2012), 071502
[arXiv:1206.4205 [hep-lat]].


\bibitem{DElia:2012ems}
M.~D'Elia,
``Lattice QCD Simulations in External Background Fields,''
Lect.\ Notes Phys.\  \textbf{871} (2013), 181-208
[arXiv:1209.0374 [hep-lat]].



\bibitem{Gursoy:2014aka}
U.~Gursoy, D.~Kharzeev and K.~Rajagopal,
``Magnetohydrodynamics, charged currents and directed flow in heavy ion collisions,''
Phys.\ Rev.\ C \textbf{89} (2014) no.5, 054905
[arXiv:1401.3805 [hep-ph]].

\bibitem{Banks:1979yr}
T.~Banks and A.~Casher,
``Chiral Symmetry Breaking in Confining Theories,''
Nucl. Phys. B \textbf{169}, 103-125 (1980)


\bibitem{Bruckmann:2013oba}
F.~Bruckmann, G.~Endrodi and T.~G.~Kovacs,
``Inverse magnetic catalysis and the Polyakov loop,''
JHEP \textbf{04} (2013), 112
[arXiv:1303.3972 [hep-lat]].

\bibitem{Andersen:2014xxa}
J.~O.~Andersen, W.~R.~Naylor and A.~Tranberg,
``Phase diagram of QCD in a magnetic field: A review,''
Rev. Mod. Phys. \textbf{88}, 025001 (2016)
[arXiv:1411.7176 [hep-ph]].

\bibitem{Miransky:2015ava}
V.~A.~Miransky and I.~A.~Shovkovy,
``Quantum field theory in a magnetic field: From quantum chromodynamics to graphene and Dirac semimetals,''
Phys. Rept. \textbf{576}, 1-209 (2015)
[arXiv:1503.00732 [hep-ph]].

\bibitem{Bandyopadhyay:2020zte}
A.~Bandyopadhyay and R.~L.~Farias,
``Inverse magnetic catalysis -- how much do we know about?,''
[arXiv:2003.11054 [hep-ph]].

\bibitem{Preis:2012fh}
F.~Preis, A.~Rebhan and A.~Schmitt,
``Inverse magnetic catalysis in field theory and gauge-gravity duality,''
Lect. Notes Phys. \textbf{871}, 51-86 (2013)
[arXiv:1208.0536 [hep-ph]].

\bibitem{Johnson:2008vna}
C.~V.~Johnson and A.~Kundu,
``External Fields and Chiral Symmetry Breaking in the Sakai-Sugimoto Model,''
JHEP \textbf{12}, 053 (2008)
[arXiv:0803.0038 [hep-th]].

\bibitem{Filev:2009xp}
V.~G.~Filev, C.~V.~Johnson and J.~P.~Shock,
``Universal Holographic Chiral Dynamics in an External Magnetic Field,''
JHEP \textbf{08}, 013 (2009)
[arXiv:0903.5345 [hep-th]].

\bibitem{Preis:2010cq}
F.~Preis, A.~Rebhan and A.~Schmitt,
``Inverse magnetic catalysis in dense holographic matter,''
JHEP \textbf{03} (2011), 033
[arXiv:1012.4785 [hep-th]].


\bibitem{Filev:2011mt}
V.~G.~Filev and D.~Zoakos,
``Towards Unquenched Holographic Magnetic Catalysis,''
JHEP \textbf{08} (2011), 022
[arXiv:1106.1330 [hep-th]].

\bibitem{Erdmenger:2011bw}
J.~Erdmenger, V.~G.~Filev and D.~Zoakos,
``Magnetic Catalysis with Massive Dynamical Flavours,''
JHEP \textbf{08} (2012), 004
[arXiv:1112.4807 [hep-th]].


\bibitem{Ballon-Bayona:2013cta}
A.~Ballon-Bayona,
``Holographic deconfinement transition in the presence of a magnetic field,''
JHEP \textbf{11}, 168 (2013)
[arXiv:1307.6498 [hep-th]].

\bibitem{Jokela:2013qya}
N.~Jokela, A.~V.~Ramallo and D.~Zoakos,
``Magnetic catalysis in flavored ABJM,''
JHEP \textbf{02} (2014), 021
[arXiv:1311.6265 [hep-th]].

\bibitem{Mamo:2015dea}
K.~A.~Mamo,
``Inverse magnetic catalysis in holographic models of QCD,''
JHEP \textbf{05}, 121 (2015)
[arXiv:1501.03262 [hep-th]].

\bibitem{Rougemont:2015oea}
R.~Rougemont, R.~Critelli and J.~Noronha,
``Holographic calculation of the QCD crossover temperature in a magnetic field,''
Phys. Rev. D \textbf{93}, no.4, 045013 (2016)
[arXiv:1505.07894 [hep-th]].

\bibitem{Dudal:2015wfn}
D.~Dudal, D.~R.~Granado and T.~G.~Mertens,
``No inverse magnetic catalysis in the QCD hard and soft wall models,''
Phys. Rev. D \textbf{93}, no.12, 125004 (2016)
[arXiv:1511.04042 [hep-th]].

\bibitem{Evans:2016jzo}
N.~Evans, C.~Miller and M.~Scott,
``Inverse Magnetic Catalysis in Bottom-Up Holographic QCD,''
Phys.\ Rev.\ D \textbf{94} (2016) no.7, 074034
[arXiv:1604.06307 [hep-ph]].

\bibitem{Fang:2016cnt}
Z.~Fang,
``Anomalous dimension, chiral phase transition and inverse magnetic catalysis in soft-wall AdS/QCD,''
Phys. Lett. B \textbf{758}, 1-8 (2016)

\bibitem{Gursoy:2016ofp}
U.~Gursoy, I.~Iatrakis, M.~Jarvinen and G.~Nijs,
``Inverse Magnetic Catalysis from improved Holographic QCD in the Veneziano limit,''
JHEP \textbf{03} (2017), 053
[arXiv:1611.06339 [hep-th]].

\bibitem{Ballon-Bayona:2017dvv}
A.~Ballon-Bayona, M.~Ihl, J.~P.~Shock and D.~Zoakos,
``A universal order parameter for Inverse Magnetic Catalysis,''
JHEP \textbf{10} (2017), 038
[arXiv:1706.05977 [hep-th]].

\bibitem{Gursoy:2017wzz}
U.~Gursoy, M.~Jarvinen and G.~Nijs,
``Holographic QCD in the Veneziano Limit at a Finite Magnetic Field and Chemical Potential,''
Phys.\ Rev.\ Lett.\  \textbf{120} (2018) no.24, 242002
[arXiv:1707.00872 [hep-th]].


\bibitem{Rodrigues:2017iqi}
D.~M.~Rodrigues, E.~Folco Capossoli and H.~Boschi-Filho,
``Magnetic catalysis and inverse magnetic catalysis in ( 2+1 )-dimensional gauge theories from holographic models,''
Phys.\ Rev.\ D \textbf{97} (2018) no.12, 126001
[arXiv:1710.07310 [hep-th]].


\bibitem{Giataganas:2017koz}
D.~Giataganas, U.~Gursoy and J.~F.~Pedraza,
``Strongly-coupled anisotropic gauge theories and holography,''
Phys.\ Rev.\ Lett.\  \textbf{121} (2018) no.12, 121601
[arXiv:1708.05691 [hep-th]].


\bibitem{Gursoy:2018ydr}
U.~Gursoy, M.~Jarvinen, G.~Nijs and J.~F.~Pedraza,
``Inverse Anisotropic Catalysis in Holographic QCD,''
JHEP \textbf{04} (2019), 071
[arXiv:1811.11724 [hep-th]].


\bibitem{Filev:2019bll}
V.~G.~Filev and R.~Rashkov,
``Critical point in a holographic defect field theory,''
JHEP \textbf{11} (2019), 027
[arXiv:1905.06472 [hep-th]].


\bibitem{Bohra:2019ebj}
H.~Bohra, D.~Dudal, A.~Hajilou and S.~Mahapatra,
``Anisotropic string tensions and inversely magnetic catalyzed deconfinement from a dynamical AdS/QCD model,''
Phys. Lett. B \textbf{801} (2020), 135184
[arXiv:1907.01852 [hep-th]].

\bibitem{He:2020fdi}
S.~He, Y.~Yang and P.~Yuan,
``Analytic Study of Magnetic Catalysis in Holographic QCD,''
[arXiv:2004.01965 [hep-th]].

\bibitem{Mateos:2011tv}
D.~Mateos and D.~Trancanelli,
``Thermodynamics and Instabilities of a Strongly Coupled Anisotropic Plasma,''
JHEP \textbf{07} (2011), 054
[arXiv:1106.1637 [hep-th]].

\bibitem{Ammon:2012qs}
M.~Ammon, V.~G.~Filev, J.~Tarrio and D.~Zoakos,
``D3/D7 Quark-Gluon Plasma with Magnetically Induced Anisotropy,''
JHEP \textbf{09} (2012), 039
[arXiv:1207.1047 [hep-th]].


\bibitem{Conde:2016hbg}
E.~Conde, H.~Lin, J.~M.~Penin, A.~V.~Ramallo and D.~Zoakos,
``D3--D5 theories with unquenched flavors,''
Nucl.\ Phys.\ B \textbf{914} (2017), 599-622
[arXiv:1607.04998 [hep-th]].

\bibitem{Penin:2017lqt}
J.~M.~Penin, A.~V.~Ramallo and D.~Zoakos,
``Anisotropic D3-D5 black holes with unquenched flavors,''
JHEP \textbf{02} (2018), 139
[arXiv:1710.00548 [hep-th]].


\bibitem{Jokela:2019tsb}
N.~Jokela, J.~M.~Penin, A.~V.~Ramallo and D.~Zoakos,
``Gravity dual of a multilayer system,''
JHEP \textbf{03} (2019), 064
[arXiv:1901.02020 [hep-th]].




  \bibitem{Garousi:2004rd} 
  M.~R.~Garousi,
  ``D-brane anti-D-brane effective action and brane interaction in open string channel,''
  JHEP {\bf 0501}, 029 (2005)
  [hep-th/0411222].
  
  
\bibitem{Coleman:1980mx} 
  S.~R.~Coleman and E.~Witten,
  ``Chiral Symmetry Breakdown in Large N Chromodynamics,''
  Phys.\ Rev.\ Lett.\  {\bf 45}, 100 (1980).
  
\bibitem{Kutasov:2000aq} 
  D.~Kutasov, M.~Marino and G.~W.~Moore,
  ``Remarks on tachyon condensation in superstring field theory,''
  hep-th/0010108.
  
\bibitem{Minahan:2000tf} 
  J.~A.~Minahan and B.~Zwiebach,
  ``Effective tachyon dynamics in superstring theory,''
  JHEP {\bf 0103}, 038 (2001)
  [hep-th/0009246].
  
\bibitem{BallonBayona:2013gx} 
  A.~Ballon-Bayona, C.~N.~Ferreira and V.~J.~V.~Otoya,
  ``DBI equations and holographic DC conductivity,''
  Phys.\ Rev.\ D {\bf 87}, no. 10, 106007 (2013)
  [arXiv:1302.0802 [hep-th]].
  
  
   \bibitem{Kinar:1998vq} 
  Y.~Kinar, E.~Schreiber and J.~Sonnenschein,
  ``Q anti-Q potential from strings in curved space-time: Classical results,''
  Nucl.\ Phys.\ B {\bf 566}, 103 (2000)
  [hep-th/9811192].
  
  
  
\bibitem{Witten:1998qj} 
  E.~Witten,
  ``Anti-de Sitter space and holography,''
  Adv.\ Theor.\ Math.\ Phys.\  {\bf 2}, 253 (1998)
  [hep-th/9802150].
  
  
  
\bibitem{Kuperstein:2004yf} 
  S.~Kuperstein and J.~Sonnenschein,
  ``Non-critical, near extremal AdS(6) background as a holographic laboratory of four dimensional YM theory,''
  JHEP {\bf 0411}, 026 (2004)
  [hep-th/0411009].


\bibitem{Mandal:2011ws}
G.~Mandal and T.~Morita,
``Gregory-Laflamme as the confinement/deconfinement transition in holographic QCD,''
JHEP \textbf{09}, 073 (2011)
[arXiv:1107.4048 [hep-th]].

  
\bibitem{Bali:2013owa}
G.~Bali, F.~Bruckmann, G.~Endrodi and A.~Schafer,
``Paramagnetic squeezing of QCD matter,''
Phys.\ Rev.\ Lett.\  \textbf{112} (2014), 042301
[arXiv:1311.2559 [hep-lat]].

\bibitem{Bali:2014kia}
G.~Bali, F.~Bruckmann, G.~Endrodi, S.~Katz and A.~Schafer,
``The QCD equation of state in background magnetic fields,''
JHEP \textbf{08}, 177 (2014)
[arXiv:1406.0269 [hep-lat]].

\bibitem{DElia:2018xwo}
M.~D'Elia, F.~Manigrasso, F.~Negro and F.~Sanfilippo,
``QCD phase diagram in a magnetic background for different values of the pion mass,''
Phys. Rev. D \textbf{98}, no.5, 054509 (2018)
[arXiv:1808.07008 [hep-lat]]

\bibitem{Bali:2020bcn}
G.~S.~Bali, G.~Endrodi and S.~Piemonte,
``Magnetic susceptibility of QCD matter and its decomposition from the lattice,''
[arXiv:2004.08778 [hep-lat]].


\bibitem{Cherman:2008eh}
A.~Cherman, T.~D.~Cohen and E.~S.~Werbos,
``The Chiral condensate in holographic models of QCD,''
Phys.\ Rev.\ C \textbf{79}, 045203 (2009)
[arXiv:0804.1096 [hep-ph]].

\bibitem{Bergman:2007pm}
O.~Bergman, S.~Seki and J.~Sonnenschein,
``Quark mass and condensate in HQCD,''
JHEP \textbf{12}, 037 (2007)
[arXiv:0708.2839 [hep-th]].

\bibitem{Aharony:2008an}
O.~Aharony and D.~Kutasov,
``Holographic Duals of Long Open Strings,''
Phys. Rev. D \textbf{78}, 026005 (2008)
[arXiv:0803.3547 [hep-th]].

\bibitem{Gellman:1968}
M.~Gell-Mann,  R.J.~Oakes,  B.~Renner,  ``Behavior of Current  Divergences  Under $SU(3)\times SU(3)$,'' 
Phys. Rev.175(1968) 2195.


\bibitem{Gasser:1984}
J.~Gasser, H.~Leutwyler, 
``Low Energy Expansion of Meson Form Factors,''
Nucl. Phys.B250(1985) 517.
 
\bibitem{Witten:1998xy}
E.~Witten,
``Baryons and branes in anti-de Sitter space,''
JHEP \textbf{07} (1998), 006
[arXiv:hep-th/9805112 [hep-th]].


\bibitem{Seo:2009kg}
Y.~Seo, J.~P.~Shock, S.~Sin and D.~Zoakos,
``Holographic Hadrons in a Confining Finite Density Medium,''
JHEP \textbf{03} (2010), 115
[arXiv:0912.4013 [hep-th]].

\bibitem{Evans:2012cx}
N.~Evans, K.~Kim, M.~Magou, Y.~Seo and S.~Sin,
``The Baryonic Phase in Holographic Descriptions of the QCD Phase Diagram,''
JHEP \textbf{09} (2012), 045
[arXiv:1204.5640 [hep-th]].

\bibitem{Hata:2007mb}
H.~Hata, T.~Sakai, S.~Sugimoto and S.~Yamato,
``Baryons from instantons in holographic QCD,''
Prog. Theor. Phys. \textbf{117}, 1157 (2007)
[arXiv:hep-th/0701280 [hep-th]].

\bibitem{Bolognesi:2013nja}
S.~Bolognesi and P.~Sutcliffe,
``The Sakai-Sugimoto soliton,''
JHEP \textbf{01}, 078 (2014)
[arXiv:1309.1396 [hep-th]].

\bibitem{Preis:2011sp}
F.~Preis, A.~Rebhan and A.~Schmitt,
``Holographic baryonic matter in a background magnetic field,''
J. Phys. G \textbf{39}, 054006 (2012)
[arXiv:1109.6904 [hep-th]].

\bibitem{Pomarol:2008aa}
A.~Pomarol and A.~Wulzer,
``Baryon Physics in Holographic QCD,''
Nucl. Phys. B \textbf{809} (2009), 347-361
[arXiv:0807.0316 [hep-ph]].

\bibitem{Ishii:2019gta}
T.~Ishii, M.~Jarvinen and G.~Nijs,
``Cool baryon and quark matter in holographic QCD,''
JHEP \textbf{07} (2019), 003
[arXiv:1903.06169 [hep-ph]].

  
   
  
\end{thebibliography}
\end{document}